\documentclass[fleqn,usenatbib]{mnras}
\usepackage{endnotes}
\usepackage{pdflscape}
\usepackage{multirow}
\usepackage{booktabs}
\usepackage{xcolor}
\usepackage{float}
\usepackage{caption}
\usepackage{CJKutf8}
\usepackage{hyperref}
\usepackage{threeparttable}
\usepackage{newtxtext,newtxmath}
\usepackage[figuresright]{rotating}
\usepackage[T1]{fontenc}
\usepackage{multicol}        
\usepackage{bm}
\DeclareRobustCommand{\VAN}[3]{#2}
\let\VANthebibliography\thebibliography
\def\thebibliography{\DeclareRobustCommand{\VAN}[3]{##3}\VANthebibliography}
\usepackage{graphicx}	
\usepackage{amsmath}

\usepackage{rotating} 
\usepackage{longtable} 
\usepackage{rotfloat} 
\title[Extremely Low Mass Ratio Contact Binaries]{Extremely Low Mass Ratio Contact Binaries. \uppercase\expandafter{\romannumeral2}. The First Photometric and Spectroscopic Investigations of Six Systems with Orbital Periods Longer than 0.5 days}

\author[Liu et al.]{
Fei Liu$^{1}$,
Kai Li$^{1}$,\thanks{E-mail: kaili@sdu.edu.cn}
Xiang Gao$^{1}$,
Jing-Yi Wang$^{1}$,
Xin Xu$^{1}$,
Yi-Fan Wang$^{1}$,
Cheng-Yu Wu$^{1}$,
Mu-Zi-Mei Li$^{1}$,
\newauthor{
Xing Gao$^{2}$,
and Guo-You Sun$^{3}$
}
\\
$^{1}$Shandong Key Laboratory of Optical Astronomy and Solar-Terrestrial Environment, School of Space Science and Physics, \\
Institute of Space Sciences, Shandong University, Weihai, Shandong, 264209, China.\\
$^{2}$Xinjiang Astronomical Observatory, 150 Science 1-Street, Urumqi 830011, Peopleʼs Republic of China.\\
$^{3}$Xingming Observatory, Urumqi 830002, Xinjiang, People’s Republic of China.
}

\date{Accepted XXX. Received YYY; in original form ZZZ}

\pubyear{XXXX}

\begin{document}
\begin{CJK}{UTF8}{gbsn}
\label{firstpage}
\pagerange{\pageref{firstpage}--\pageref{lastpage}}
\maketitle

\begin{abstract}
The photometric and spectroscopic studies of six contact binaries were performed for the first time. The orbital periods of all the six targets are longer than 0.5d, and we discovered that their mass ratios are smaller than 0.15. So, they are extremely low mass-ratio contact binaries. Only one target is a W-subtype contact binary (ASASSN-V J105032.88+420829.0), while the others are A-subtype contact binaries. From orbital period analysis, ASASSN-V J075442.44+555623.2 shows no orbital period change. Three of the six targets demonstrate a secular period increase, and two targets for a secular period decrease. We investigated the LAMOST spectra employing the spectral subtraction method. All six contact binaries show no chromospheric emission line, implying no chromospheric activity. Their absolute parameters, initial masses, ages, energy transfer parameters, and instability parameters were calculated. The bolometric luminosity ratios ($(L_2/L_1)_{bol}$), the energy transfer parameters ($\beta$), the contact degrees ($f$), and the mass ratios ($q$) were collected for a sample of 218 contact binaries and we analyzed and discussed some correlations. The results by analyzing the relation between $\beta$, $f$ and $q$ indicate that the energy transfer parameter between the two components of extremely low mass-ratio contact binaries is independent of the contact degree. And the predicted cutoff mass ratio was estimated as 0.021 by analyzing the relation between $f$ and $q$.
\end{abstract}
\begin{keywords}
binaries: close-binaries:eclipsing-stars:evolution-stars:individual.
\end{keywords}



\section{Introduction}
\label{sec:Introduction}

Contact binaries are systems composed of two components with a convective envelope \citep{1968ApJ...151.1123L}, indicating that the surface potentials of each component are the same ($\Omega=\Omega_{1}=\Omega_{2}$). Both components in a contact binary usually have spectral types F, G, and K, although contact binaries with much earlier spectral types are known \citep{1993ASSL..177..111R}. The light curves of W UMa type binaries show a continuous brightness variation owing to the highly distorted shapes of each component and mutual eclipses \citep{1999anmi.conf...59M}. In most cases, the effective temperatures of both components are roughly the same. Thus, the depths of the primary and secondary eclipses are often approximately the same \citep{1941ApJ....93..133K, 1967AJ.....72Q.813L}. The periods of most contact binaries are less than 0.7 days \citep{2001icbs.book.....H}. Contact binaries are categorized into the A-subtype and the W-subtype \citep{BINNENDIJK1970217}. The A-subtype refers to the situation in which the more massive component has a higher temperature than the less massive component. On the contrary, the more massive component has a lower temperature than the less massive component in a W-subtype system.   

Contact binaries are believed to originate from short-period detached binaries through angular momentum loss due to magnetic braking \citep{1988ASIC..241..345G, 1994ASPC...56..228B} and evolve to coalesce into a fast rotating single star ultimately \citep{2006Ap&SS.304...25Q}. Photometric and spectroscopic observations are critical for deriving the physical parameters and investigating the structure and evolution of contact binaries. The mass ratio is one of the most important physical parameters for representing the evolutionary state. The cutoff mass ratio of contact binaries is still controversial regarding the merging conditions \citep{1995ApJ...444L..41R, 10.1111/j.1365-2966.2006.10462.x, 2022AJ....164..202L, 2024A&A...692L...4L, 10.1093/mnras/stac2811, 10.1093/mnras/stad417, 10.1093/mnras/stad026}. The radial velocity and light curves are the two main methods to determine a contact binary's mass ratio. However, it is tough to obtain the radial velocities of both components for extremely low mass-ratio contact binaries (ELMRCBs) due to the faintness of the less massive component and the blend of spectra \citep{2019AJ....158..186K}. Based on statistical analysis of contact binaries observed in photometry and spectroscopy simultaneously, \citet{2003CoSka..33...38P} proposed that mass ratios of photometry equal that of spectroscopy approximately for totally eclipsing contact binaries. Other statistical studies and simulated results derived the same conclusions (e.g., \citealt{2021ApJS..254...10L, 2021AJ....162...13L}). So, we can obtain relatively accurate parameters with the photometric light curves for totally eclipsing contact binaries that lack radial velocity observations. 

Magnetic activities in the chromosphere and photosphere are common in late-type stars, which may have noticeable effects on their convective motion \citep{2009A&ARv..17..251S}. Contact binaries are usually composed of two late-type components, which exhibit chromospheric emission lines normally, such as H$\alpha$, H$_{\beta}$, H$\gamma$, Ca \uppercase\expandafter{\romannumeral2} H \& K, and Ca \uppercase\expandafter{\romannumeral2} IRT, etc \citep{2018A&A...615A.120P, 2021MNRAS.506.4251Z}. Besides, magnetic activities may result in the O’Connell effect, which refers to the apparent light curve asymmetry \citep{1951PRCO....2...85O}. There are four main explanations for the mechanism of the O’Connell effect: cool spot with magnetic activity, hot spot due to mass transfer, circumstellar material \citep{2003ChJAA...3..142L}, and asymmetry of circumfluence due to the Coriolis effect \citep{1990ApJ...355..271Z}. 

The relation between the mass ratio $q$ and the orbital period $P$ is essential to establish evolutionary theories of contact binaries. Based on the conservation of mass and angular momentum, \citet{2019AAS...23344805M} found that the cutoff mass ratio $q_{min}$ is related to the orbital period for a contact binary system. \citet{2022ApJS..262...12K} proposed that the $q_{min}$ is getting greater when the orbital period of a contact binary is longer regarding the Darwin instability \citep{1893Obs....16..172D}. \citet{2022ApJS..262...12K} suggested that true contact binaries become rare at these longer periods ($P > 0.5d$). So extremely low mass-ratio contact binaries with orbital periods longer than 0.5 days (ELMRCBs with $P > 0.5d$), whose mass ratios are smaller than 0.15 ($q < 0.15$, \citealt{2022AJ....164..202L}), are important systems. However, the evolutionary mechanism of these systems has yet to be determined due to the small sample sizes. 

ELMRCBs are of interest because they may be the progenitors of blue stragglers, or FK Com-type single rapidly rotating stars \citep{2012JASS...29..145E, 2015AJ....150...69Y, 2022AJ....164..202L}. \citet{2001AJ..122...1007} noted a correlation between the light curve amplitude and the mass ratio where a smaller mass ratio implies a smaller light curve amplitude. So observations on a number of small amplitude totally eclipsing contact binaries, which were selected from the variable star catalog of All Sky Automated Survey for SuperNovae (ASAS-SN\footnote{\url{https://asas-sn.osu.edu}}, \citealt{2014ApJ...788...48S}), have been carried out by us since 2019. This paper is the second contribution in the series of \textit{Extremely Low Mass Ratio Contact Binaries} (\citealt{2022AJ....164..202L}, hereinafter Paper \uppercase\expandafter{\romannumeral1}). In this paper, we selected six contact binaries with orbital periods longer than 0.5d and small light variability amplitudes. Their basic parameters are listed in Table \ref{tab:Brief Parametric Introduction}.

\begin{table*}
	\centering
	\caption{Basic information of the six targets.}
	\label{tab:Brief Parametric Introduction}
	\begin{tabular}{lcccccccc}
		\hline
	      Target & Hereinafter & Other Name &   Period  &  Mean VMag  & Amplitude&  B-V  & J-K   & Reference\\
	              &           &            &     (d)     &  (mag)      &   (mag)  & (mag) & (mag) &      \\
		\hline
            ASASSN-V J063344.02+534647.3 & J063344 & CSS\_J063343.9+534648 & 0.5702129 &  12.90 & 0.21 & 0.644 & 0.241  & (1)\\
	    ASASSN-V J073647.29+492044.5 & J073647 & GS Lyn                & 0.5584836 &  13.29 & 0.29 & 0.589 & 0.237  & (1)\\
            ASASSN-V J075442.44+555623.2 & J075442 & CSS\_J075442.6+555622 & 0.5072806 &  14.30 & 0.25 & 0.433 & 0.259  & (1) \\
	    ASASSN-V J094123.98+234533.6 & J094123 & ASAS J094124+2345.6   & 0.6181188 &  12.81 & 0.30 & 0.484 & 0.231  & (1)   \\
            ASASSN-V J105032.88+420829.0 & J105032 & V0377 UMa             & 0.5769032 &  12.49 & 0.31 & 0.384 & 0.212  & (1)\\
	    ASASSN-V J163001.36+544555.5 & J163001 & T-Dra0-02224          & 0.5669059 &  12.85 & 0.24 & 0.452 & 0.244  & (1) \\
		\hline
	\end{tabular}
         \begin{tablenotes}
        \item References.(1) From The ASAS-SN Catalog of Variable Stars II: \citet{2019MNRAS.486.1907J}.
        \end{tablenotes}
\end{table*}

\section{Observation} \label{sec:Observation}
\subsection{Photometric Observations}
We observed J063344 using the 60 cm telescope at the Xinglong Station of National Astronomical Observatories (XL60) with the Andor DU934P camera. Its viewing field is $18\,arcmin \times18\,arcmin$. We also carried out observations on J073647, J075442, J094123, J105032, and J163001 using the Ningbo Bureau of Education and Xinjiang Observatory Telescope (NEXT) with a focal ratio of F/8. Equipped with a $60 cm$ caliber and an FLI PL23042 back-illuminated CCD whose pixel size is $2k\times2k$, the viewing field of NEXT is up to $22\,arcmin \times22\,arcmin$. Table \ref{tab:Observation details of the six targets} lists the detailed information of our observations. 

\begin{figure*}
	\centering
	\begin{minipage}[t]{0.32\linewidth}
		\centering
		\includegraphics[width=2.2in]{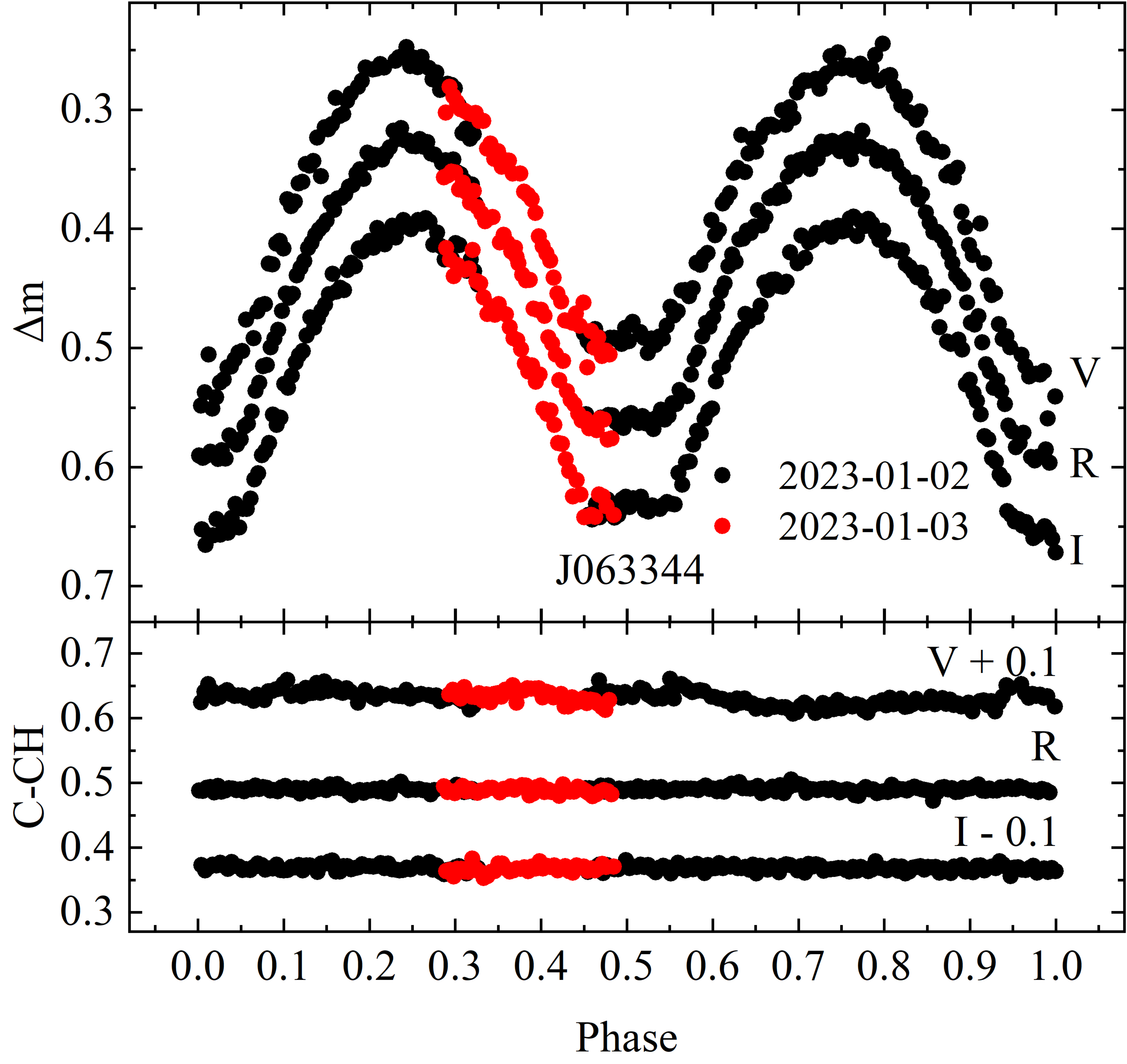}
	\end{minipage}%
	\begin{minipage}[t]{0.32\linewidth}
		\centering
		\includegraphics[width=2.2in]{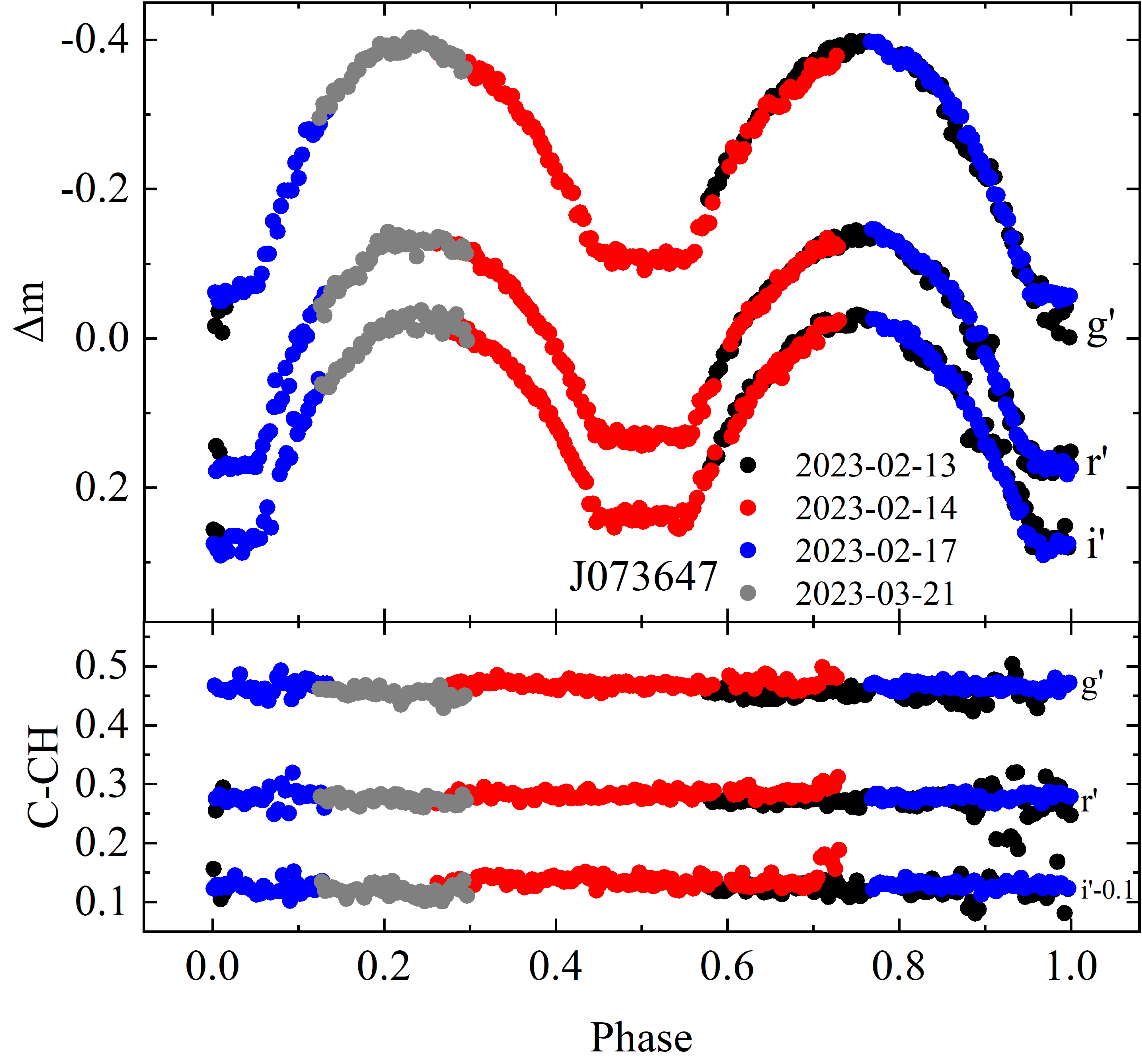}
	\end{minipage}
	\begin{minipage}[t]{0.32\linewidth}
		\centering
		\includegraphics[width=2.2in]{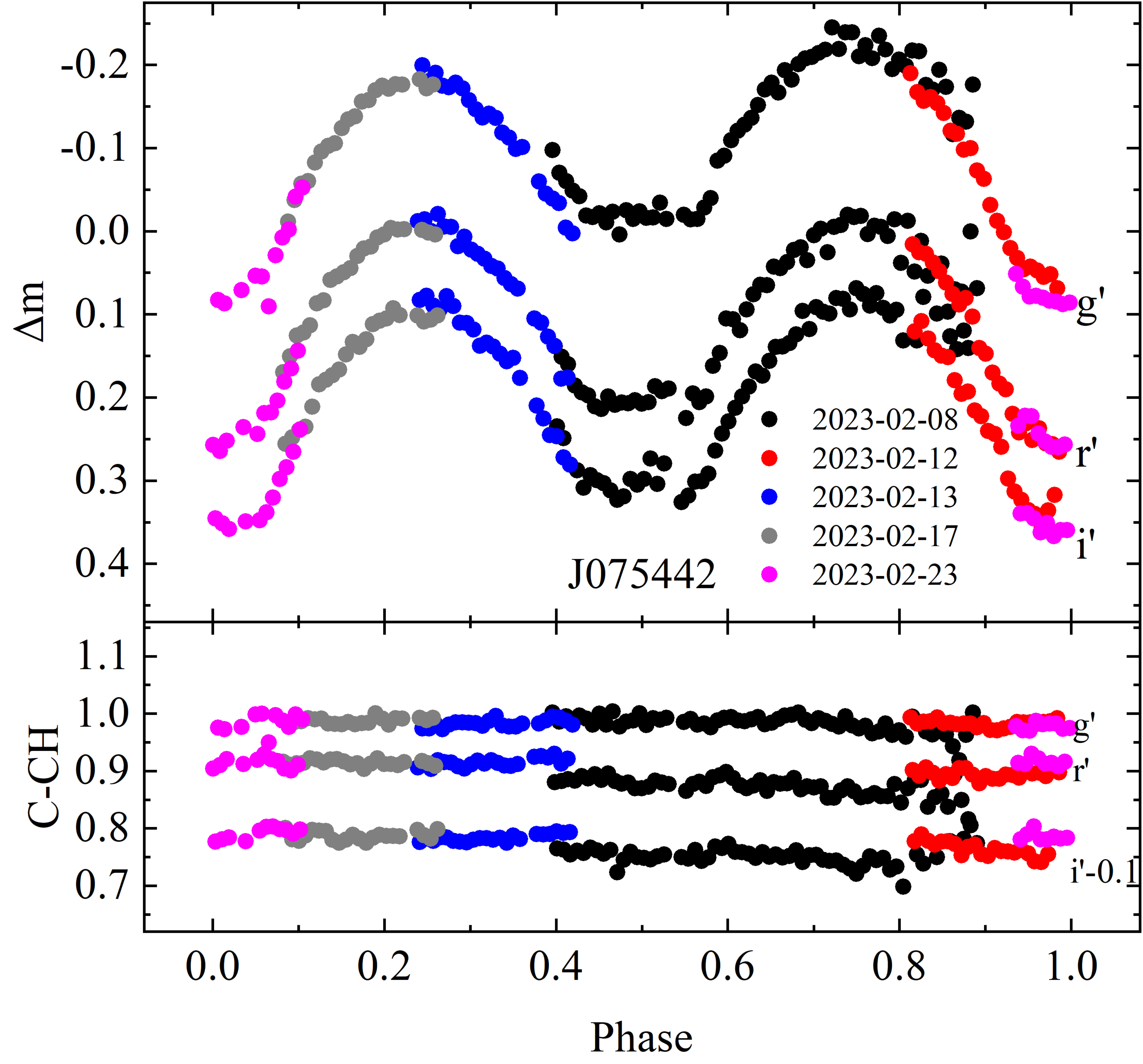}
	\end{minipage}
    
       \begin{minipage}[t]{0.32\linewidth}
		\centering
		\includegraphics[width=2.2in]{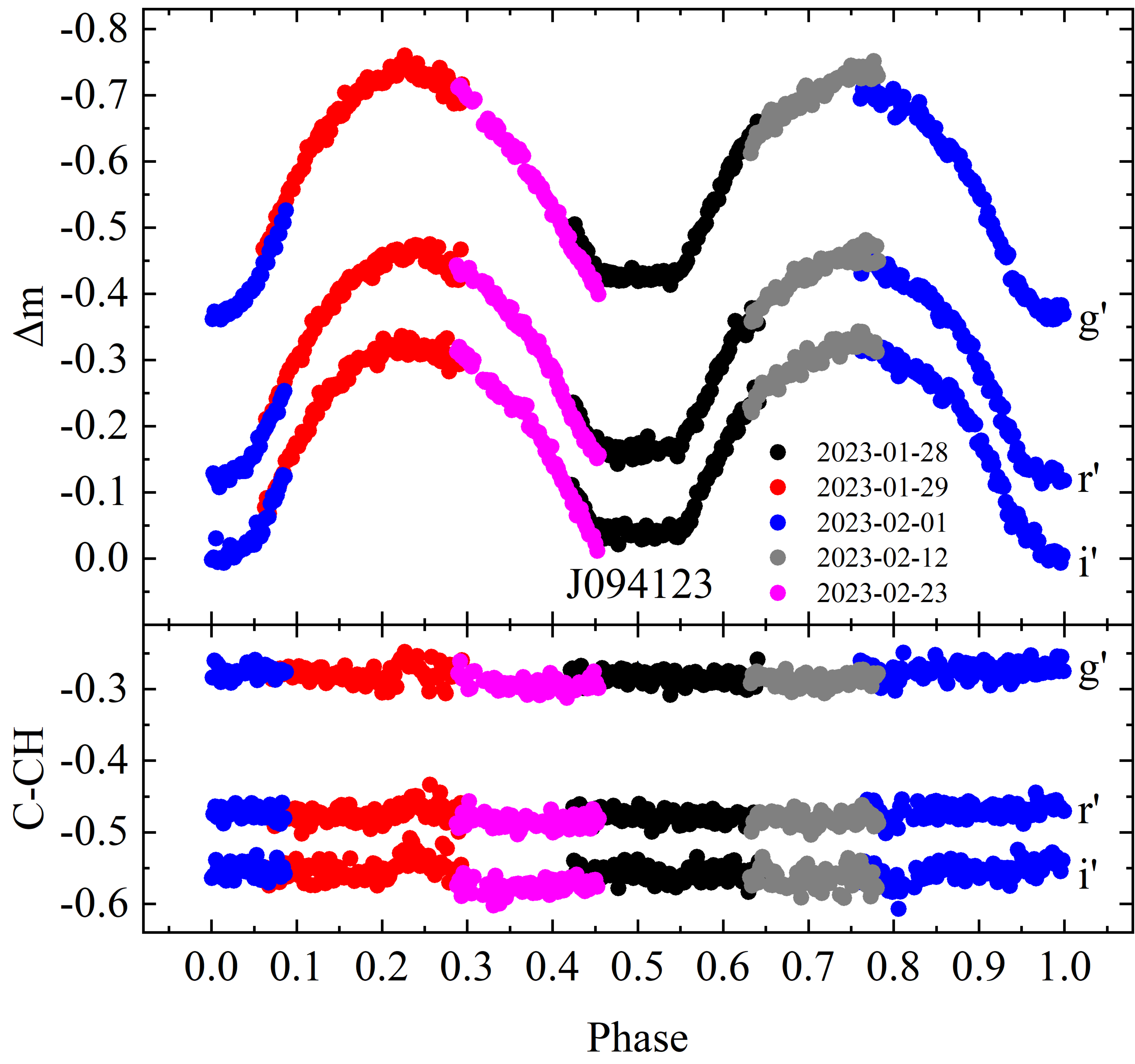}
	\end{minipage}%
	\begin{minipage}[t]{0.32\linewidth}
		\centering
		\includegraphics[width=2.2in]{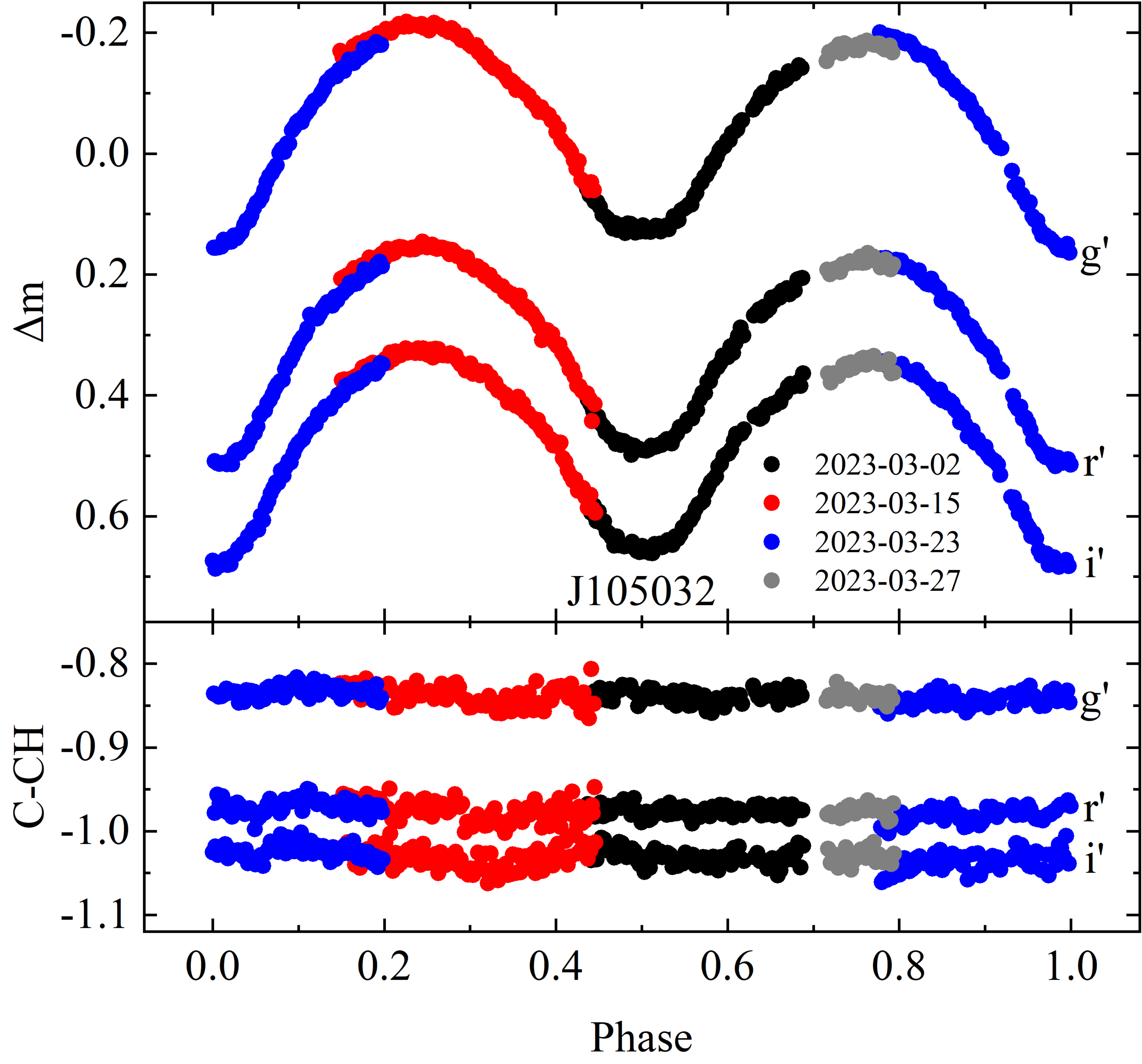}
	\end{minipage}
	\begin{minipage}[t]{0.32\linewidth}
		\centering
		\includegraphics[width=2.2in]{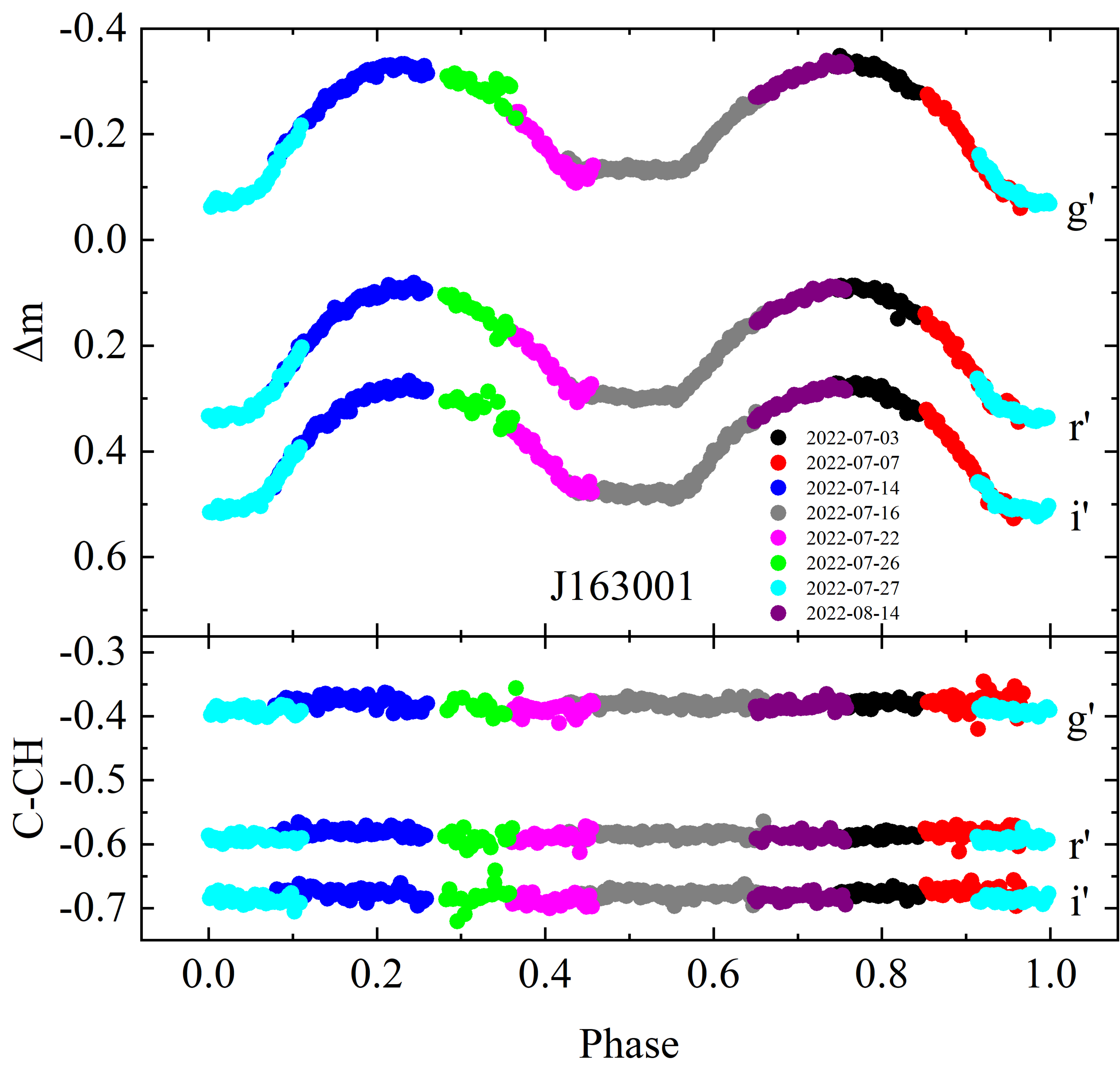}
	\end{minipage}
    \caption{Multi-band light curves of the six targets. The different colors represent observational data from different days, forming a complete light curve within one cycle in the phase diagram.}
    \label{fig:V-C and C-CH of six targets}
\end{figure*}

We used standard differential photometry to derive the light variations for our targets. Consequently, a set of comparison and check stars selected from the Two Micron All-Sky Survey (2MASS, \citealt{2003ApJ...598L...1M, 2006AJ....131.1163S}) for each target is tabulated in Table \ref{tab:Comparison stars and Check stars}. All CCD images were processed with C-MUNIPACK\footnote{\url{http://c-munipack.sourceforge.net/}}, where we performed bias subtraction, flat correction, and aperture photometry in separate bands for each target. Subsequently, we calculated the differential magnitudes between the target and its selected comparison star ($\Delta m$) and between the comparison star and the check star ($C-CH$). Photometric data are tabulated in Table \ref{tab:our photometric data of XL60 and NEXT}. Figure \ref{fig:V-C and C-CH of six targets} depicts the observed light curves of our six targets in multi-bands. The feature of total eclipse is apparent and we calculated the duration time of the total eclipses, $\Delta t = 67min$ for J063344, $\Delta t = 79min$ for J073647, $\Delta t = 95min$ for J075442, $\Delta t = 78min$ for J094123, $\Delta t = 47min$ for J105032, and $\Delta t = 108min$ for J163001.

Note that Transiting Exoplanet Survey Satellite (TESS, \citealt{2014SPIE.9143E..20R}) observed all the six targets. J063344 was observed by Sectors 20, 60, and 73. Both J073647 and J075442 were observed by Sectors 20, 47, and 60. J094123 was observed by Sectors 21, 46, 48, and 72. J105032 was observed by Sectors 21 and 48. J163001 was observed by 14 sectors: Sectors 16, 23, 24, 25, 49, 50, 51, 52, 56, 59, 76, 77, 78, and 79. In addition, several other photometric surveys have observed the six systems in photometry either, such as ASAS-SN, Catalina Real-time Transient Survey (CRTS, \citealt{Drake_2009}), Super Wide Angle Search for Planets (SuperWASP, \citealt{2010A&A...520L..10B}), and Zwicky Transient Facility (ZTF, \citealt{2019PASP..131a8002B}).

\begin{table*}
	\centering
	\caption{Observation details of the six targets.}
	\label{tab:Observation details of the six targets}
	\begin{tabular}{cccccc}
\hline
Target & Telescope & Year &Date                                                  & Band                  & Exposure Time  \\
\hline
J063344 & XL60      & 2023 &01-02, 01-03                                          & V, R, I               &  80s, 50s, 50s \\
J073647 & NEXT      & 2023 &02-13, 02-14, 02-17, 03-21                            & $g^{\prime}, r^{\prime}, i^{\prime}$ &  50s, 45s, 55s \\
J075442 & NEXT      & 2023 &02-08, 02-12, 02-13, 02-17, 02-23                     & $g^{\prime}, r^{\prime}, i^{\prime}$ &  95s, 85s, 110s\\
J094123 & NEXT      & 2023 &01-28, 01-29, 02-01, 02-12, 02-23                     & $g^{\prime}, r^{\prime}, i^{\prime}$ &  26s, 22s, 30s \\
J105032 & NEXT      & 2023 &03-02, 03-15, 03-23, 03-27                            & $g^{\prime}, r^{\prime}, i^{\prime}$ &  32s, 29s, 35s \\
J163001 & NEXT      & 2022 &07-03, 07-07, 07-14, 07-16, 07-22, 07-26, 07-27, 08-14& $g^{\prime}, r^{\prime}, i^{\prime}$ &  35s, 30s, 40s \\
\hline
\end{tabular}
\end{table*}

\begin{table*}
\begin{threeparttable}
	\centering
	\caption{The information of the comparison stars and check stars for the six targets.}
	\label{tab:Comparison stars and Check stars}
	\begin{tabular}{cccccccc}
\hline
Target & Comparison star &B-V  & $\theta_{1}$\tnote{1} & Check star &B-V  & $\theta_{2}$\tnote{1} & Reference\tnote{2}\\
       &                 &(mag)& (${^\prime}$)          &            &(mag)& (${^\prime}$)          &          \\
\hline
J063344 &2MASS 06335517+5347534& 0.714 & 2.0 &2MASS 06334148+5346545& 0.749 & 0.4 & (1) \\
J073647 &2MASS 07370077+4918163& 0.838 & 3.3 &2MASS 07365728+4926122& 0.656 & 5.7 & (1) \\
J075442 &2MASS 07535963+5555527& 0.578 & 6.0 &2MASS 07535213+5551351& 0.508 & 8.5 & (1) \\
J094123 &2MASS 09412713+2340293& 0.746 & 5.1 &2MASS 09412798+2344039& 0.541 & 1.8 & (1) \\
J105032 &2MASS 10504187+4208418& 0.749 & 1.7 &2MASS 10501028+4208148& 0.581 & 4.2 & (1) \\
J163001 &2MASS 16293848+5441203& 0.856 & 5.7 &2MASS 16302596+5442227& 0.673 & 5.0 & (1) \\
\hline
\end{tabular}
\begin{tablenotes}[para,flushleft]
\item[1] $\theta_{1, 2}$ represents the angular separation of the comparison star and check star from the target, respectively. \\
\item[2] References. (1) \citet{2016yCat.2336....0H}
\end{tablenotes}
\end{threeparttable}
\end{table*}

\subsection{Spectroscopic Observations}

The Large Sky Area Multi-Object Fiber Spectroscopic Telescope (LAMOST, \citealt{2012RAA....12.1197C}) is equipped with an active reflecting Schmidt telescope of 4 meters and a view field of 5$^{\circ}$. Its 4000 fibers make it possible to obtain 4000 spectra simultaneously. LAMOST has been operating for over 12 years and has made over 20 million spectroscopic observations public. Its scientific endeavors focus on the structure and evolution of the Galaxy and the cross-identification of multi-waveband properties in celestial objects. LAMOST has observed all the six targets in the low-resolution mode. The data of DR10 v1.0 is public on the website\footnote{\url{https://www.lamost.org/dr10/v1.0/search}}. Table \ref{tab:Spectral Observations, LAMOST} lists an overview of the spectral parameters of the six targets.

\begin{table*}
	\centering
       \footnotesize
	\caption{Spectral observations of our six targets from LAMOST.}
	\label{tab:Spectral Observations, LAMOST}
	\begin{tabular}{ccccccccc}
        \toprule
		 Target   & Observational Date & Exposure Time & SNR$_g$ &   Spectral Type & $T_{eff}$ & log g &   Radial Velocity  & [Fe/H]\\
		          &                    &     (s)       &          &       &   (K)     & (cgs)   &    (km\slash{s})   &     \\
		\hline
         \multirow{4}{*}{J063344} &     2016-02-18     &    1800.00 & 92   &      F0    & $6877\pm17$ & $4.196\pm0.024$ &       $75.74\pm2.37$ &$-0.406 \pm 0.014$\\
		                         &     2017-02-20     &    4500.00 & 10   &      F0    & $7280\pm255$ & $4.684\pm0.428$ &       $56.87\pm16.44$ &$-0.482 \pm 0.279$\\
		                       &     2017-02-28     &    4500.00 & 140   &      F0    & $6950\pm19$ & $4.171\pm0.025$ &       $48.34\pm3.04$ &$-0.415 \pm 0.014$\\
                                  &     2020-11-09     &    3600.00 & 240   &      F0    & $6903\pm9$ & $4.170\pm0.014$ &       $60.17\pm10.97$ &$-0.383 \pm 0.007$\\
            \hline
         \multirow{4}{*}{J073647} &     2015-01-30     &    1800.00 & 85   &      F0    & $6759\pm24$ & $4.143\pm0.033$ &       $-21.21\pm3.24$ &$-0.095 \pm 0.020$\\
		                         &     2015-03-28     &    1800.00 & 83   &      F0    & $6845\pm18$ & $4.150\pm0.025$ &       $-24.01\pm2.42$ &$-0.083 \pm 0.015$\\
		                       &     2016-02-19     &    1800.00 & 82   &      F0    & $6857\pm21$ & $4.133\pm0.029$ &       $13.12\pm2.82$  &$-0.076 \pm 0.017$\\
                                  &     2016-03-13     &    4500.00 & 173   &      F3    & $6697\pm14$ & $4.255\pm0.018$ &       $4.14\pm2.36$  &$-0.025 \pm 0.009$\\
            \hline            
	\multirow{2}{*}{J075442}  &     2014-03-23     &    4195.21 & 24   &      A7    & $6906\pm229$ & $4.149\pm0.365$ &       $15.30\pm15.60$ &$-0.412 \pm 0.216$\\  
		                       &     2017-02-09     &    1500.00 & 28   &      F3    & -            & -               &       -               &-\\
            \hline             
	\multirow{4}{*}{J094123}  &     2014-12-01     &    1800.00 & 115   &      A9    & $6786\pm21$ & $4.246\pm0.034$ &       $-20.82\pm8.70$ &$-0.366 \pm 0.019$\\
		                         &     2016-06-08     &    1800.00 & 11   &      F0    & $7004\pm198$ & $4.340\pm0.331$&       $-9.54\pm13.04$  &$-0.300 \pm 0.216$\\
		                       &     2018-12-12     &    1038.78 & 91   &      A9    & $6800\pm32$ & $4.247\pm0.053$ &       $6.73\pm11.48$   &$-0.345 \pm 0.030$\\
		                       &     2021-11-30     &    1800.00 & 87   &      F0    & $6789\pm25$ & $4.244\pm0.035$ &       $-15.19\pm3.44$  &$-0.352 \pm 0.021$\\
            \hline
	\multirow{2}{*}{J105032}  &     2014-03-23     &    1200.00 & 97   &      A7    & $7183\pm18$ & $4.095\pm0.026$ &       $5.64\pm2.61$    &$-0.307 \pm 0.015$\\  
		                       &     2017-03-09     &    1800.00 & 177   &      F0    & $7105\pm13$ & $4.089\pm0.017$ &       $43.45\pm2.32$  &$-0.286 \pm 0.009$\\
            \hline 
	\multirow{2}{*}{J163001}  &     2013-05-07     &    1800.00 & 43   &      F0    & $6780\pm38$ & $4.148\pm0.054$ &       $-122.54\pm5.17$ &$-0.464 \pm 0.033$\\  
		                       &     2016-04-24     &    1800.00 & 60   &      F0    & $6821\pm25$ & $4.159\pm0.036$ &       $-154.61\pm3.34$ &$-0.589 \pm 0.021$\\
		\toprule
	\end{tabular}
\end{table*}

\section{Photometric Solution}
\label{sec:Photometric Solution}

\begin{figure*}
        \centering
	\begin{minipage}[m]{0.3\linewidth}
	    \centering
	      \includegraphics[width=2.1in]{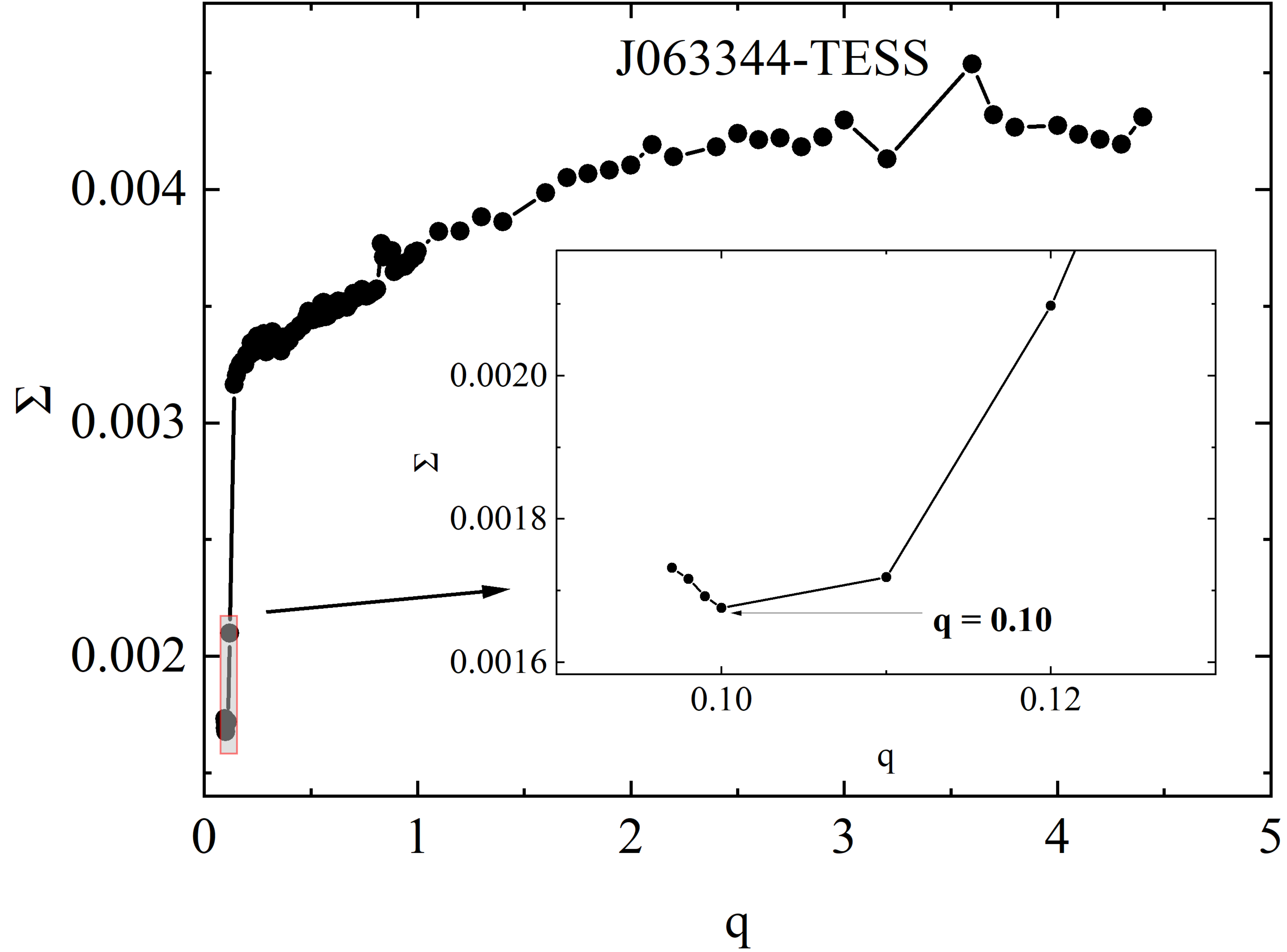}
        \end{minipage}%
        \begin{minipage}[m]{0.3\linewidth}
	    \centering
	    \includegraphics[width=2.1in]{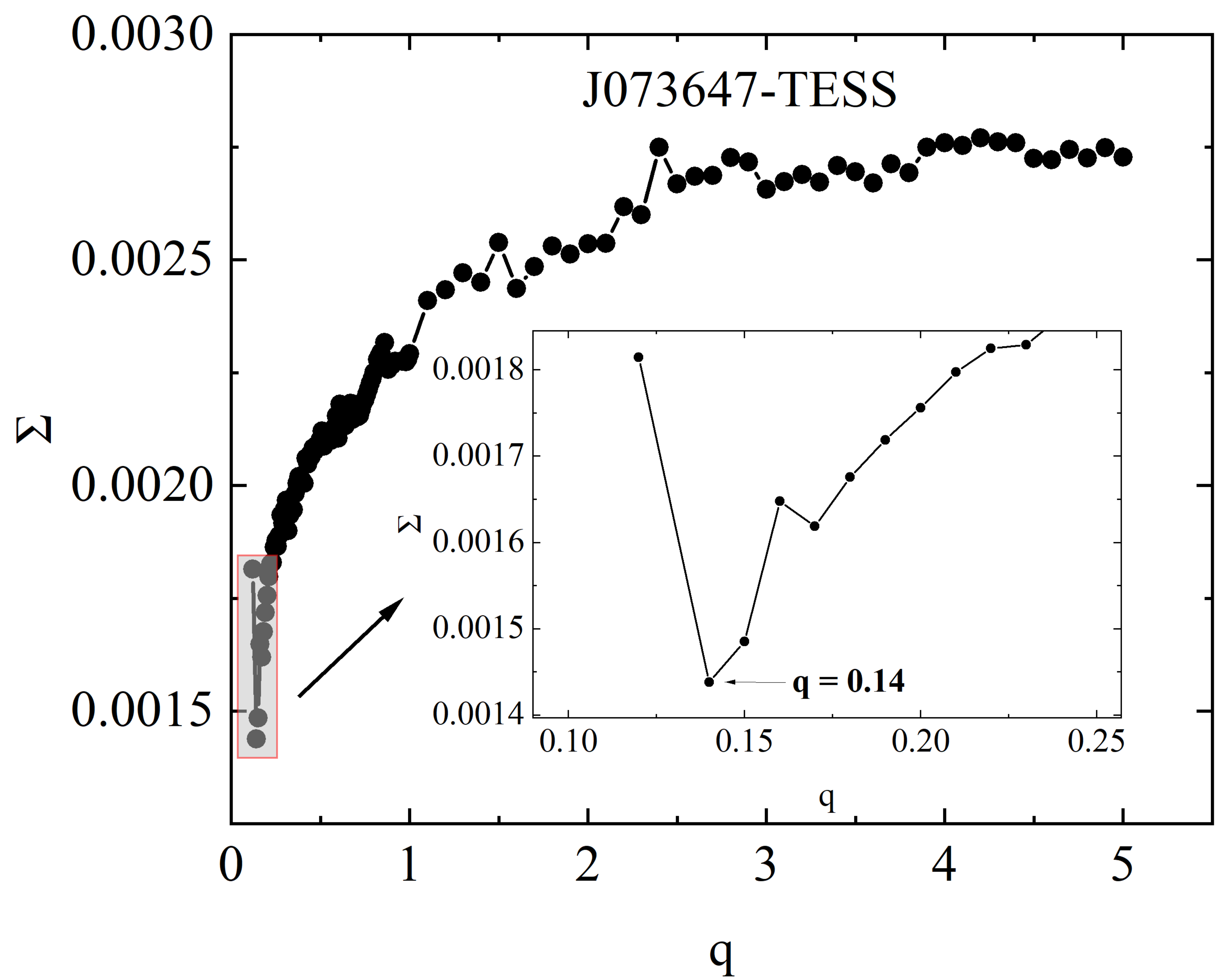}
        \end{minipage}%
        \begin{minipage}[m]{0.3\linewidth}
	    \centering
	    \includegraphics[width=2.1in]{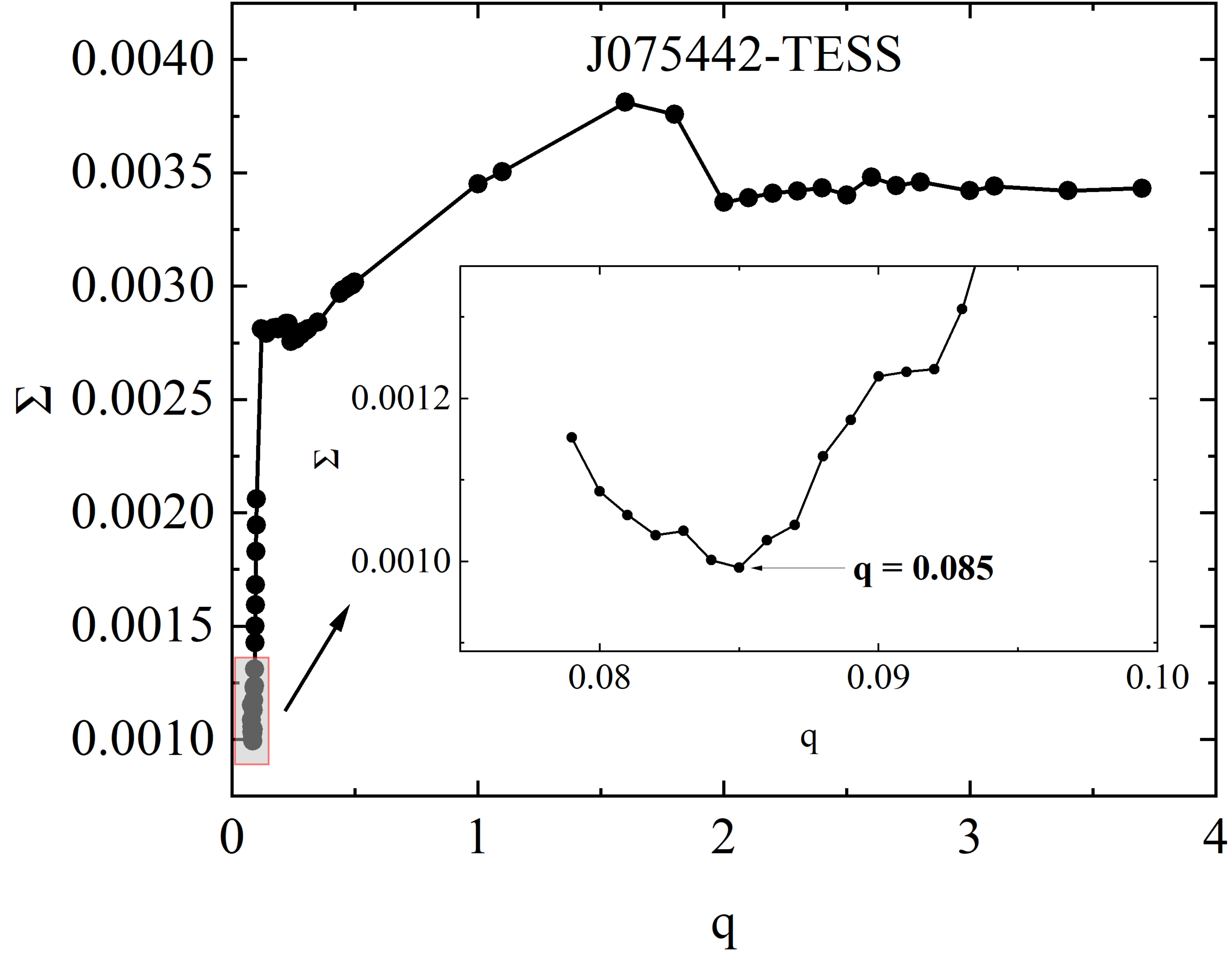}
        \end{minipage}%

	 \begin{minipage}[m]{0.3\linewidth}
	    \centering
	      \includegraphics[width=2.1in]{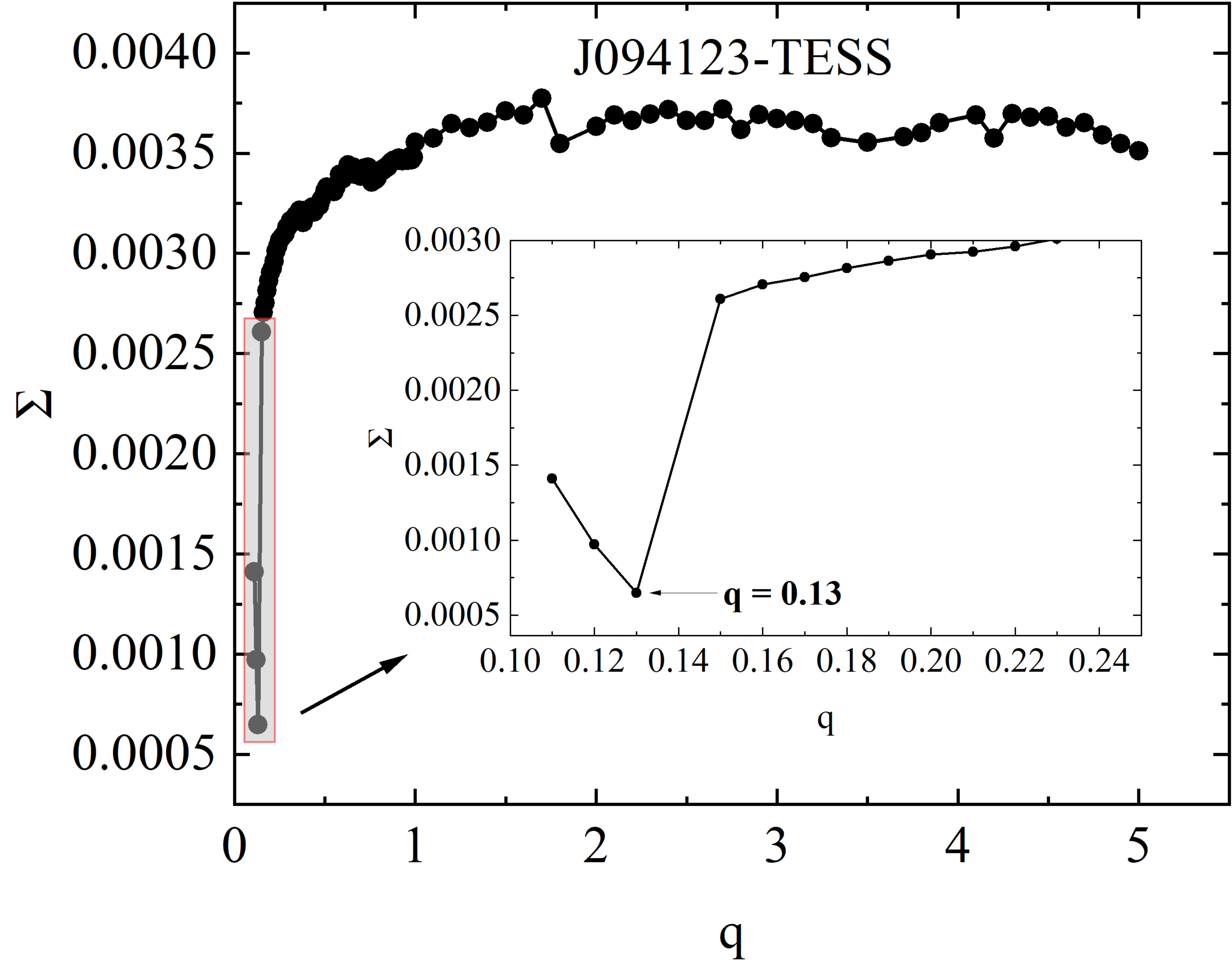}
        \end{minipage}%
        \begin{minipage}[m]{0.3\linewidth}
	    \centering
	    \includegraphics[width=2.1in]{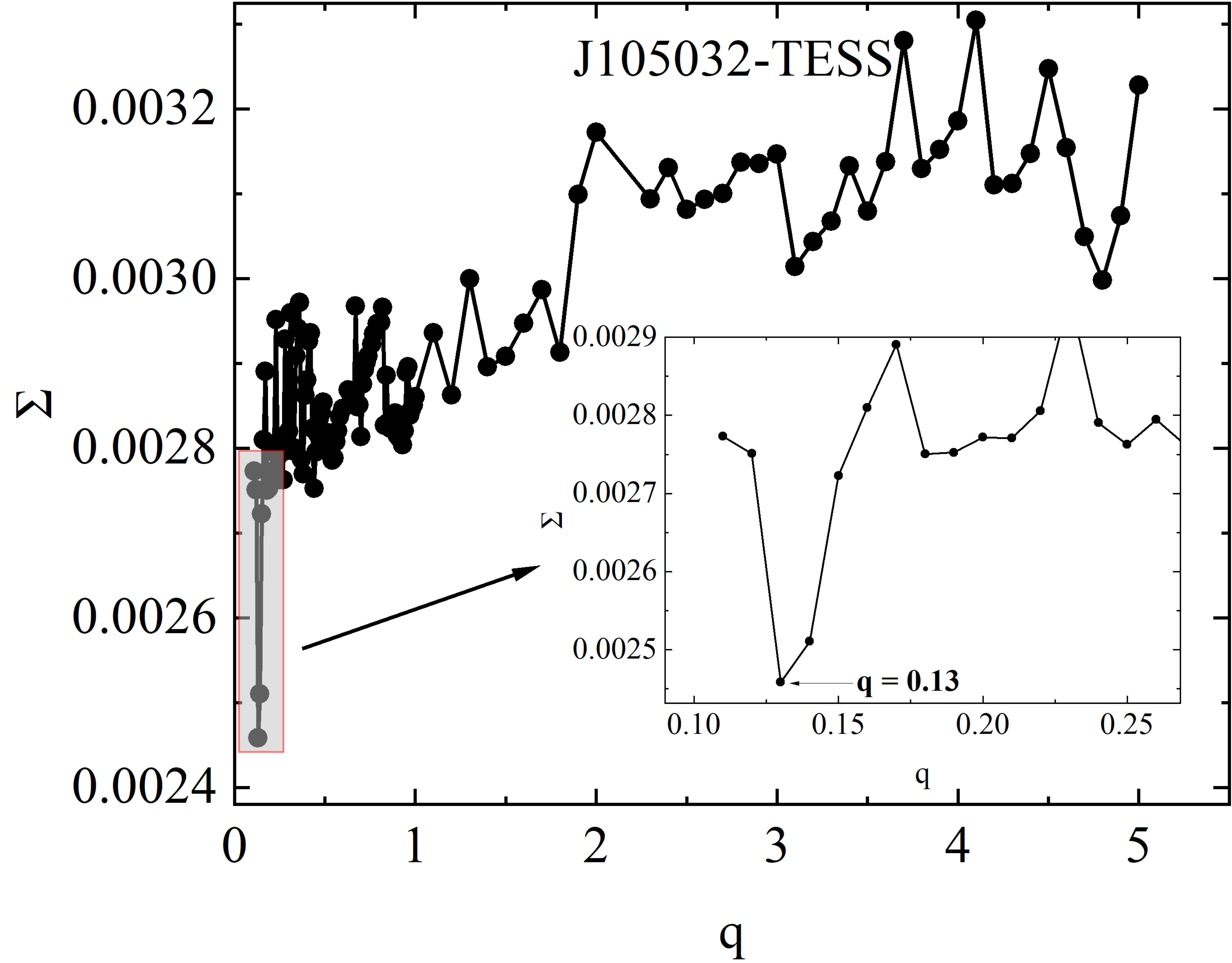}
        \end{minipage}%
        \begin{minipage}[m]{0.3\linewidth}
	    \centering
	    \includegraphics[width=2.1in]{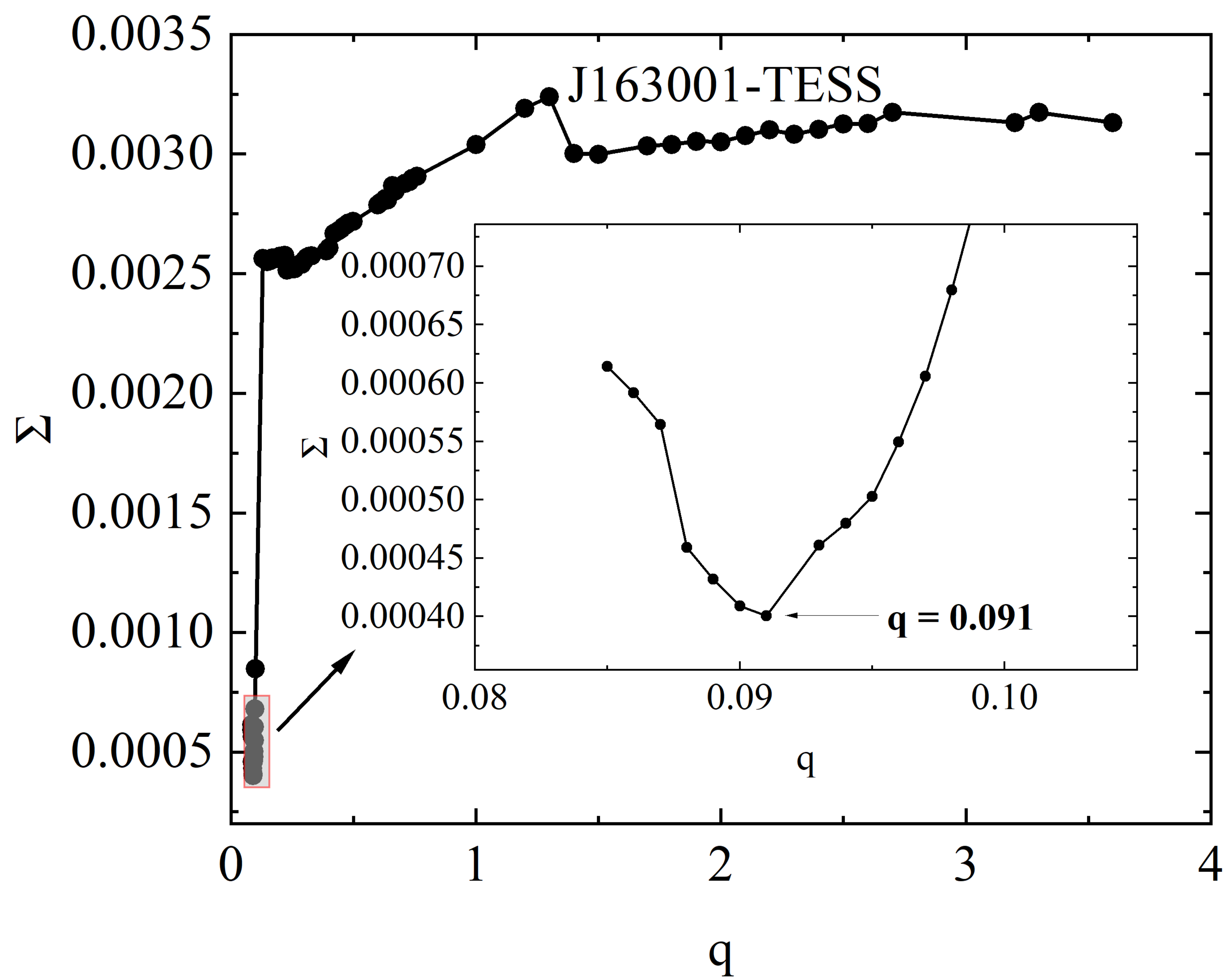}
        \end{minipage}%

        \begin{minipage}[m]{0.3\linewidth}
	    \centering
	      \includegraphics[width=2.1in]{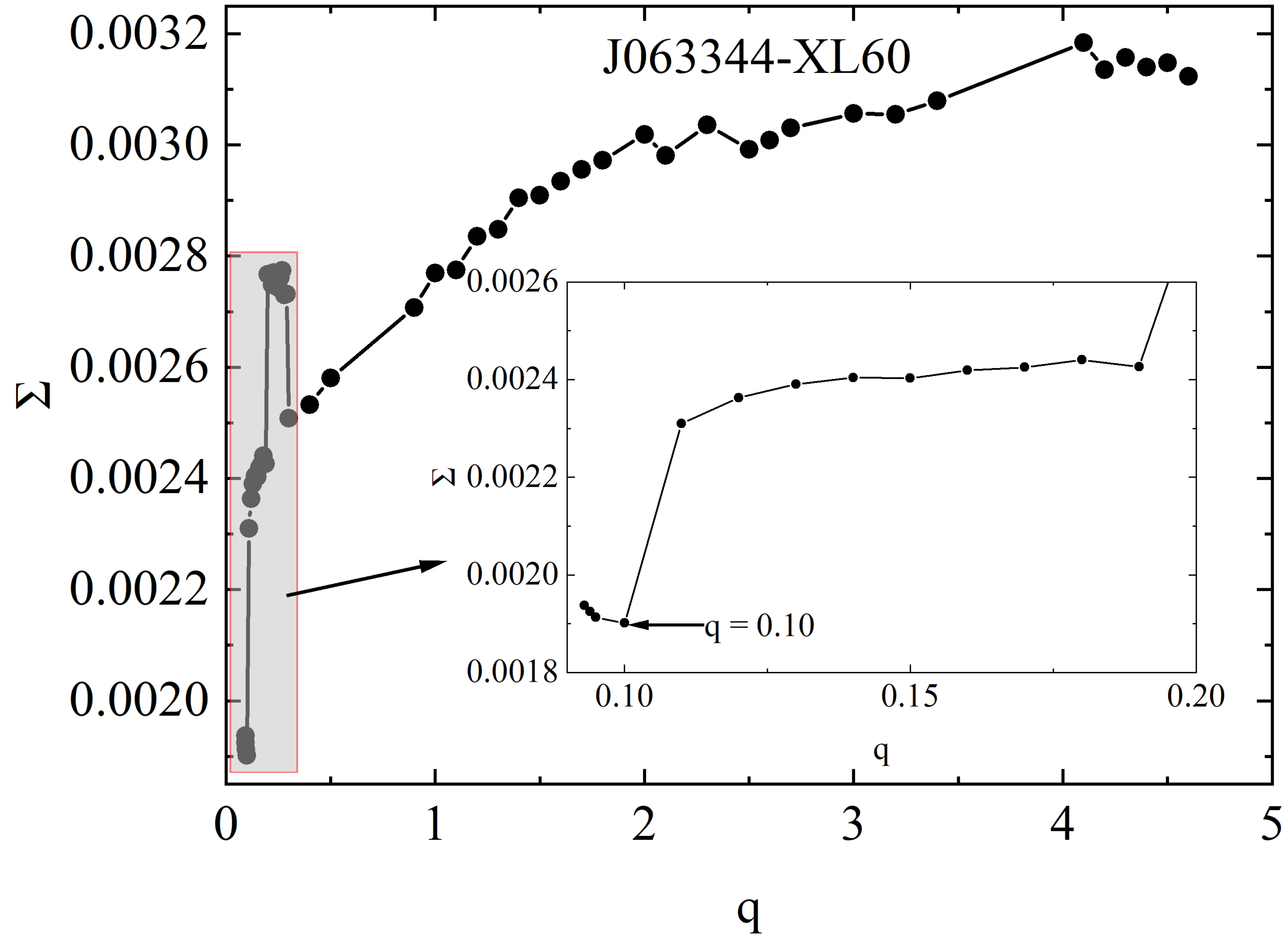}
        \end{minipage}%
        \begin{minipage}[m]{0.3\linewidth}
	    \centering
	    \includegraphics[width=2.1in]{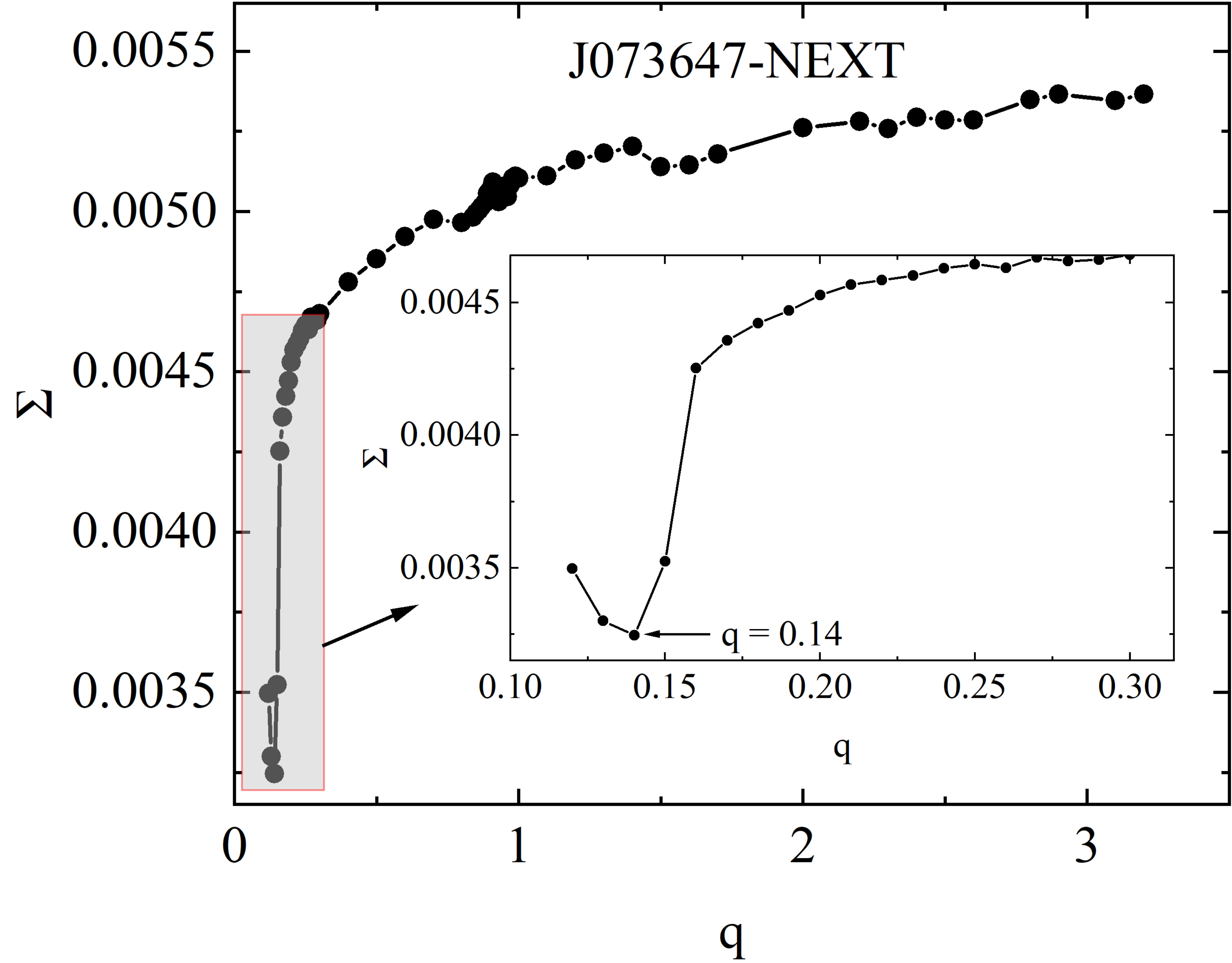}
        \end{minipage}%
        \begin{minipage}[m]{0.3\linewidth}
	    \centering
	    \includegraphics[width=2.1in]{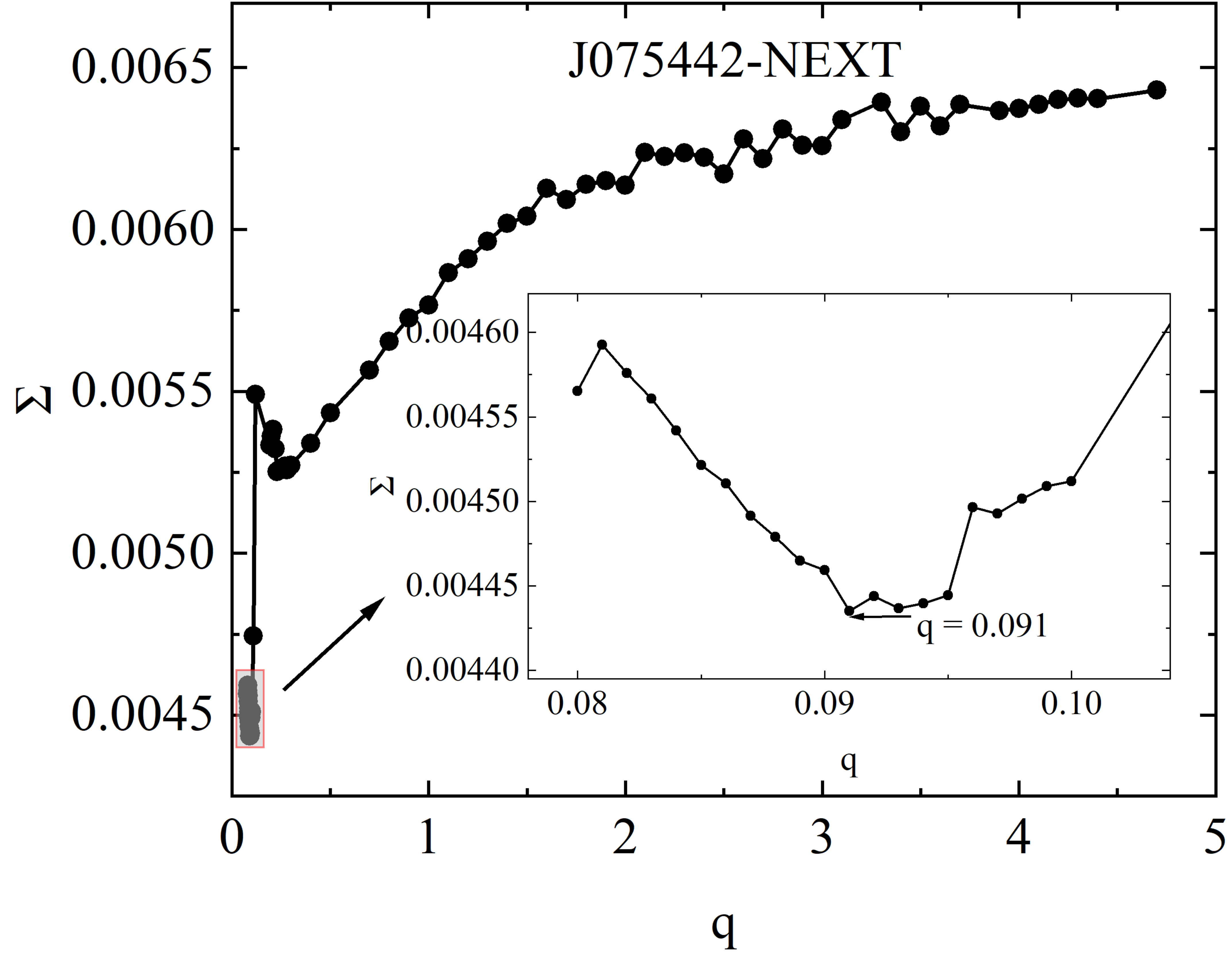}
        \end{minipage}%

	 \begin{minipage}[m]{0.3\linewidth}
	    \centering
	      \includegraphics[width=2.1in]{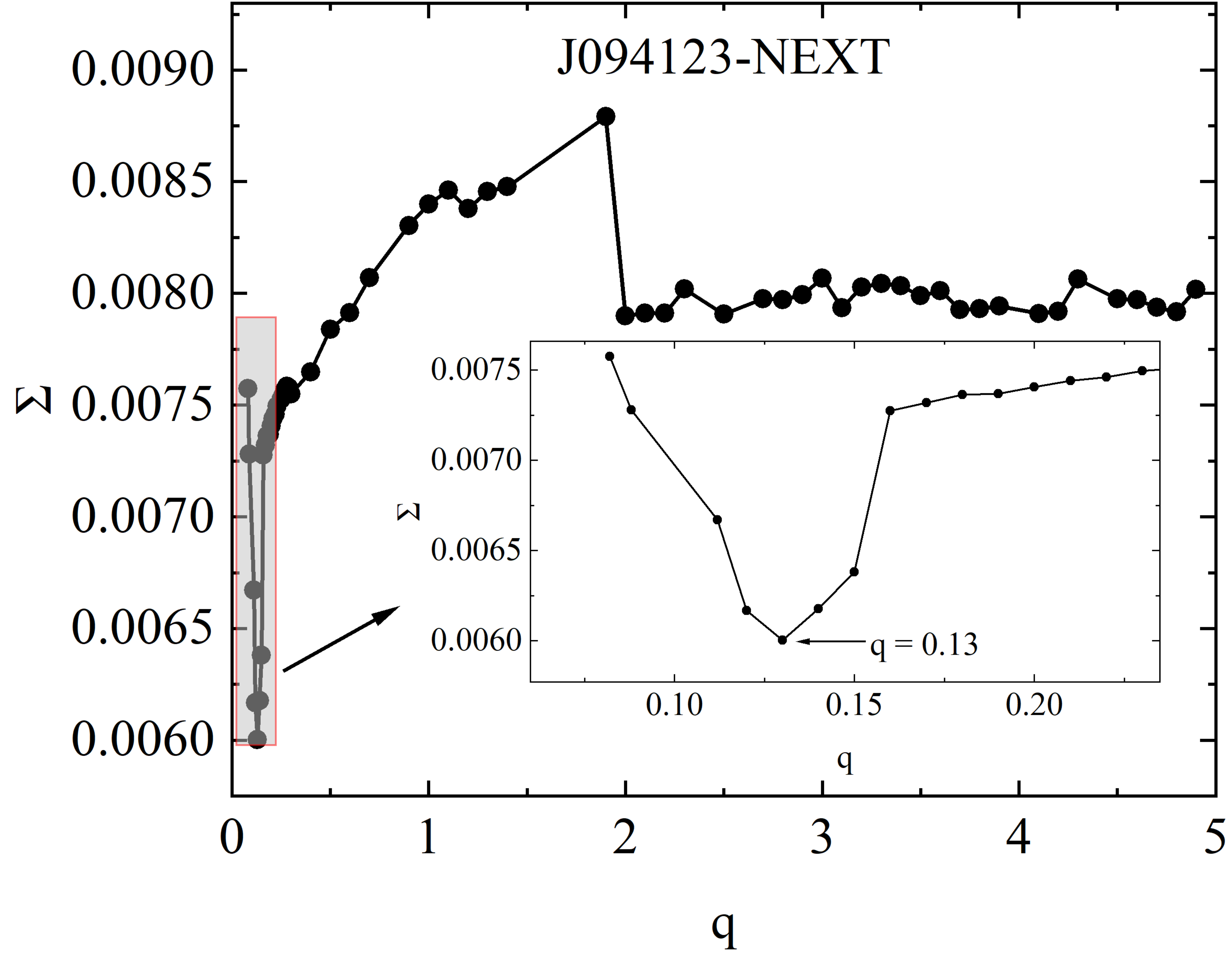}
        \end{minipage}%
        \begin{minipage}[m]{0.3\linewidth}
	    \centering
	    \includegraphics[width=2.1in]{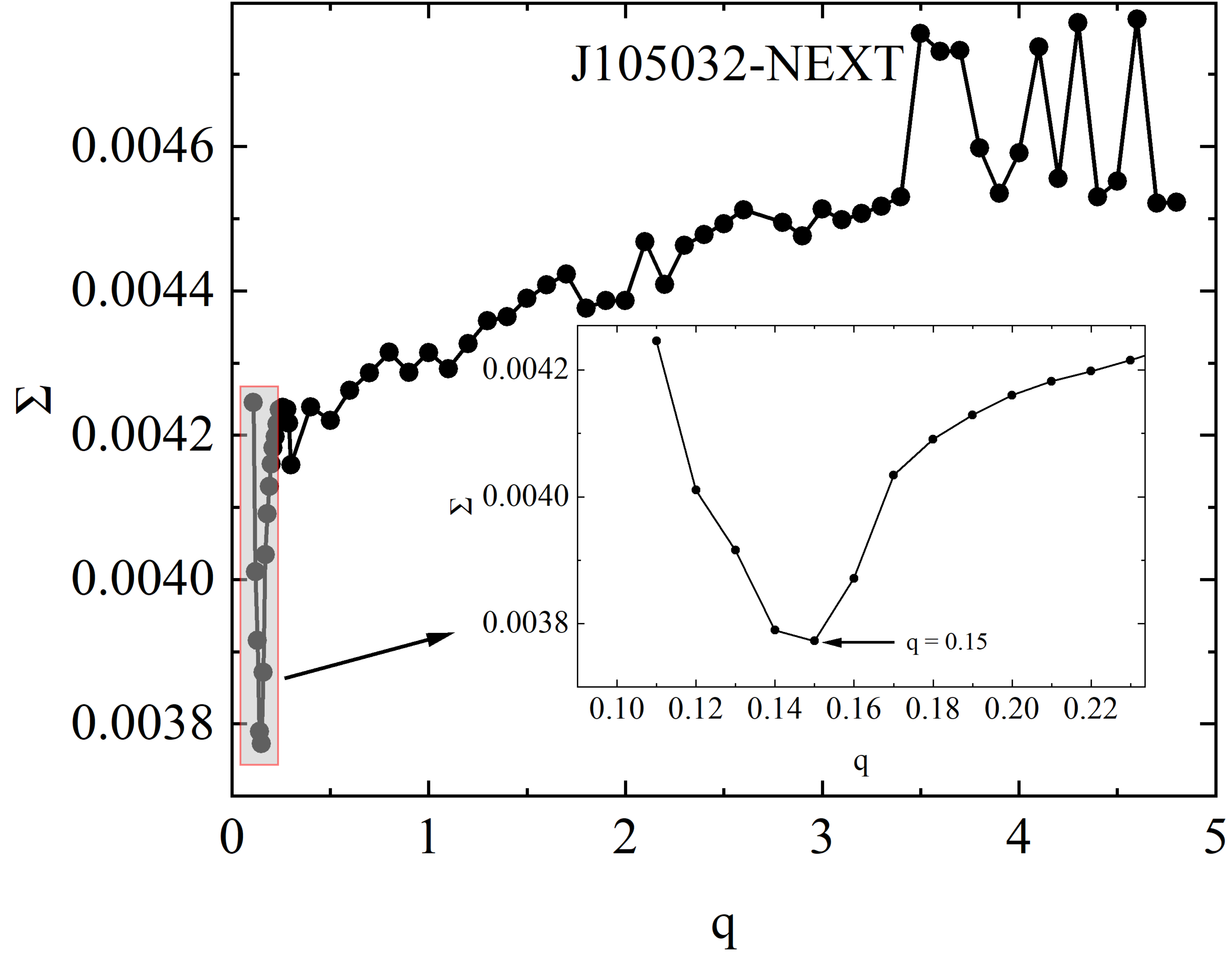}
        \end{minipage}%
        \begin{minipage}[m]{0.3\linewidth}
	    \centering
	    \includegraphics[width=2.1in]{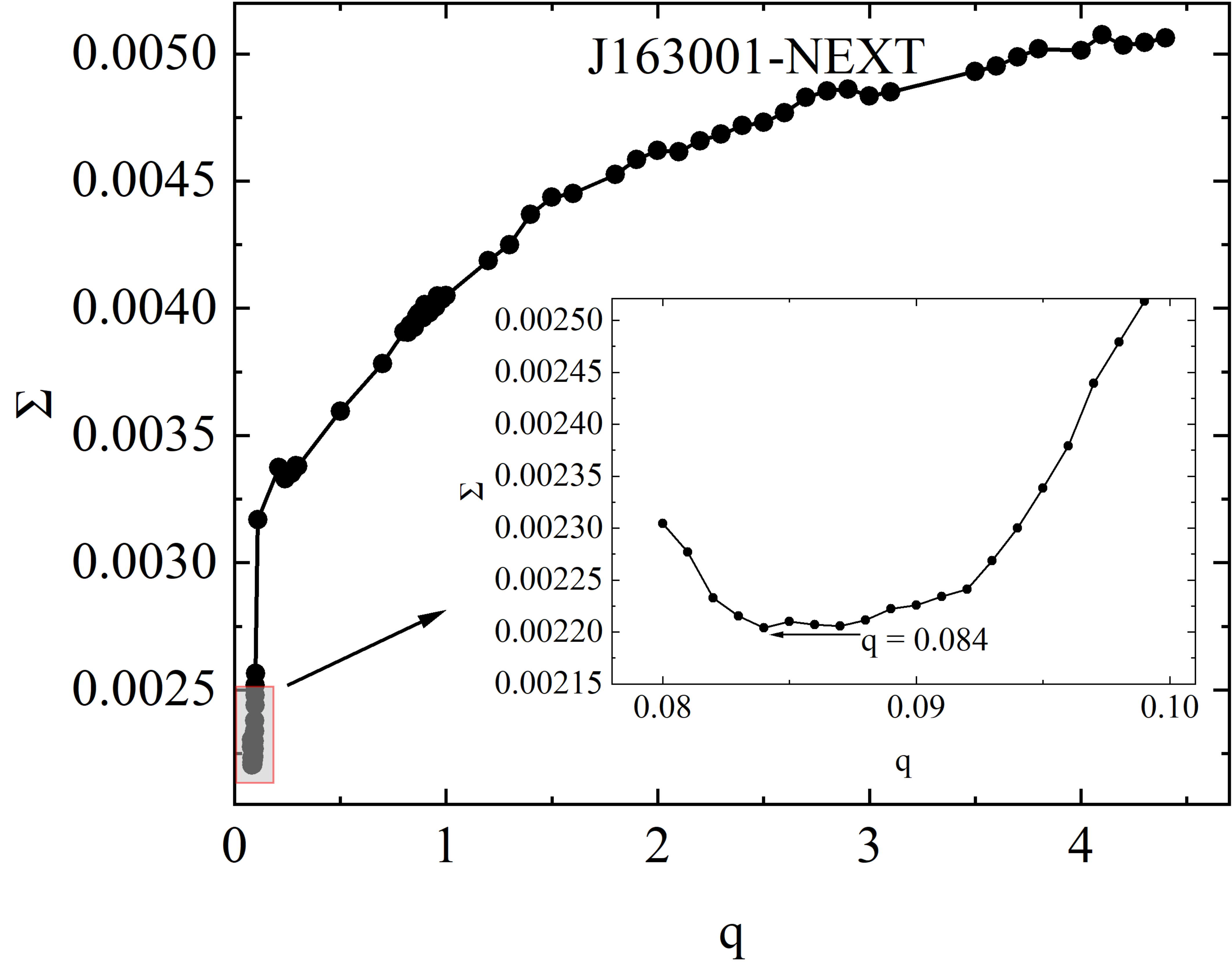}
        \end{minipage}%
    \caption{The relation between mass ratio $q$ and mean residual $\Sigma$ of the six targets.}
    \label{fig:q-search}
\end{figure*}

The previous statistical studies conclude that the mass ratios obtained by photometry are approximately equal to those obtained by spectroscopy ($q_{ph} \sim q_{sp}$) for totally eclipsing contact binaries \citep{2003CoSka..33...38P, 2021AJ....162...13L}. Although all the six targets lack radial velocity observations, the features of total eclipses support the reliability of parameters obtained by photometric light curves. For TESS data, the SAP-FLUX was normalized and transformed to magnitude. We adopted two strategies to improve the efficiency of photometric solutions. Firstly, the light curves were analyzed together when their shapes were alike. And we combined the data into 200 normal points to improve the efficiency of photometric analysis. We employed the Wilson–Devinney program (W-D) of the 2013 version to acquire light curve solutions \citep{1971ApJ...166..605W, 1979ApJ...234.1054W, Wilson_1994}. For each target, the initial temperature of the primary component was calculated by the average value of $T_{eff}$ from LAMOST and set as a fixed parameter in the W-D program. All six targets were studied for the first time, so we searched for the mass ratios ($q$) using the TESS data and our data to obtain $q$ according to the minimum mean residuals ($\Sigma$), respectively. During the q-search, the step was set as 0.001 when $q < 0.1$, the step was set as 0.01 when $0.1 \leq q \leq 1.0$, and the step was set as 0.1 when $q > 1.0$. The diagrams are displayed in Figure \ref{fig:q-search}, implying that the q-search results of TESS data and our data are generally consistent.

\begin{figure*}
        \centering
	\begin{minipage}[m]{0.3\linewidth}
	    \centering
	      \includegraphics[width=2in]{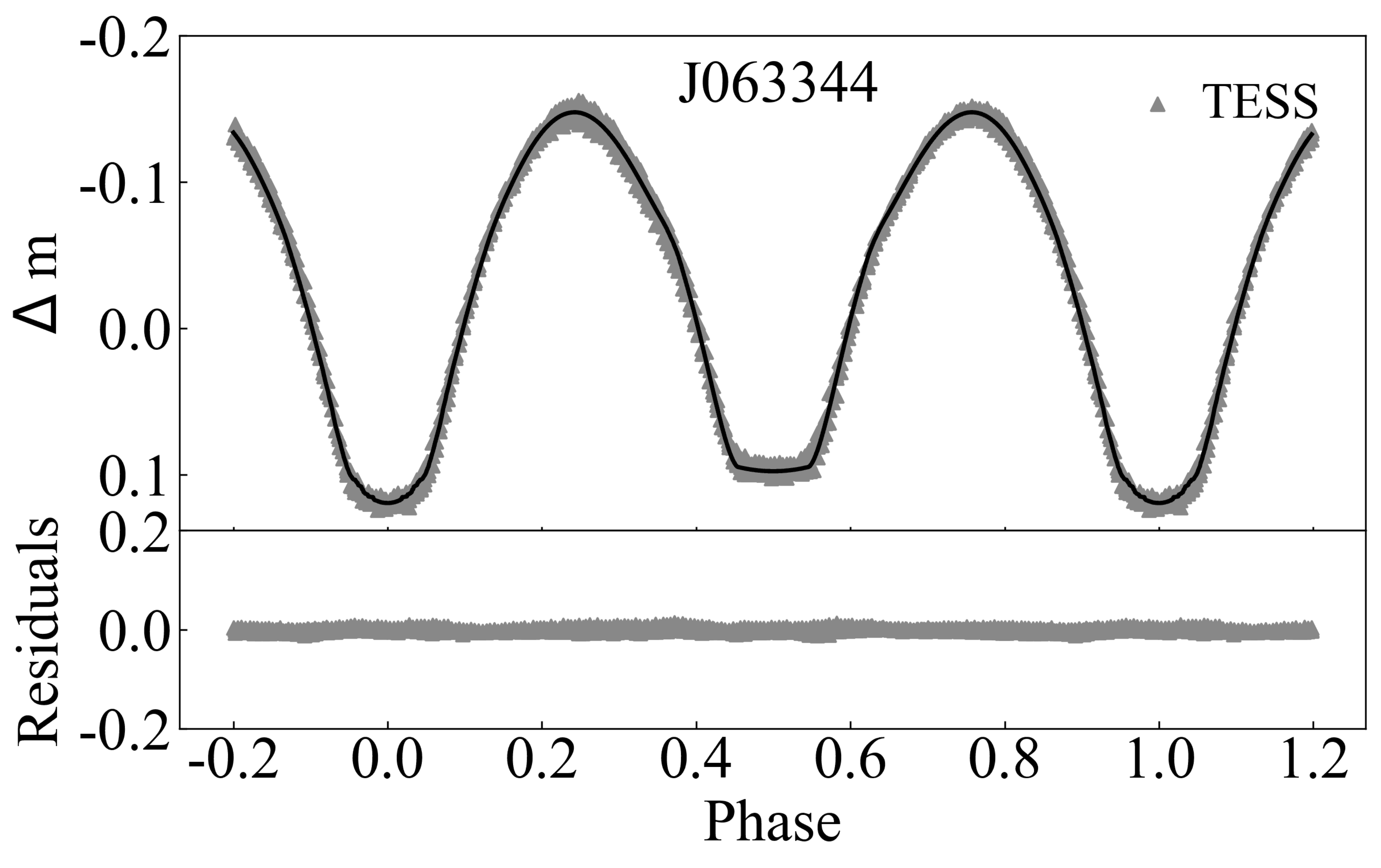}
        \end{minipage}%
        \begin{minipage}[m]{0.3\linewidth}
	    \centering
	    \includegraphics[width=2in]{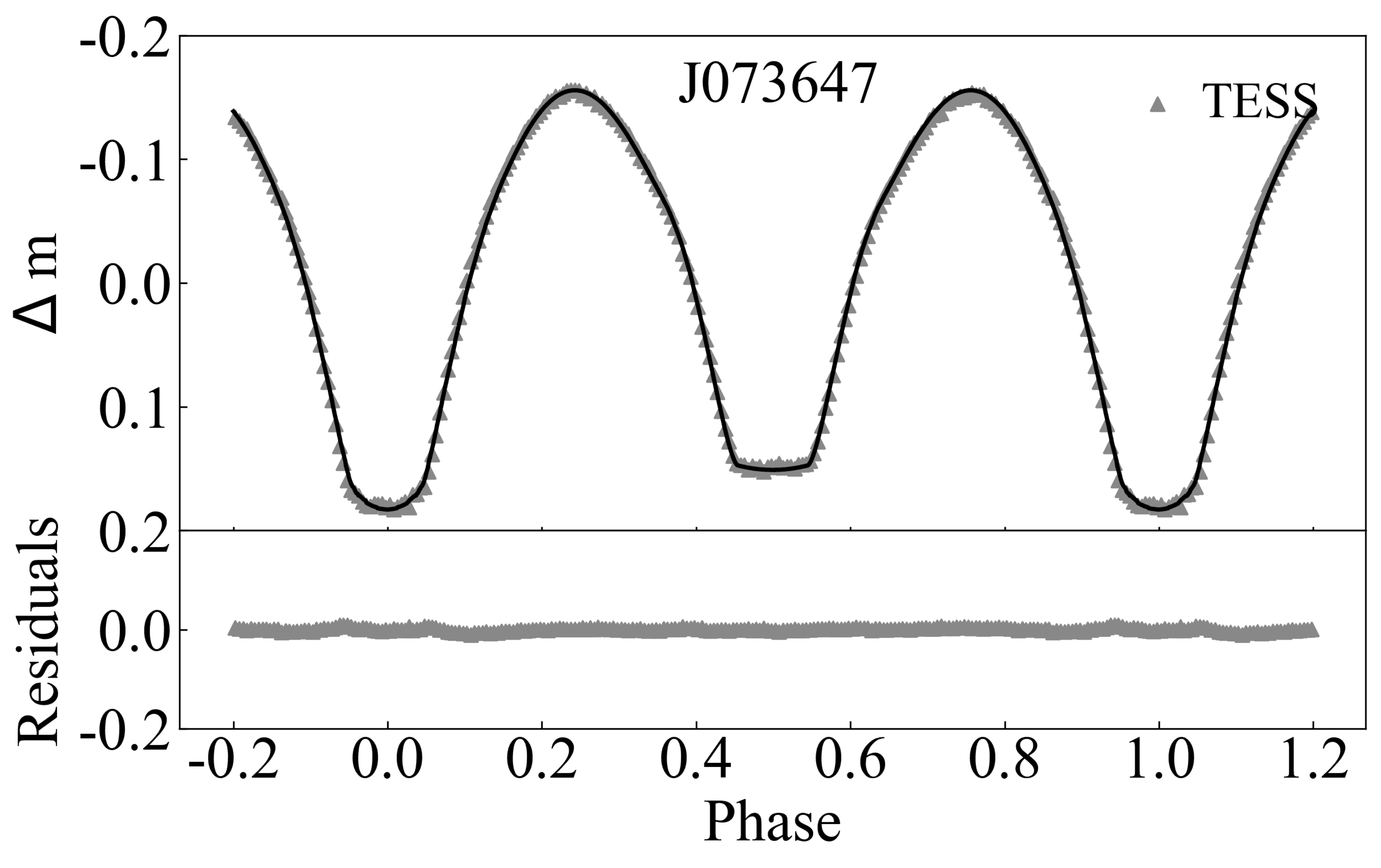}
        \end{minipage}%
        \begin{minipage}[m]{0.3\linewidth}
	    \centering
	    \includegraphics[width=2in]{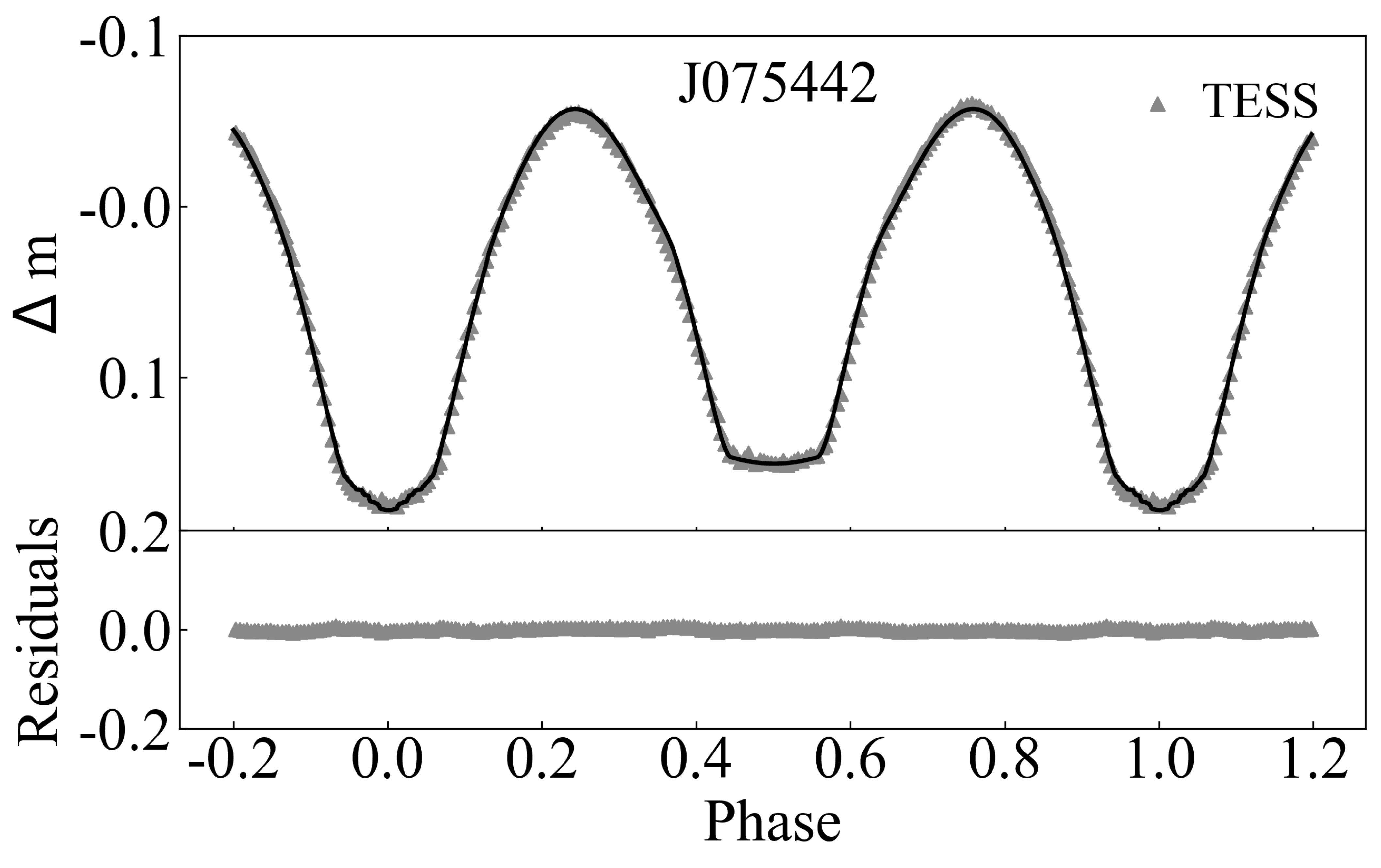}
        \end{minipage}%

	\begin{minipage}[m]{0.3\linewidth}
	    \centering
	      \includegraphics[width=2in]{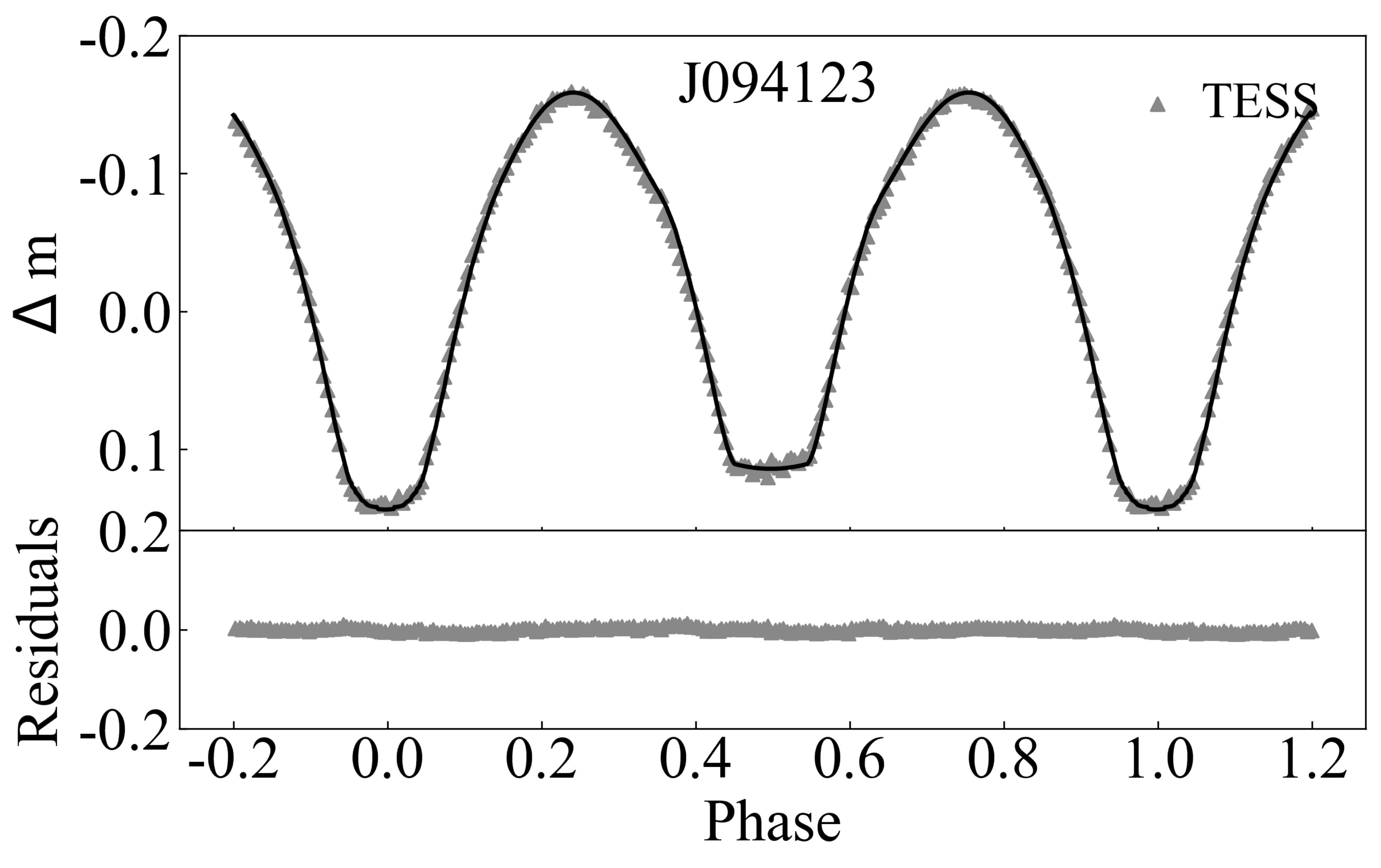}
        \end{minipage}%
        \begin{minipage}[m]{0.3\linewidth}
	    \centering
	    \includegraphics[width=2in]{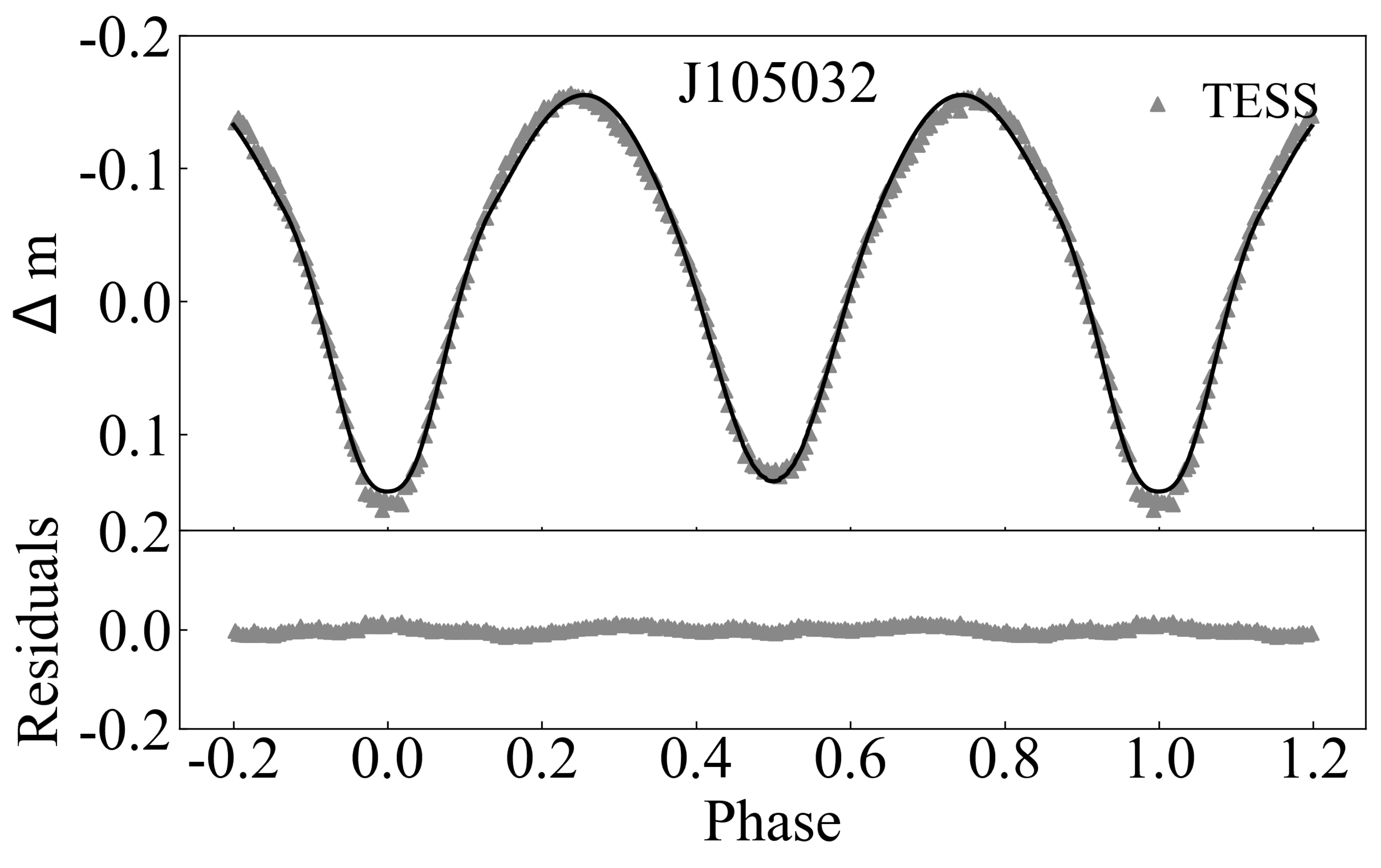}
        \end{minipage}%
        \begin{minipage}[m]{0.3\linewidth}
	    \centering
	    \includegraphics[width=2in]{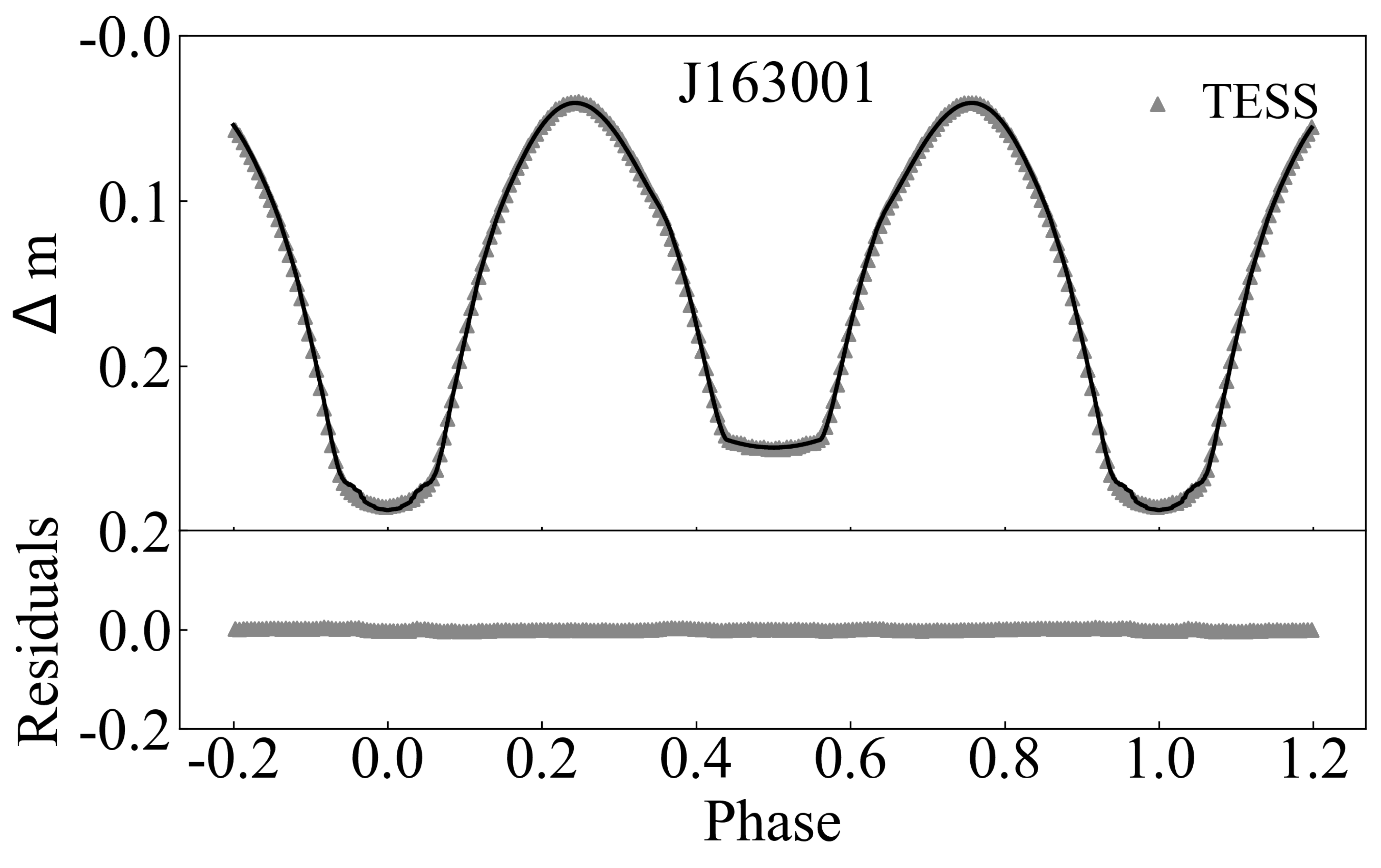}
        \end{minipage}%
    \caption{Comparison between the photometric data and the theoretical light curves for TESS data. The different symbols represent the different bands. The lower panels show the residuals.}
    \label{fig:LCs of TESS}
\end{figure*}

The LAMOST spectra of low-resolution are blended spectra of the two components, so the $T_{eff}$ is the composite temperature of the primary and secondary components. We utilized the following equations to calculate the precise temperatures for the primary and secondary components individually \citep{2003A&A...404..333Z, 2013AJ....146..157C},
\begin{equation}
\label{eq: Derive the individual T2}
\begin{split}
&k=r_2/r_1,\\
&T_1=(((1+k^2)T_{eff}^4)/(1+k^2(T_{2i}/T_{1i})^4))^{0.25},\\
&T_2=T_1(T_{2i}/T_{1i}),
\end{split}
\end{equation}
where $T_{2i}/T_{1i}$ is the temperature ratio obtained in W-D program, and $r_1$ and $r_2$ are the primary and secondary components' equal-volume radii, respectively. Then, we substituted the calculated $T_1$ and $T_2$ to obtain the more precise physical parameters as the ultimate results. We adopted the photometric results derived from the TESS data as the primary basis for our study due to the best quality of TESS data. Figure \ref{fig:LCs of TESS} demonstrates the photometric data from TESS and the theoretical light curves. Table \ref{tab:Photometric solutions of TESS} displays the photometric solutions of TESS data. Note that the light curves of J073647, J075442, and J105032 exhibit some visible asymmetries; however, the difference between the two maxima is nearly zero. The O’Connell effect in these three targets is weak. The attempts to model it using the spot model in the W-D program have not produced satisfactory results. Therefore, the O’Connell effect was not taken into account in our photometric solutions.

\begin{table*}
\begin{threeparttable}
	\centering
	\caption{Photometric solutions of TESS} data.
	\label{tab:Photometric solutions of TESS}
\begin{tabular}{ccccccccc}
\hline
Target         & J063344 &     J073647     &     J075442   &  J094123   & J105032 & J163001\\
\hline 
$T_1$(K)\tnote{1}	&	$	7015	\pm	84	$	&	$	6807	\pm	21	$	&	$	6927	\pm	252	$	&	$	6866	\pm	78	$	&	$	7162	\pm	24	$	&	$	6838	\pm	36	$	\\
$T_2$(K)\tnote{1}	&	$	6919	\pm	148	$	&	$	6698	\pm	34	$	&	$	6726	\pm	453	$	&	$	6721	\pm	134	$	&	$	7627	\pm	54	$	&	$	6481	\pm	59	$	\\
$q$($M_2$/$M_1$)	&	$	0.106 	\pm	0.001 	$	&	$	0.150 	\pm	0.001 	$	&	$	0.083 	\pm	0.001 	$	&	$	0.134 	\pm	0.001 	$	&	$	0.117 	\pm	0.001 	$	&	$	0.094 	\pm	0.001 	$	\\
i($^{\circ}$)	&	$	76.9 	\pm	0.2 	$	&	$	81.5 	\pm	0.2 	$	&	$	78.4 	\pm	0.2 	$	&	$	80.2 	\pm	0.2 	$	&	$	70.8 	\pm	0.6 	$	&	$	82.6 	\pm	0.1 	$	\\
$\Omega$	&	$	1.939 	\pm	0.002 	$	&	$	2.056 	\pm	0.002 	$	&	$	1.871 	\pm	0.001 	$	&	$	2.030 	\pm	0.003 	$	&	$	1.965 	\pm	0.003 	$	&	$	1.916 	\pm	0.001 	$	\\
$r_1$	&	$	0.592 	\pm	0.001 	$	&	$	0.565 	\pm	0.001 	$	&	$	0.611 	\pm	0.001 	$	&	$	0.568 	\pm	0.001 	$	&	$	0.586 	\pm	0.001 	$	&	$	0.596 	\pm	0.001 	$	\\
$r_2$	&	$	0.228 	\pm	0.004 	$	&	$	0.252 	\pm	0.003 	$	&	$	0.216 	\pm	0.002 	$	&	$	0.236 	\pm	0.004 	$	&	$	0.238 	\pm	0.002 	$	&	$	0.216 	\pm	0.002 	$	\\
$L_2/L_1$	&	$	0.139 	\pm	0.001 	$	&	$	0.184 	\pm	0.001 	$	&	$	0.110 	\pm	0.001 	$	&	$	0.161 	\pm	0.001 	$	&	$	0.187 	\pm	0.005 	$	&	$	0.109 	\pm	0.001 	$	\\
$f$($\%$)\tnote{2}	&	$	55.6 	\pm	2.9 	$	&	$	48.8 	\pm	2.5 	$	&	$	63.8 	\pm	1.1 	$	&	$	33.6 	\pm	3.0 	$	&	$	58.5 	\pm	3.8 	$	&	$	42.2 	\pm	1.9 	$	\\
\hline
\end{tabular}
\begin{tablenotes}[para,flushleft]
\item[1] The errors for $T_1$ were performed with the errors of $T_{eff}$ being taken into account. And the errors for $T_2$ were performed with the errors of $T_1$ being taken into account.\\
\item[2] For contact binaries, $\Omega_{out} \leq \Omega \leq \Omega_{inner}$, where $\Omega_{out}$ and $\Omega_{inner}$ are the dimensionless potentials at the outer and inner Lagrangian surface respectively \citep{2009ebs..book.....K}. Contact degree ($f$) was calculated by the equation $f=\frac{\Omega_{inner}-\Omega}{\Omega_{inner}-\Omega_{out}}$ \citep{2009ebs..book.....K}.
\end{tablenotes}
\end{threeparttable}
\end{table*}

\begin{figure*}
        \centering
	\begin{minipage}[m]{0.23\linewidth}
	    \centering
	      \includegraphics[width=1.6in]{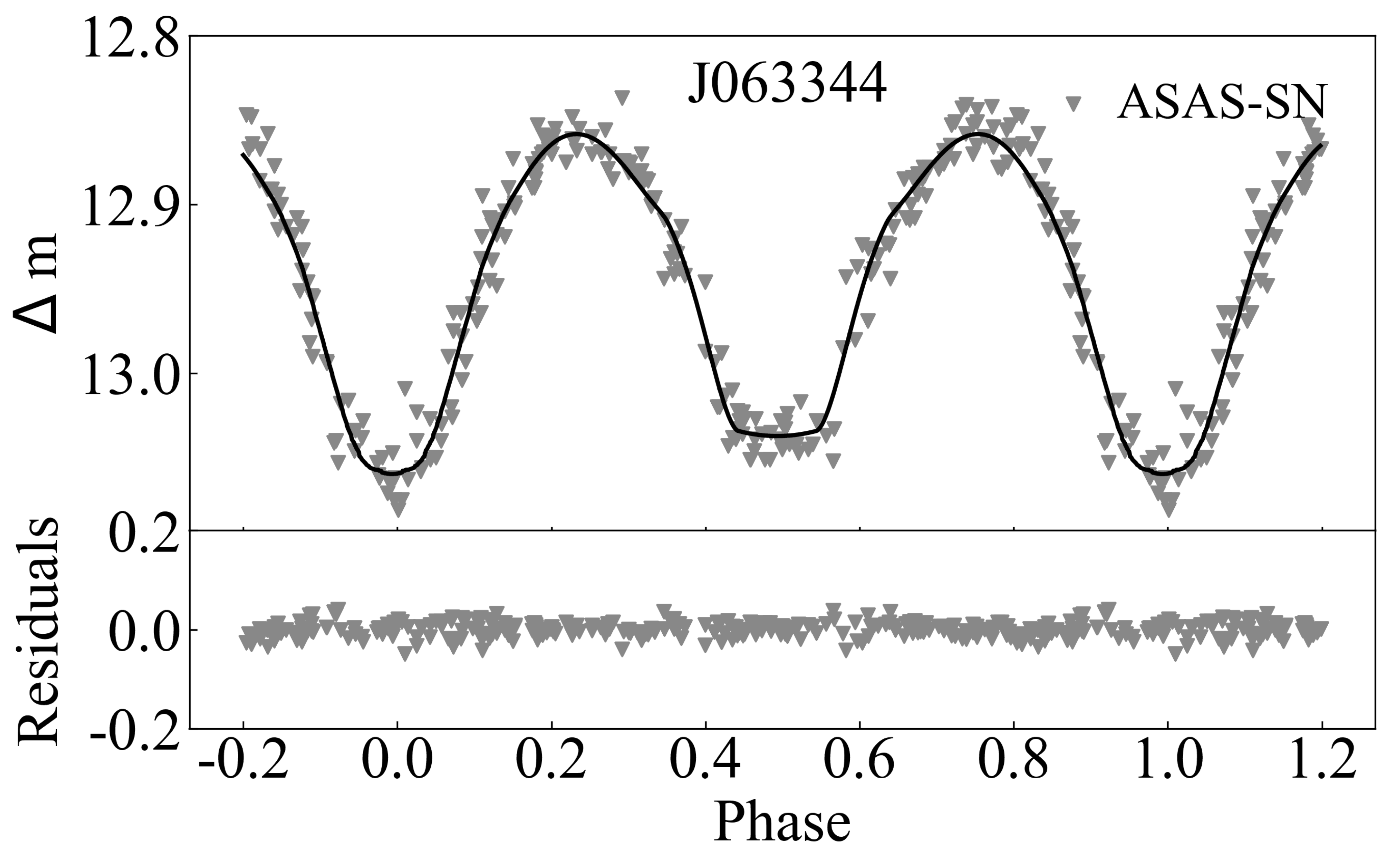}
        \end{minipage}%
        \begin{minipage}[m]{0.23\linewidth}
	    \centering
	    \includegraphics[width=1.6in]{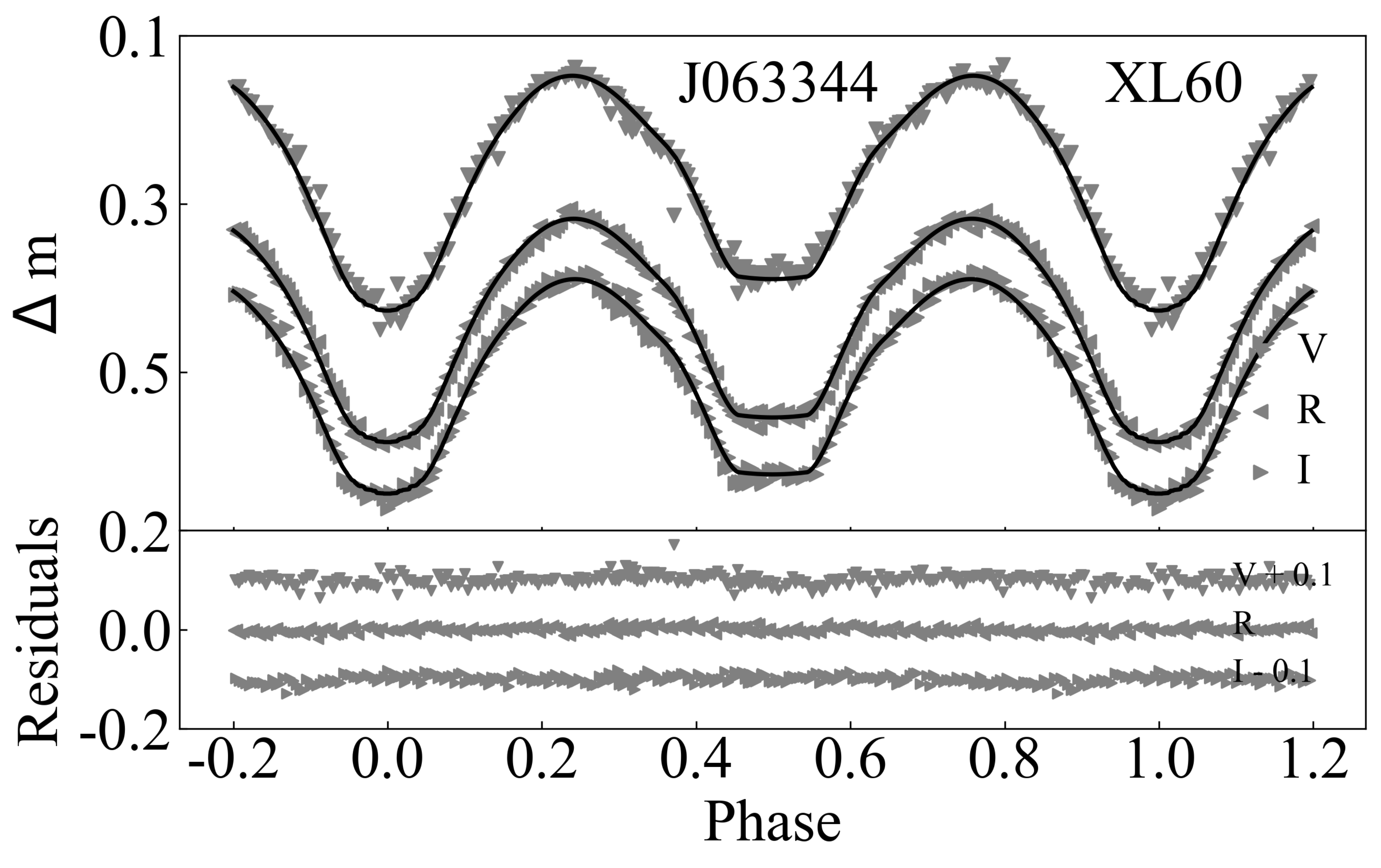}
        \end{minipage}%
	  \begin{minipage}[m]{0.23\linewidth}
	    \centering
	      \includegraphics[width=1.6in]{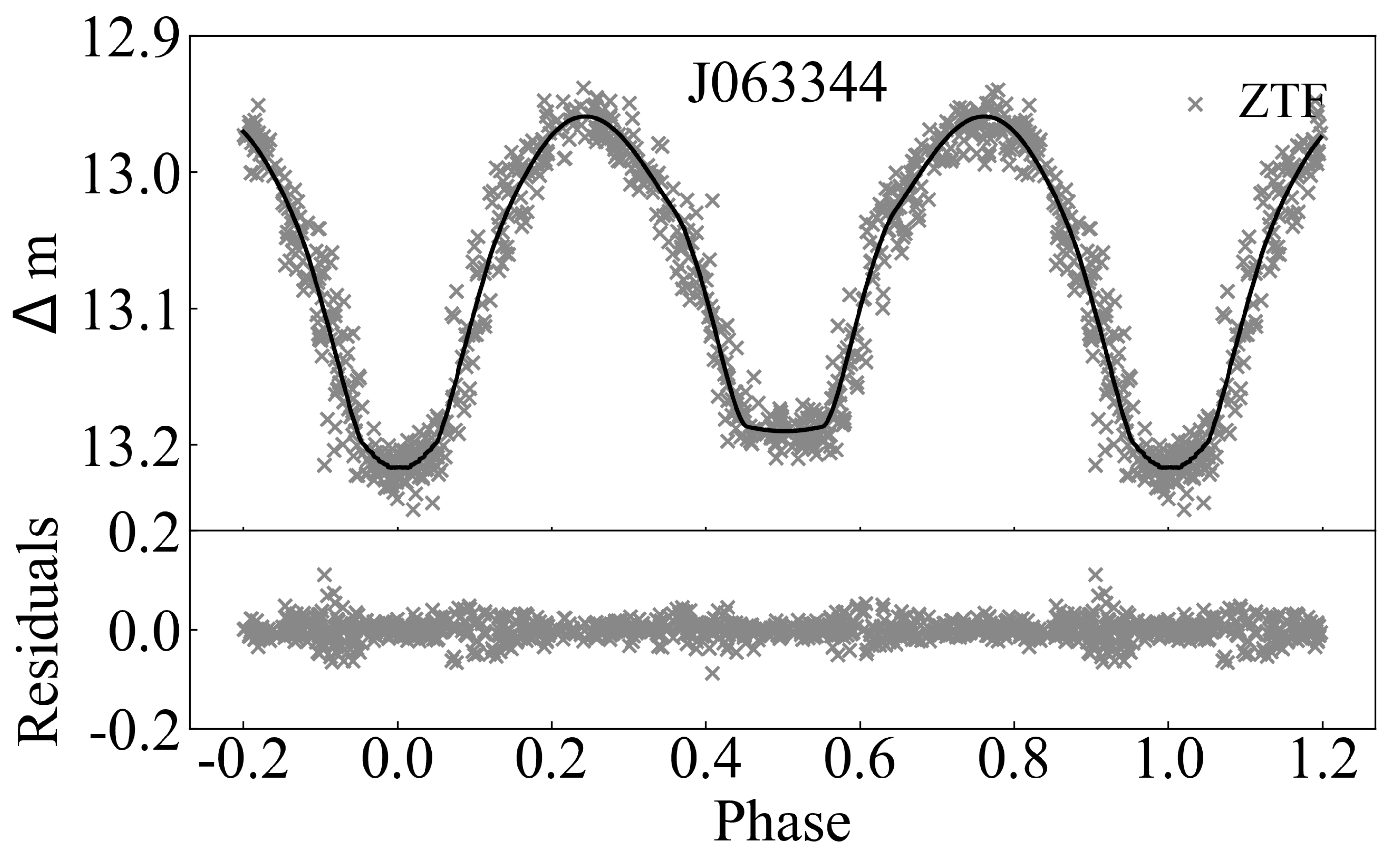}
        \end{minipage}%
        \begin{minipage}[m]{0.23\linewidth}
	    \centering
	    \includegraphics[width=1.6in]{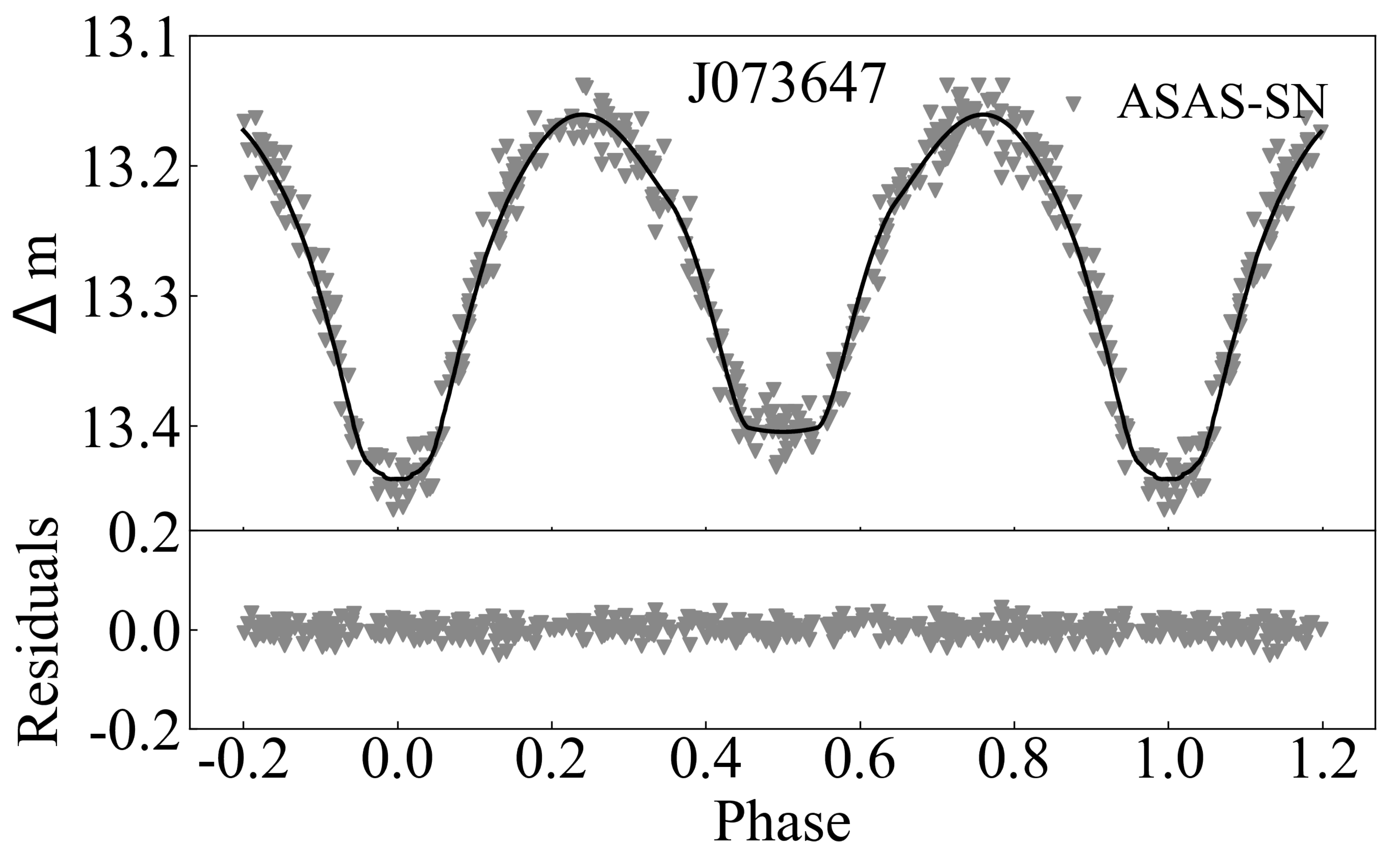}
        \end{minipage}%
        
        \begin{minipage}[m]{0.23\linewidth}
	    \centering
	    \includegraphics[width=1.6in]{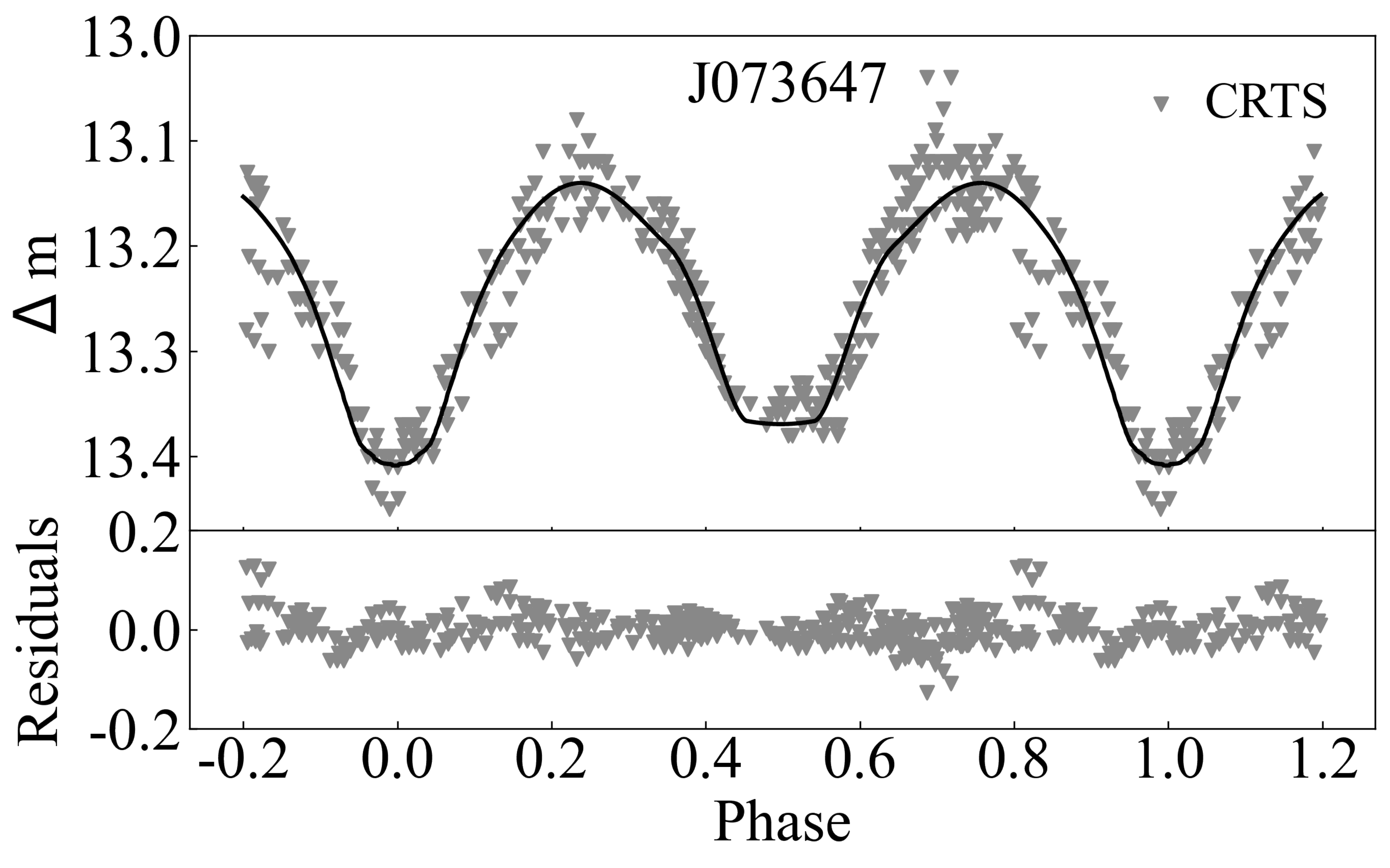}
        \end{minipage}%
	\begin{minipage}[m]{0.23\linewidth}
	    \centering
	      \includegraphics[width=1.6in]{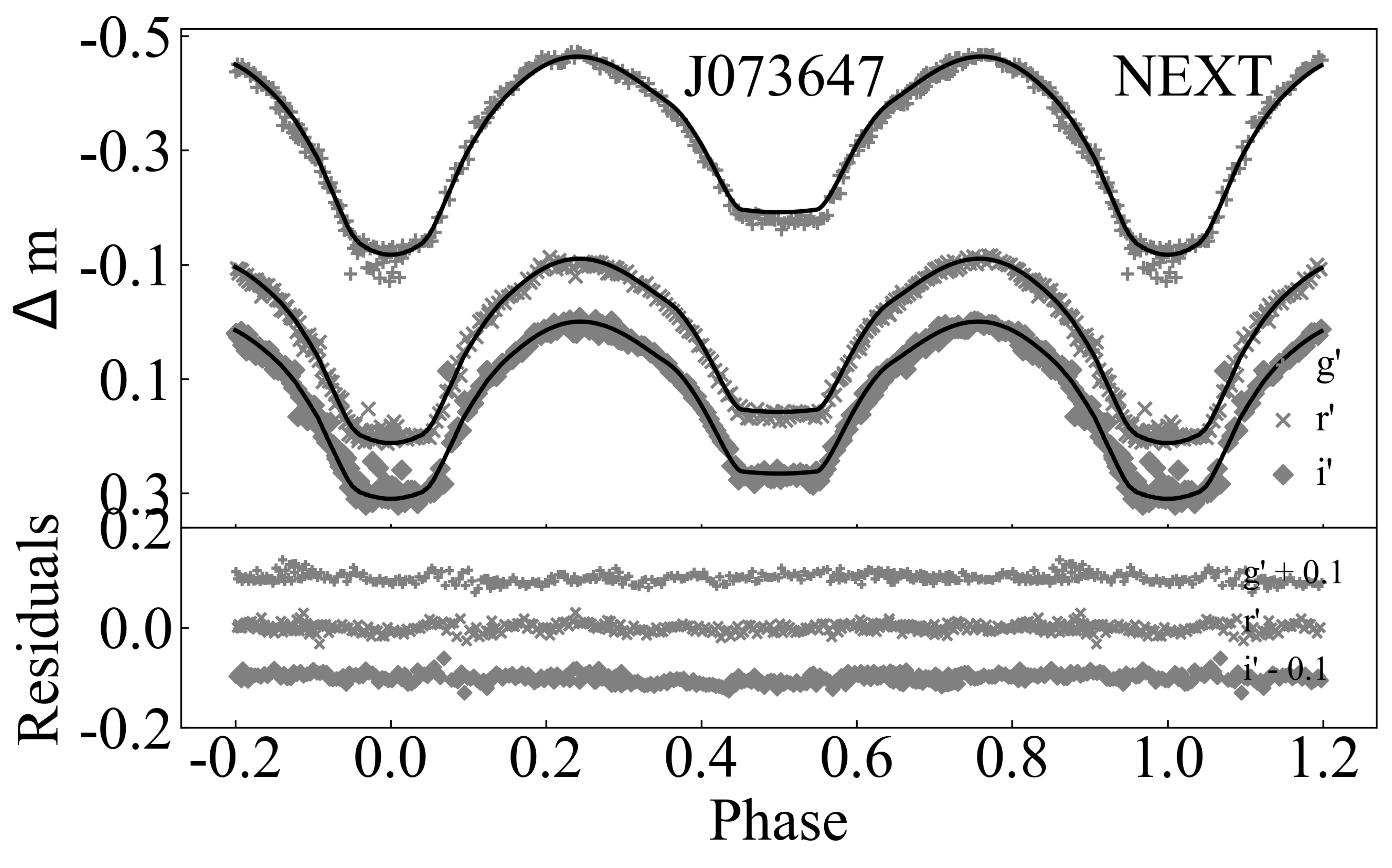}
        \end{minipage}%
        \begin{minipage}[m]{0.23\linewidth}
	    \centering
	    \includegraphics[width=1.6in]{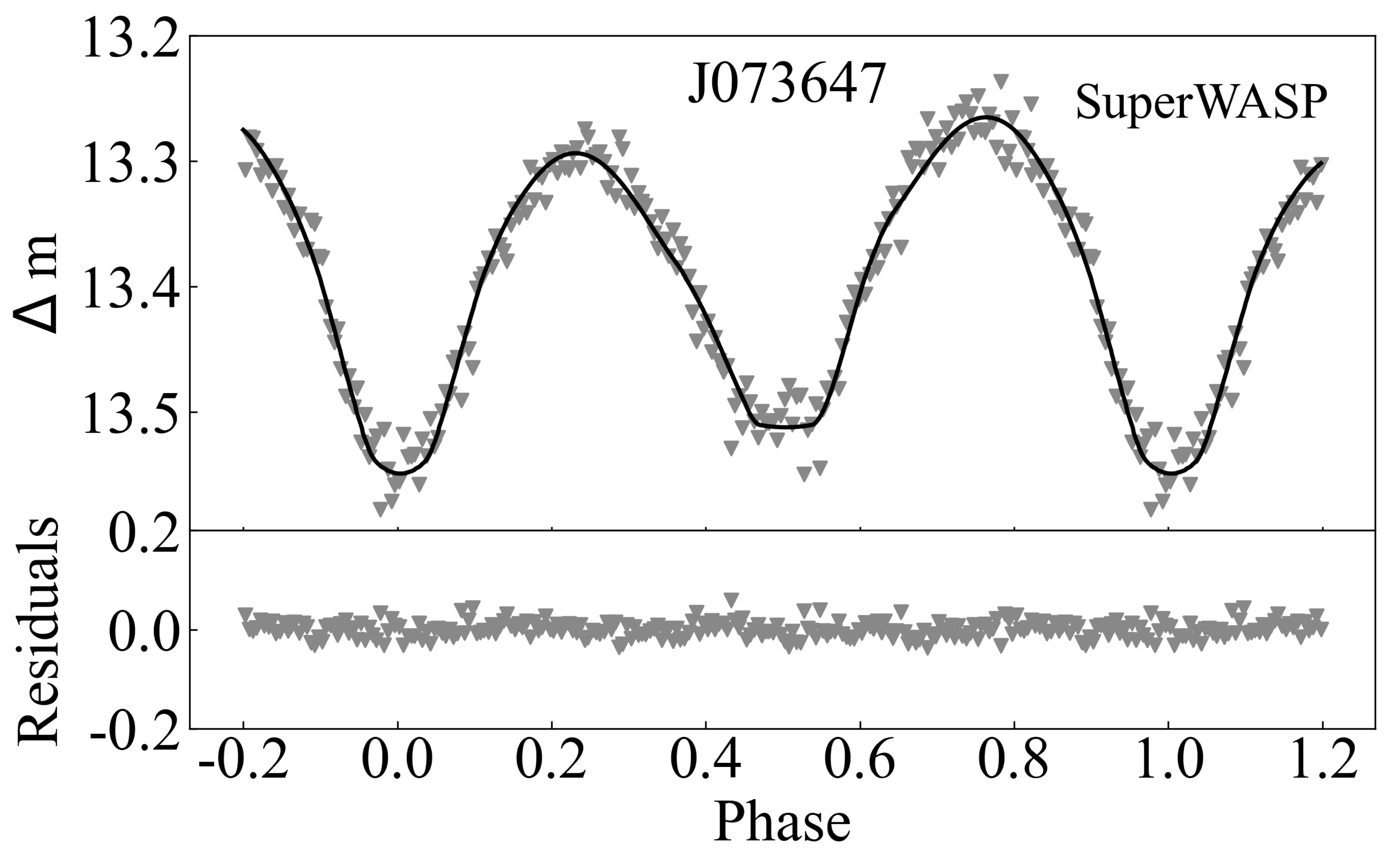}
        \end{minipage}%
        \begin{minipage}[m]{0.23\linewidth}
	    \centering
	    \includegraphics[width=1.6in]{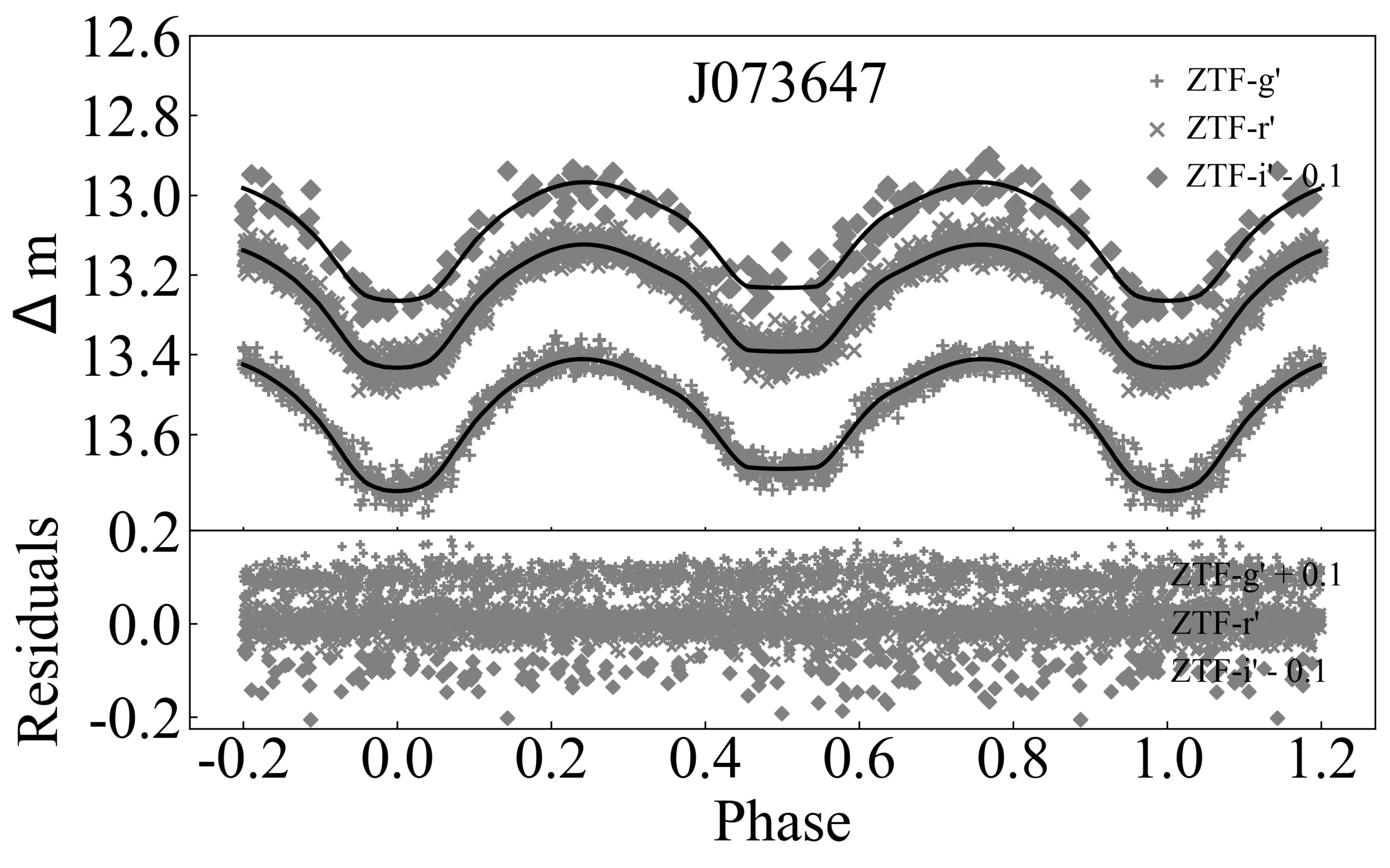}
        \end{minipage}%
        
	\begin{minipage}[m]{0.23\linewidth}
	    \centering
	      \includegraphics[width=1.6in]{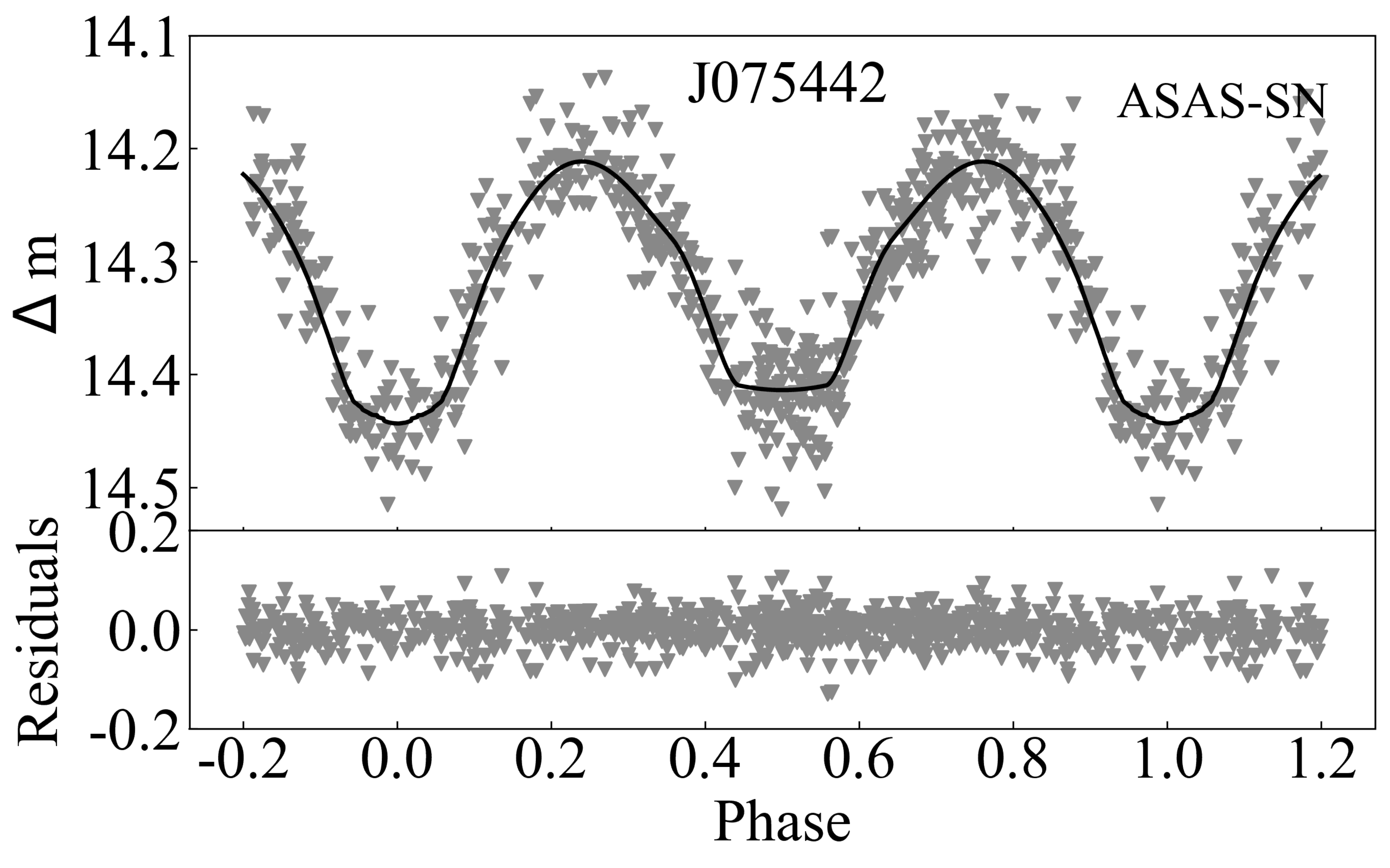}
        \end{minipage}%
        \begin{minipage}[m]{0.23\linewidth}
	    \centering
	    \includegraphics[width=1.6in]{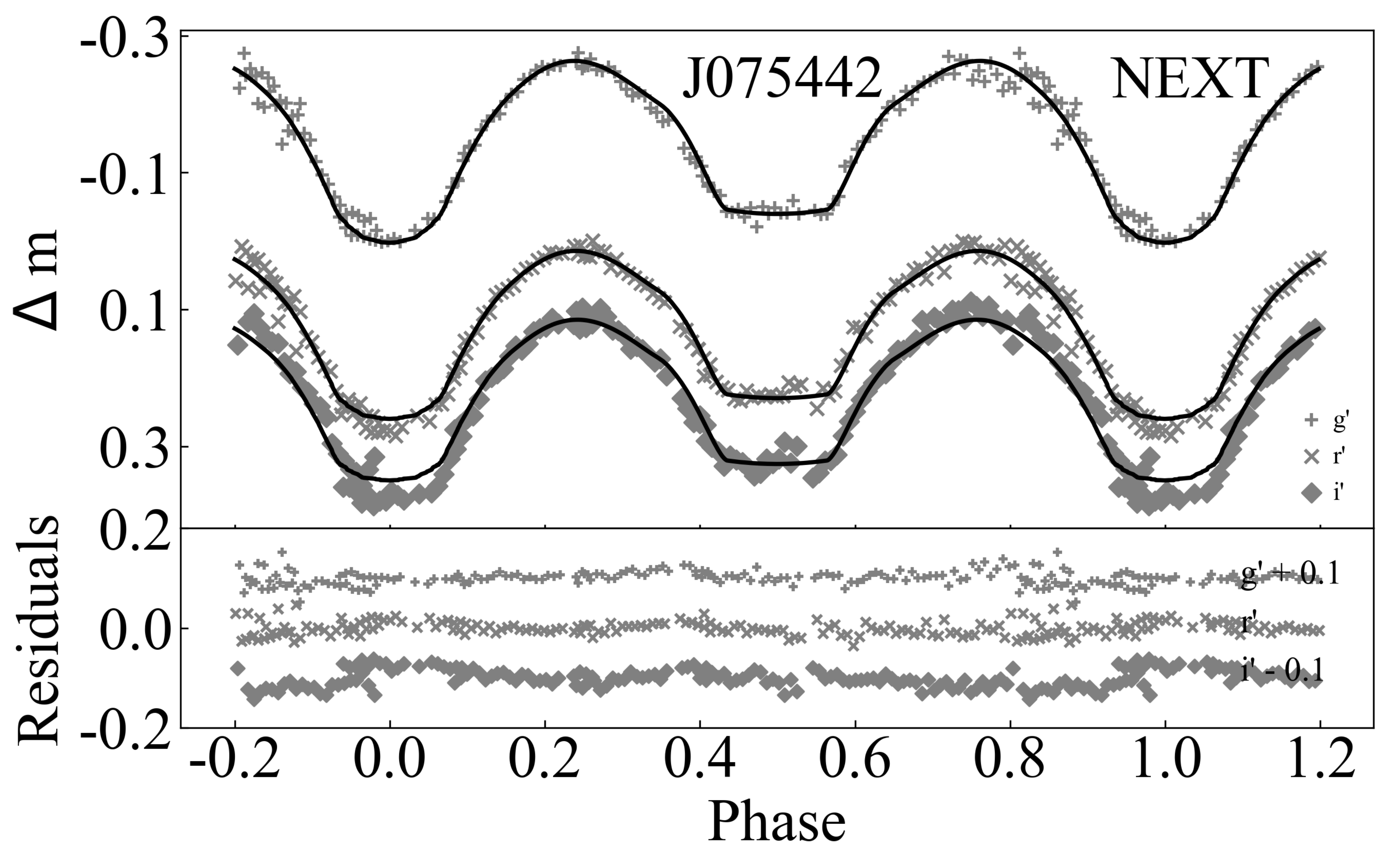}
        \end{minipage}%
        \begin{minipage}[m]{0.23\linewidth}
	    \centering
	    \includegraphics[width=1.6in]{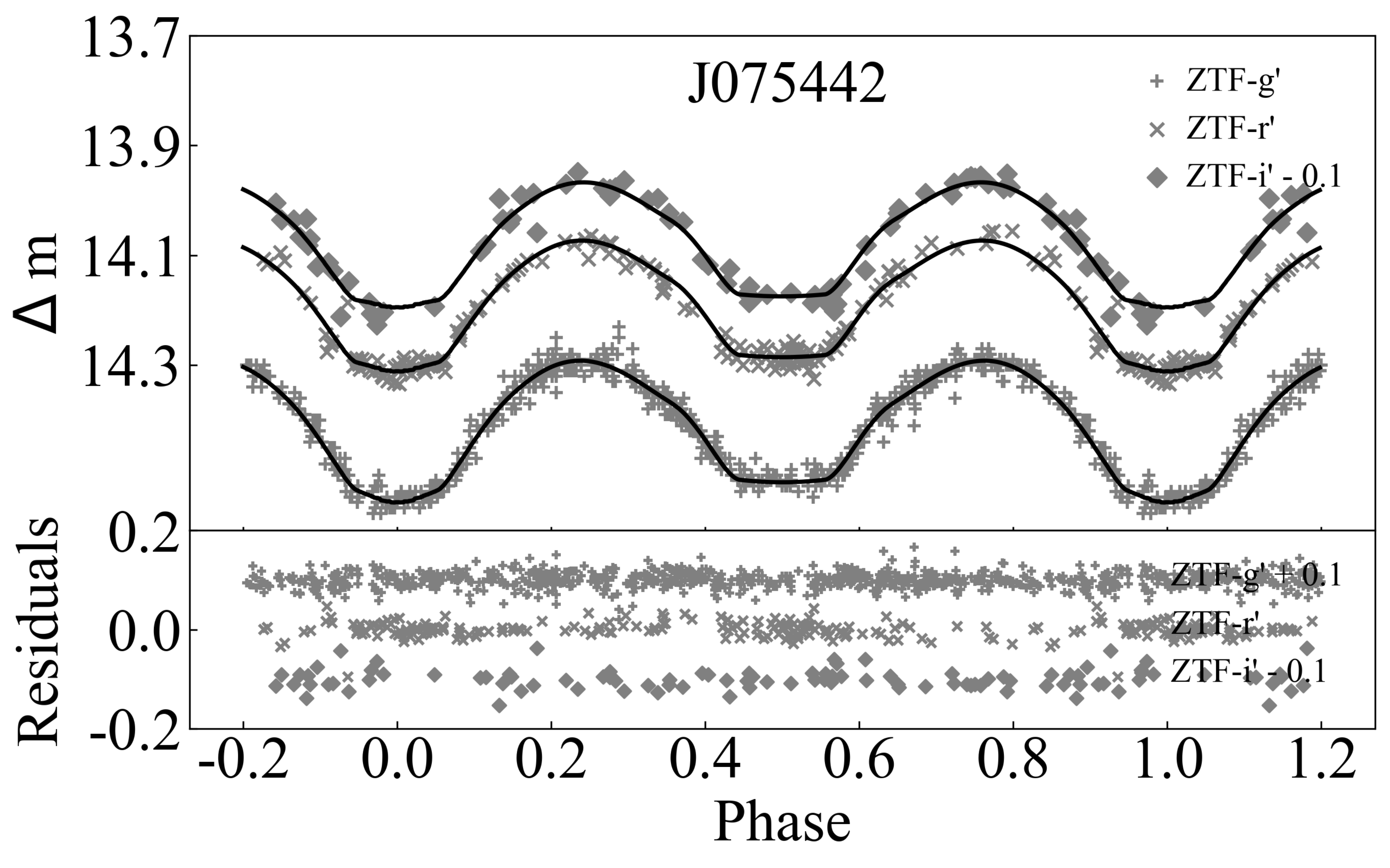}
        \end{minipage}%
	\begin{minipage}[m]{0.23\linewidth}
	    \centering
	      \includegraphics[width=1.6in]{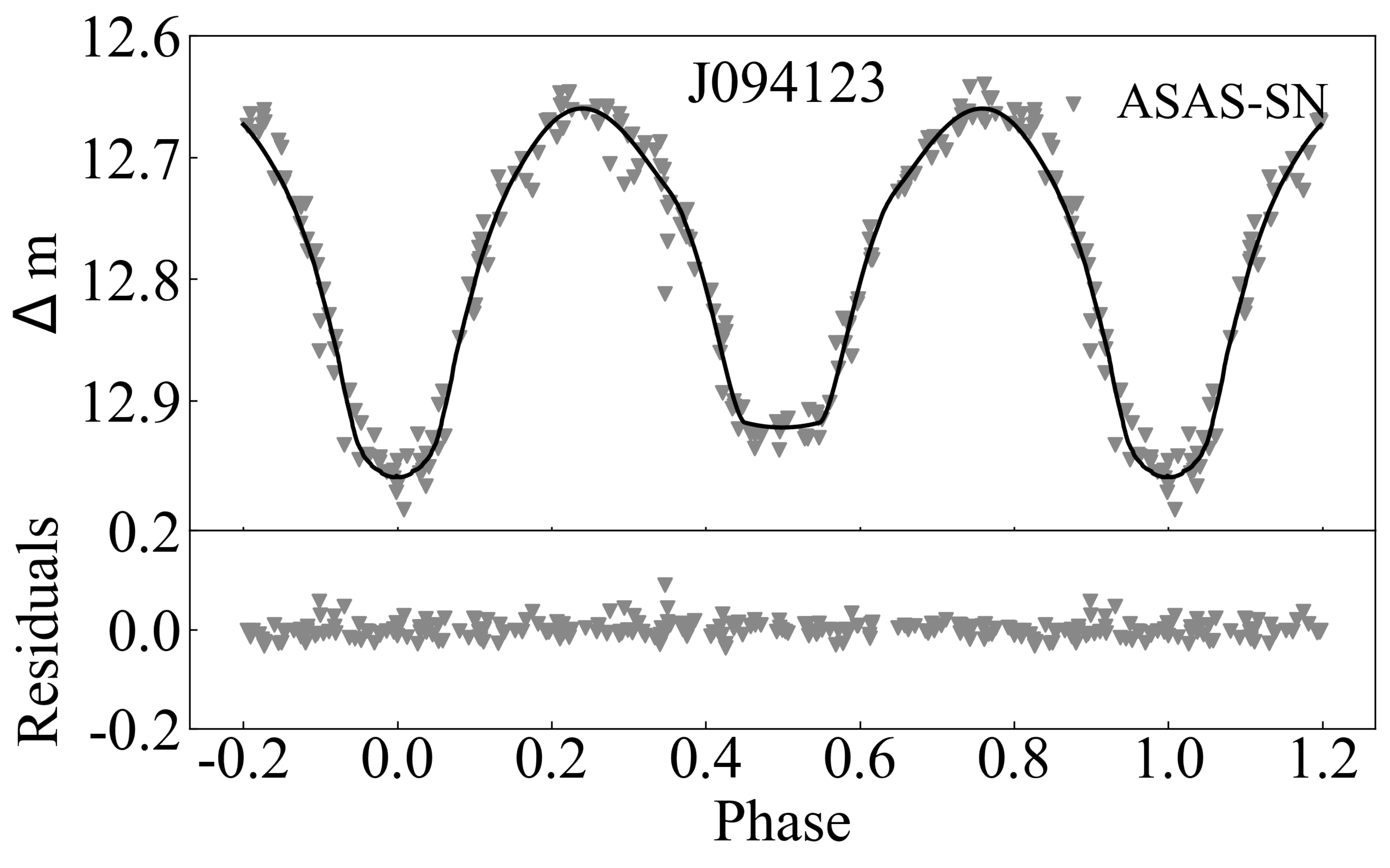}
        \end{minipage}%
        
        \begin{minipage}[m]{0.23\linewidth}
	    \centering
	    \includegraphics[width=1.6in]{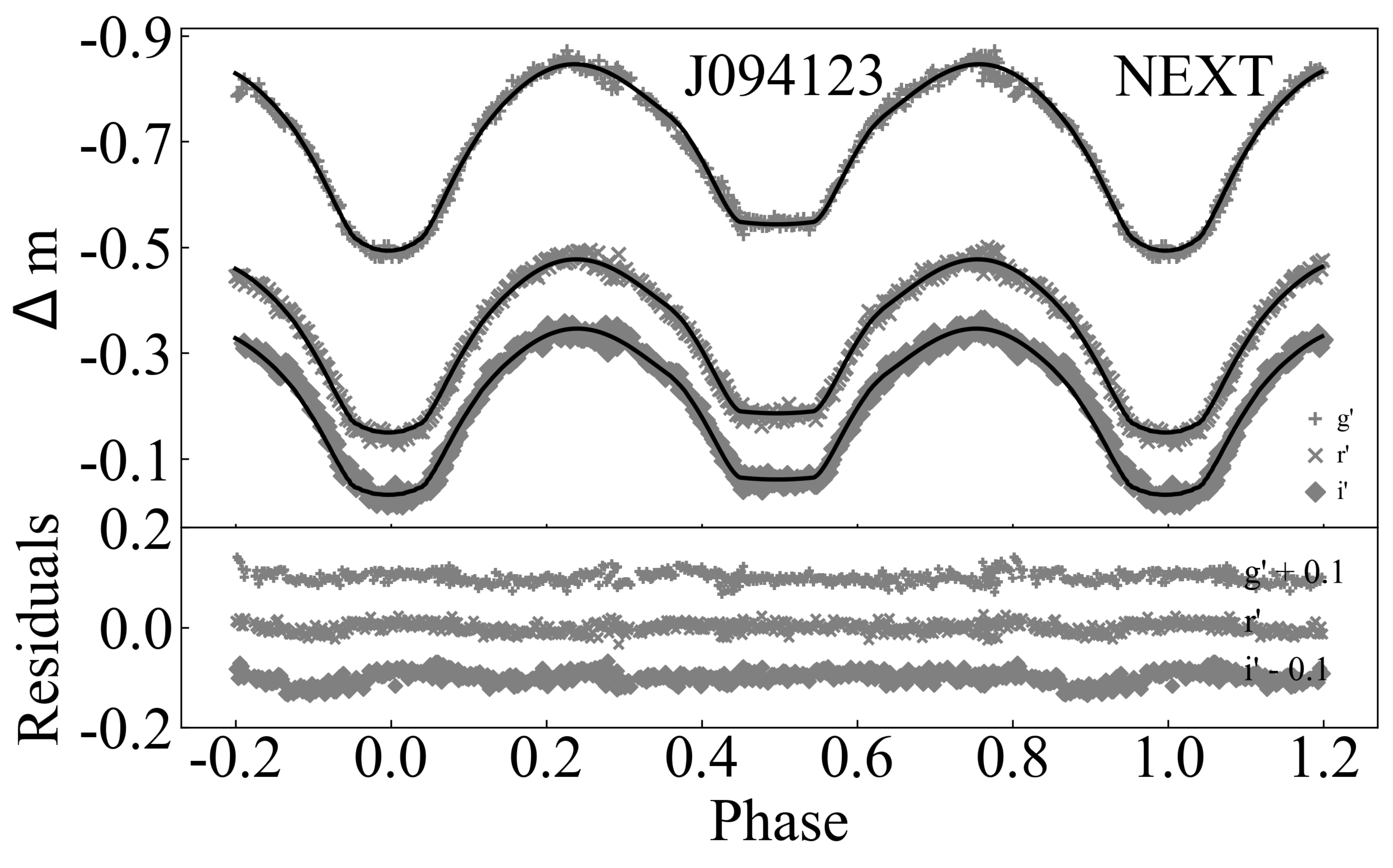}
        \end{minipage}%
        \begin{minipage}[m]{0.23\linewidth}
	    \centering
	    \includegraphics[width=1.6in]{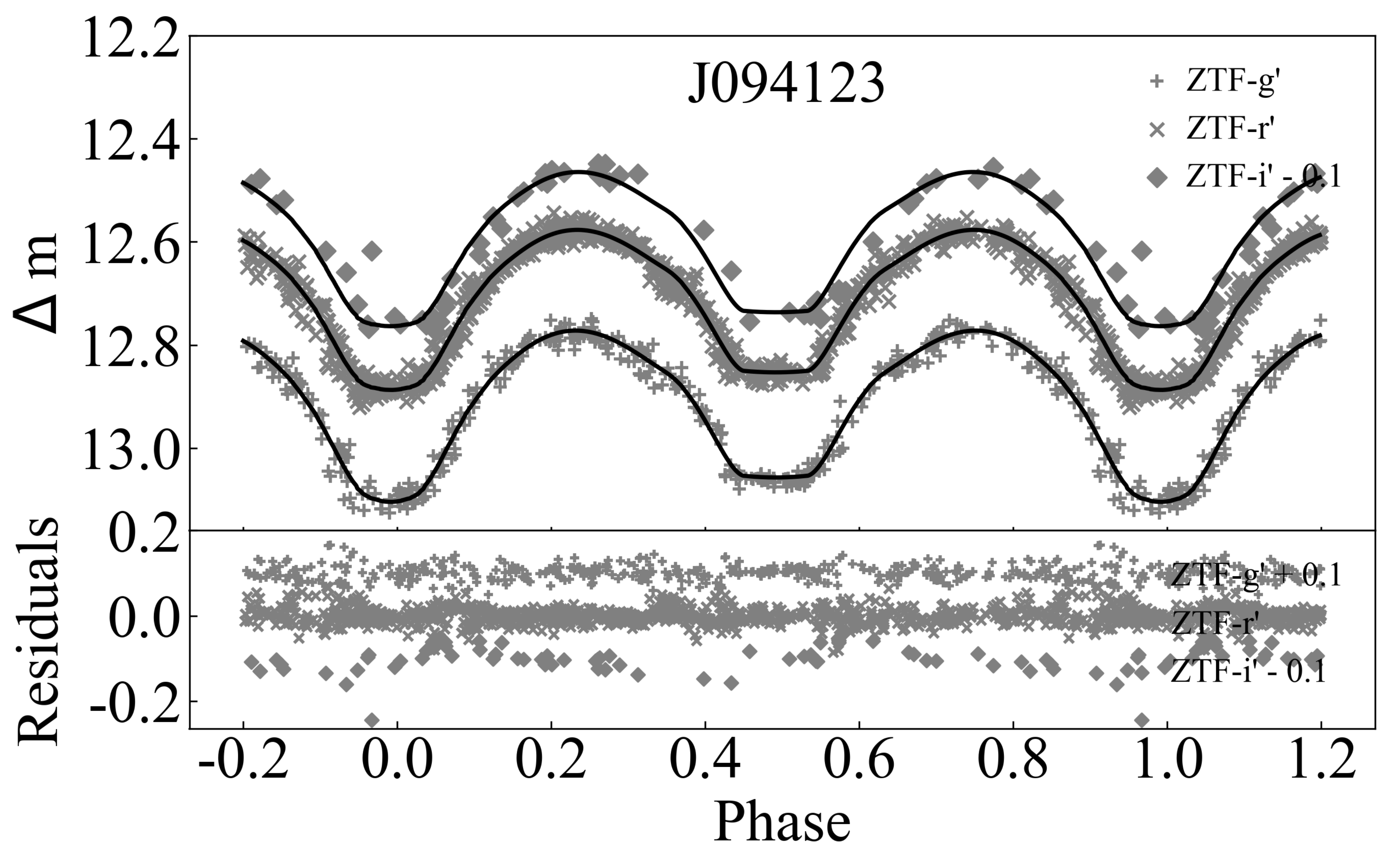}
        \end{minipage}%
	\begin{minipage}[m]{0.23\linewidth}
	    \centering
	      \includegraphics[width=1.6in]{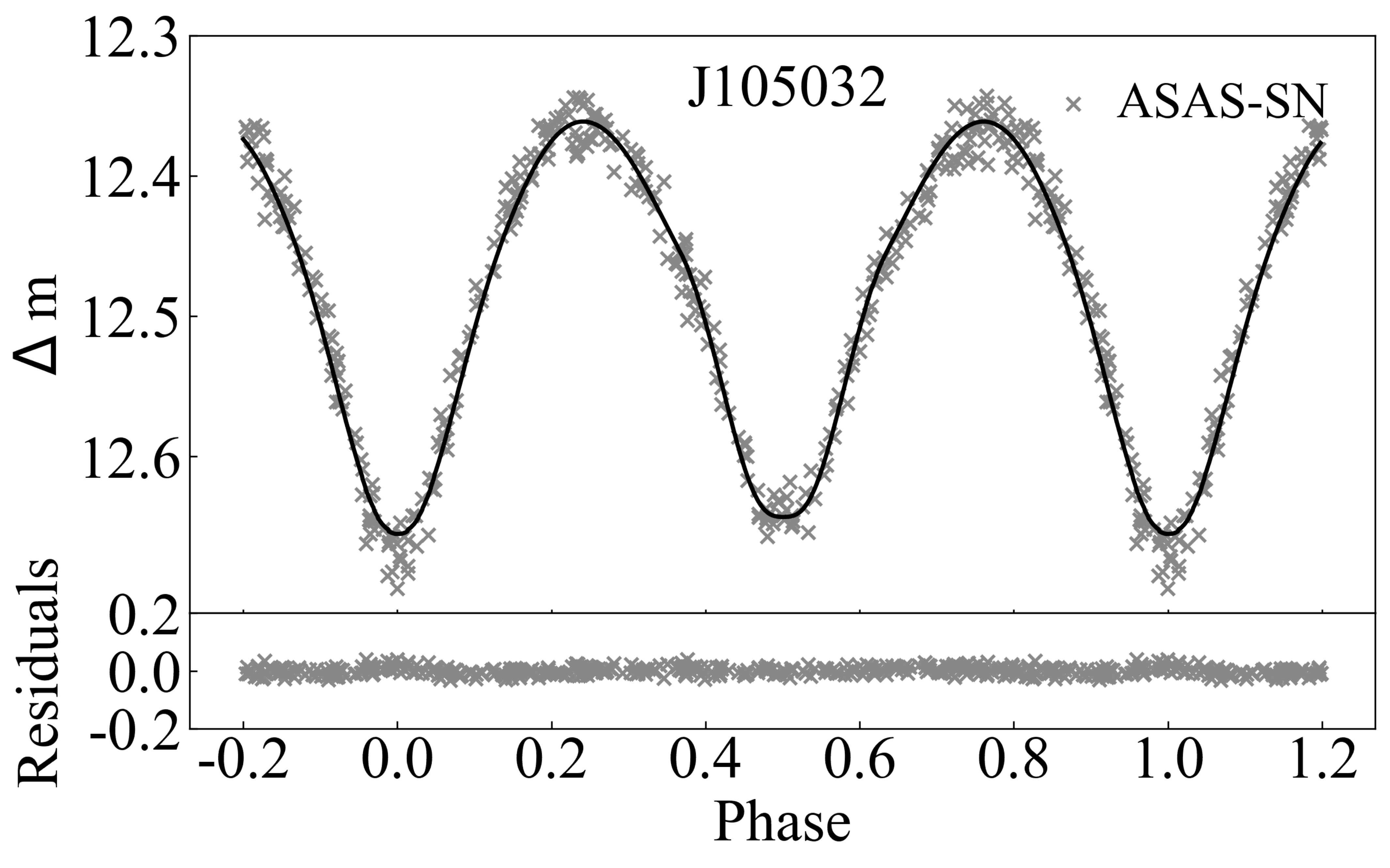}
        \end{minipage}%
        \begin{minipage}[m]{0.23\linewidth}
	    \centering
	    \includegraphics[width=1.6in]{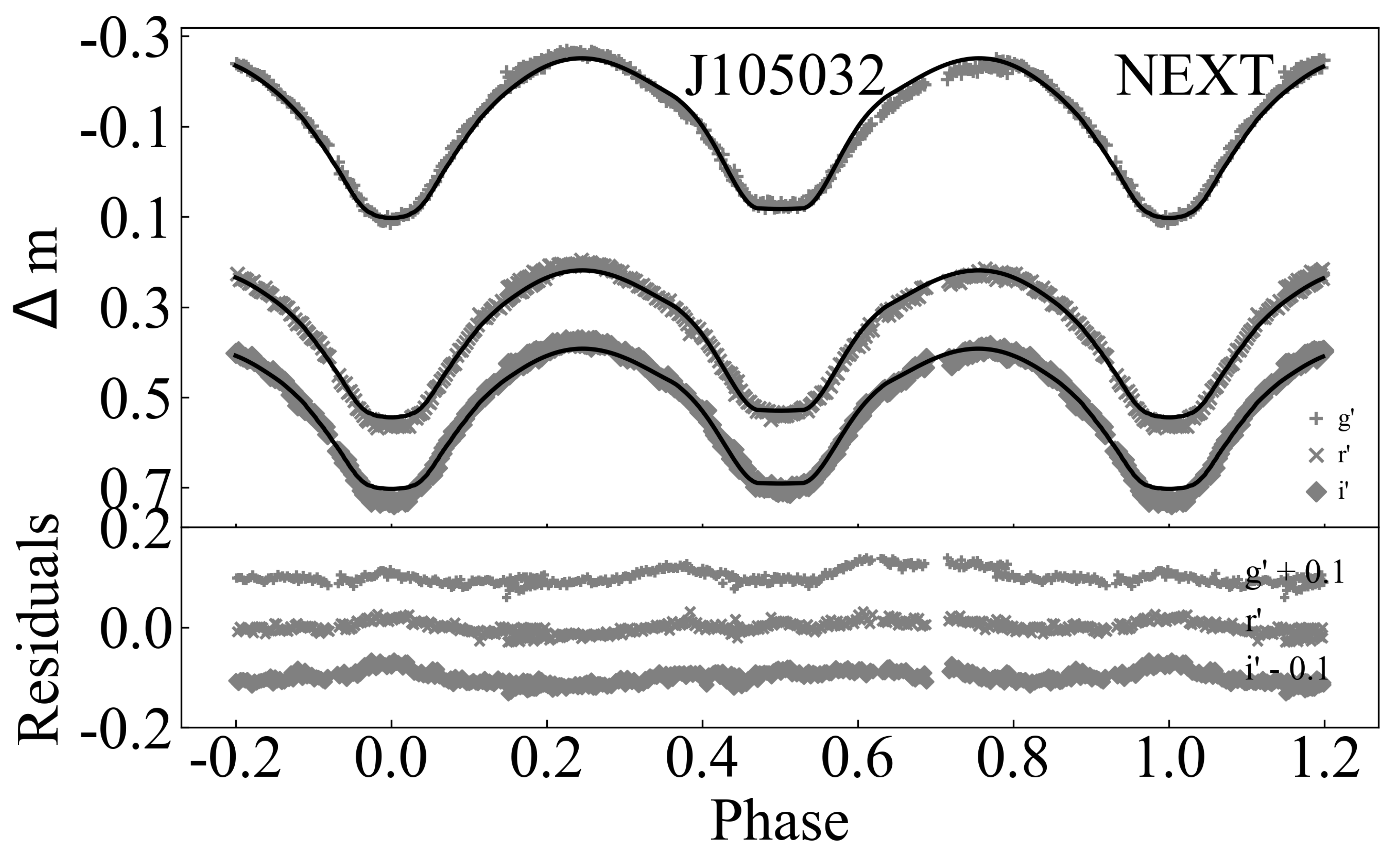}
        \end{minipage}%
        
        \begin{minipage}[m]{0.23\linewidth}
	    \centering
	    \includegraphics[width=1.6in]{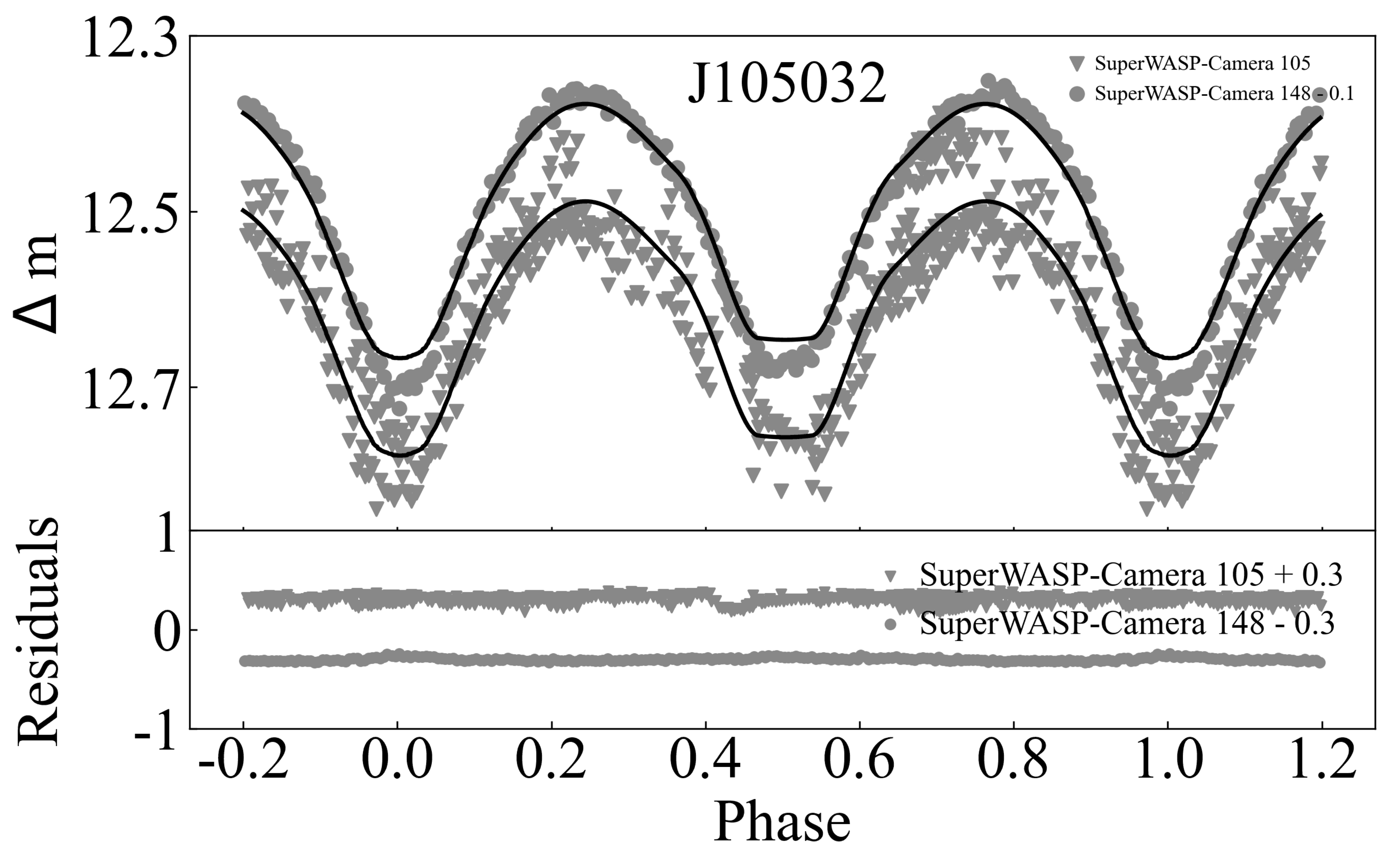}
        \end{minipage}%
	\begin{minipage}[m]{0.23\linewidth}
	    \centering
	      \includegraphics[width=1.6in]{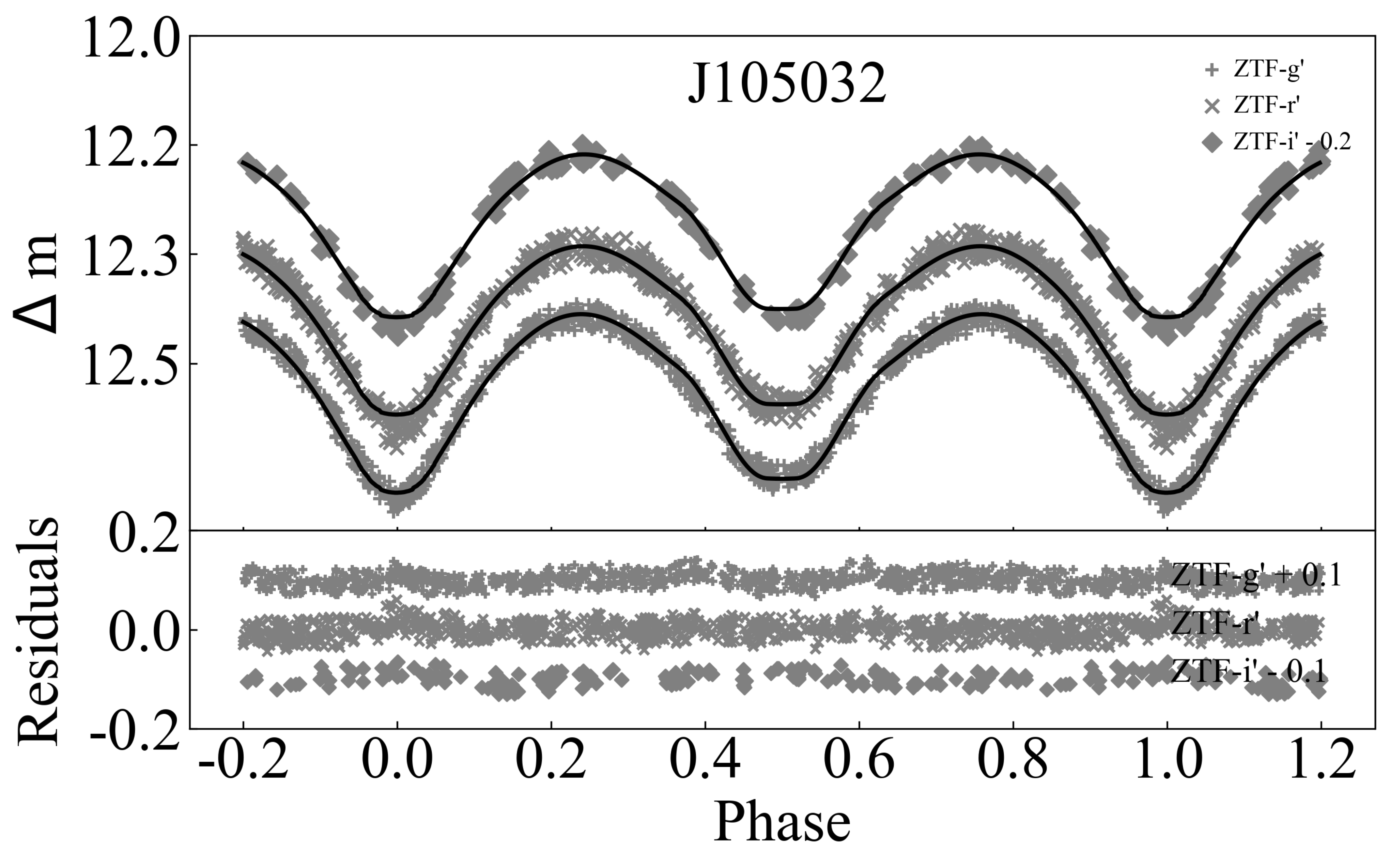}
        \end{minipage}%
        \begin{minipage}[m]{0.23\linewidth}
	    \centering
	    \includegraphics[width=1.6in]{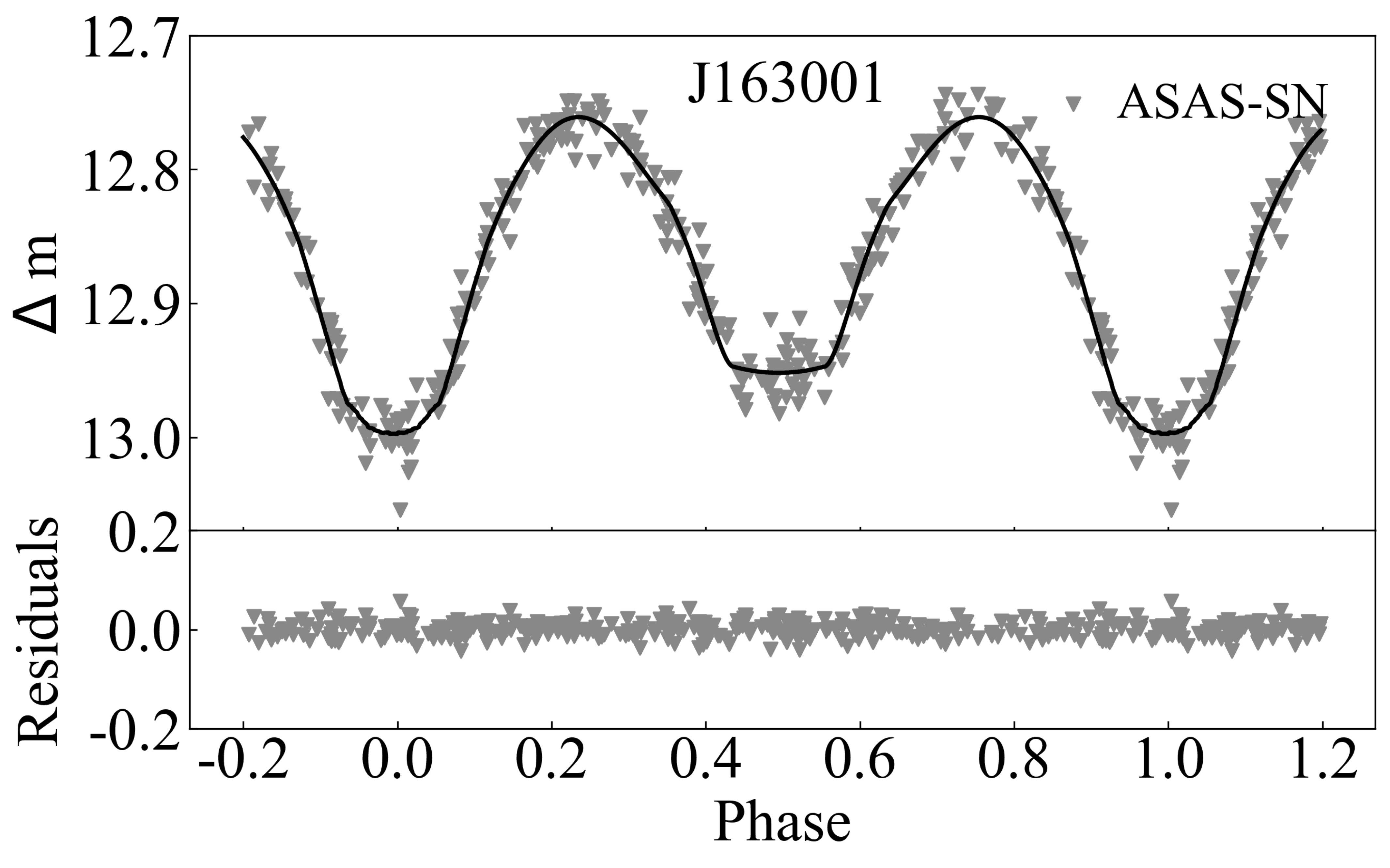}
        \end{minipage}%
        \begin{minipage}[m]{0.23\linewidth}
	    \centering
	    \includegraphics[width=1.6in]{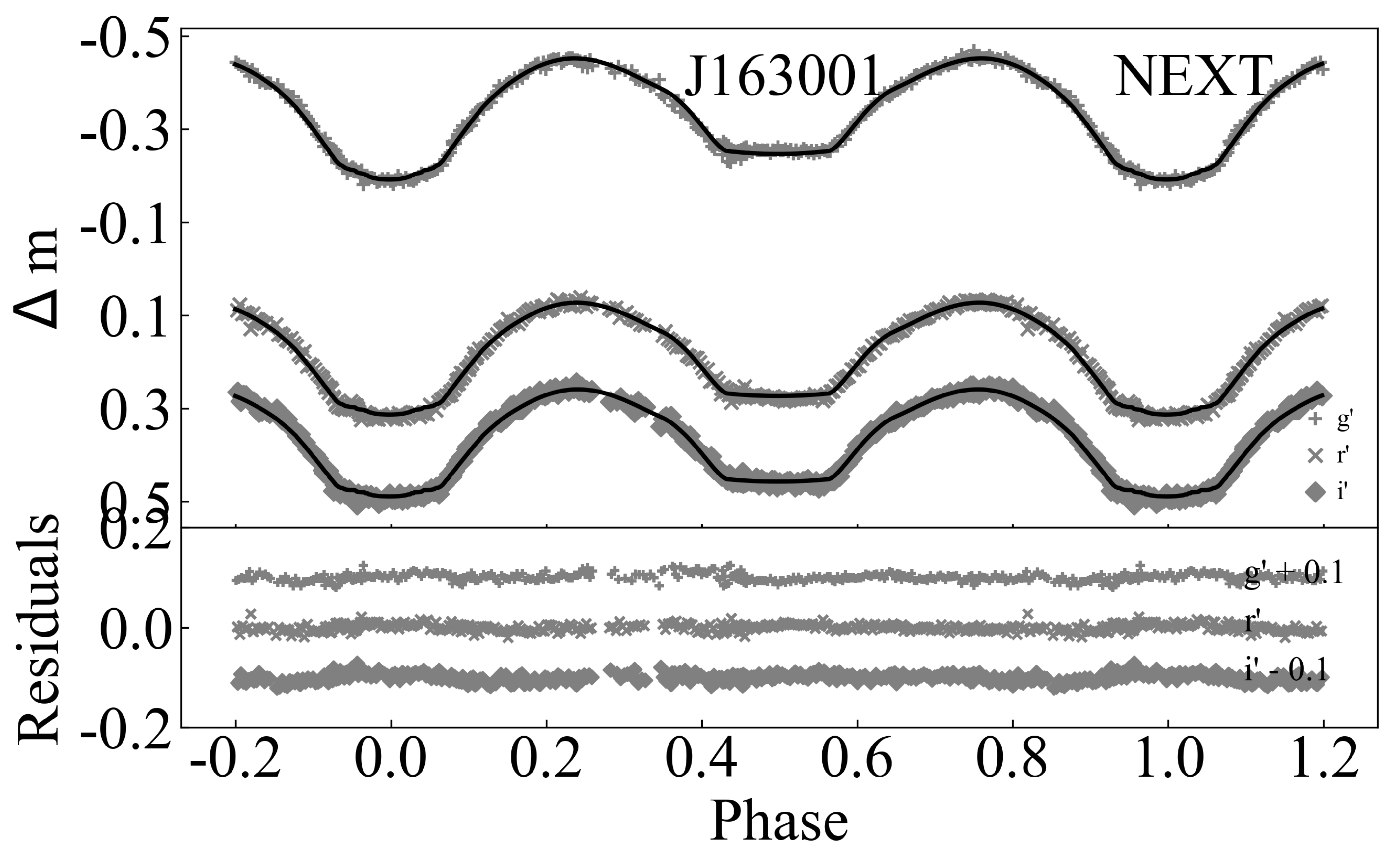}
        \end{minipage}%
        
	\begin{minipage}[m]{0.23\linewidth}
	    \centering
	      \includegraphics[width=1.6in]{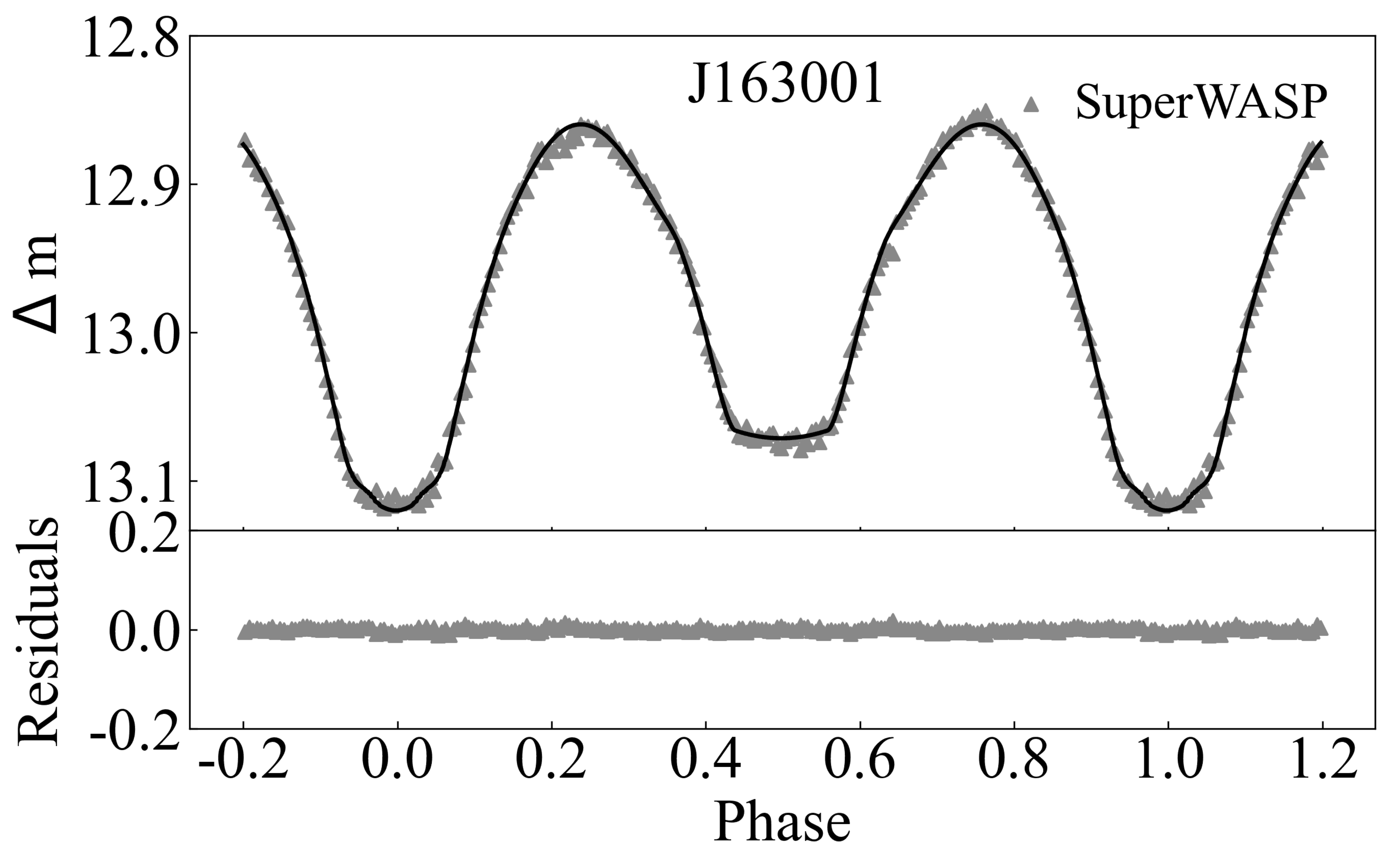}
        \end{minipage}%
        \begin{minipage}[m]{0.23\linewidth}
	    \centering
	    \includegraphics[width=1.6in]{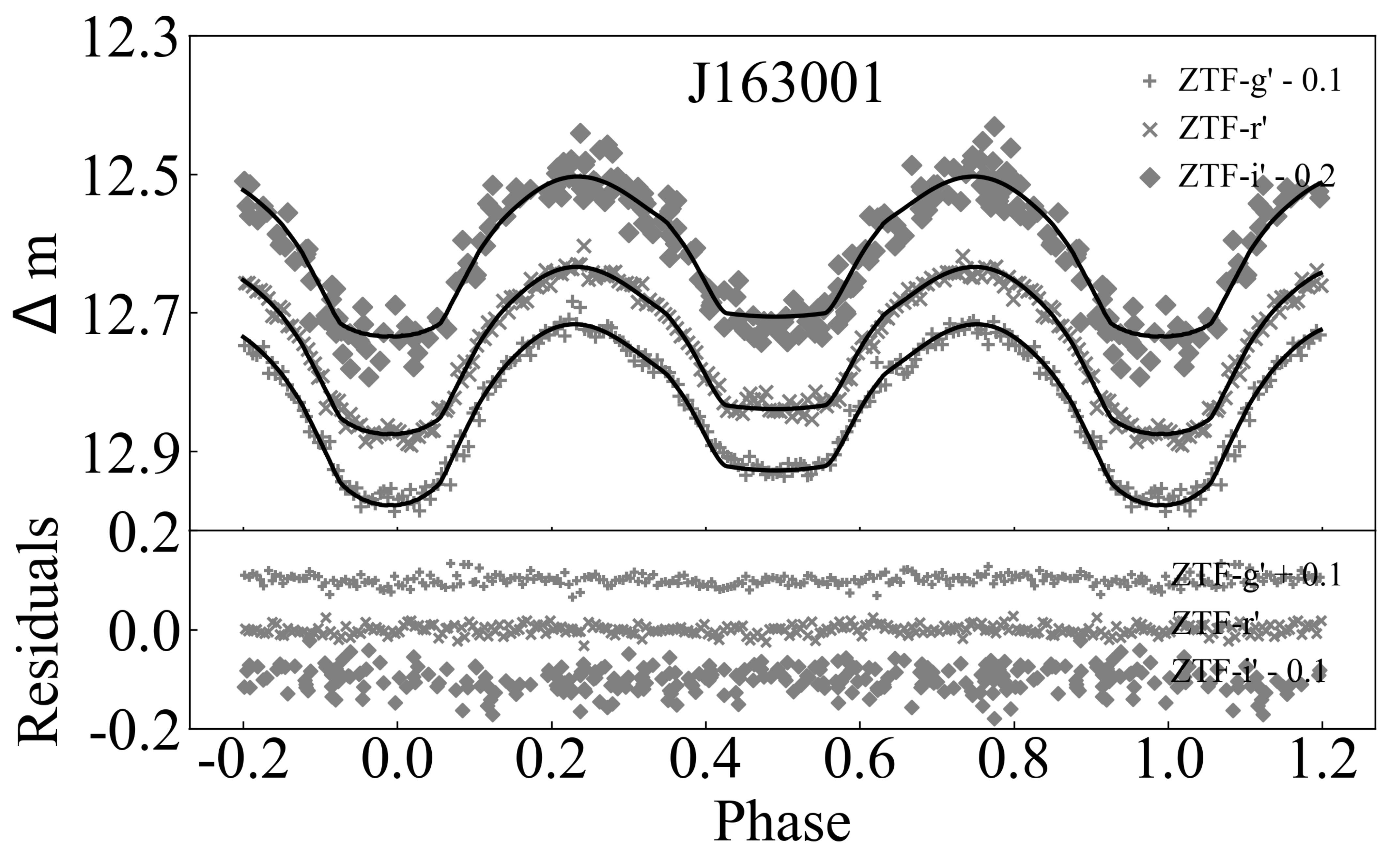}
        \end{minipage}%
    \caption{Comparison between the photometric data and the theoretical light curves for other telescopes. The different symbols represent the different bands. The bottom panels show the residuals.}
    \label{fig:LCs of other telescopes}
\end{figure*}

Due to the large dispersion of data, which cannot form complete light curves, we discarded the data of CRTS for J063344, J075442, J094123, J105032, and J163001, and the data of SuperWASP for J063344 and J075442. Figure \ref{fig:LCs of other telescopes} displays the fitted light curves of other telescopes. The results are listed in Table \ref{tab:Photometric solutions of J063344}, Table \ref{tab:Photometric solutions of J073647}, Table \ref{tab:Photometric solutions of J075442}, Table \ref{tab:Photometric solutions of J094123}, Table \ref{tab:Photometric solutions of J105032}, and Table \ref{tab:Photometric solutions of J163001}. Note that only the SuperWASP data exhibits the O'Connell effect for J073647. We simulated a cool spot on the secondary component to obtain better photometric solutions.  

\section{Orbital Period Analysis}
\label{sec:Orbital Period Analysis}

\begin{figure*}
        \centering
	\begin{minipage}[m]{0.3\linewidth}
	    \centering
	      \includegraphics[width=2in]{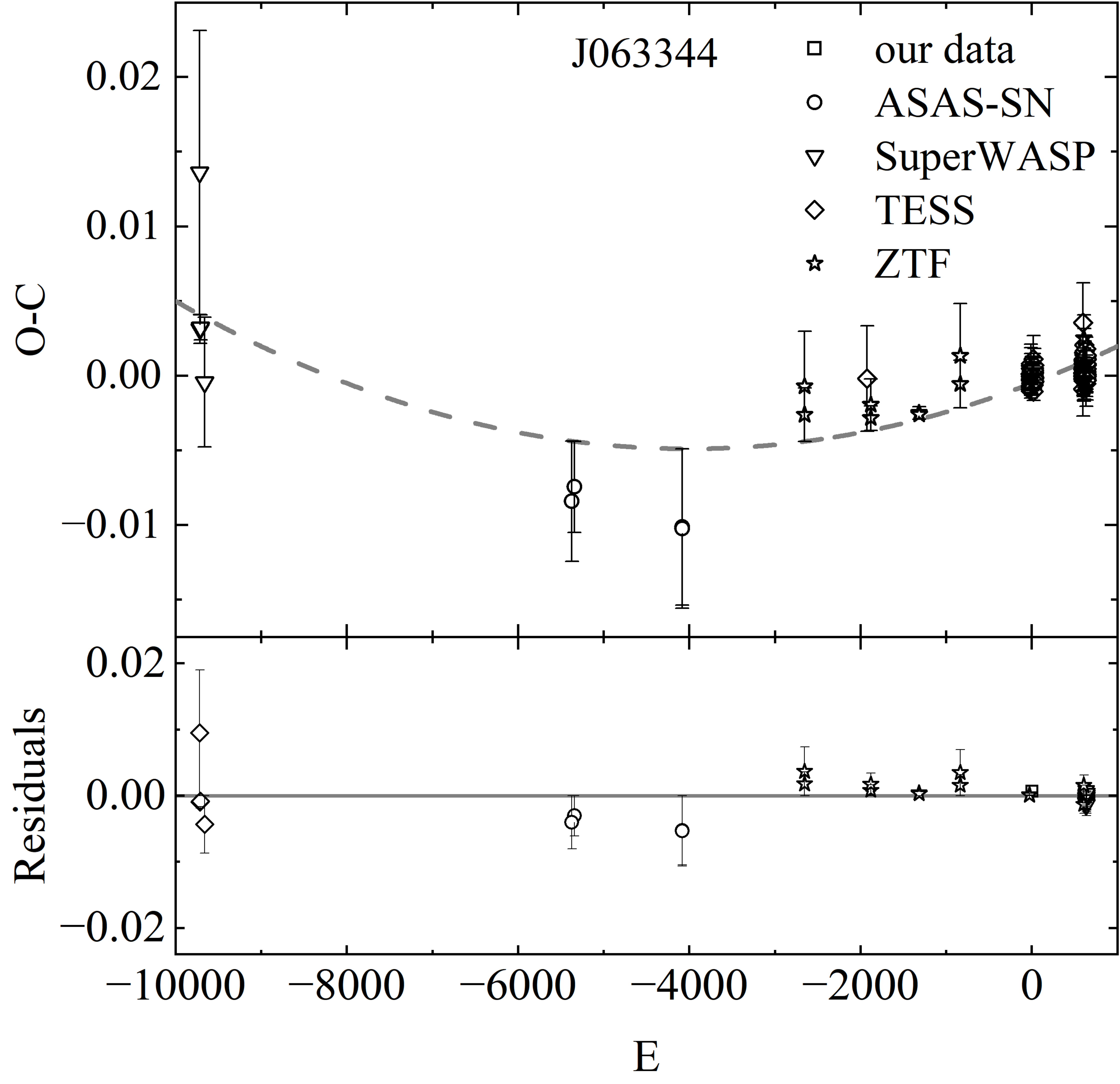}
        \end{minipage}%
        \begin{minipage}[m]{0.3\linewidth}
	    \centering
	    \includegraphics[width=2in]{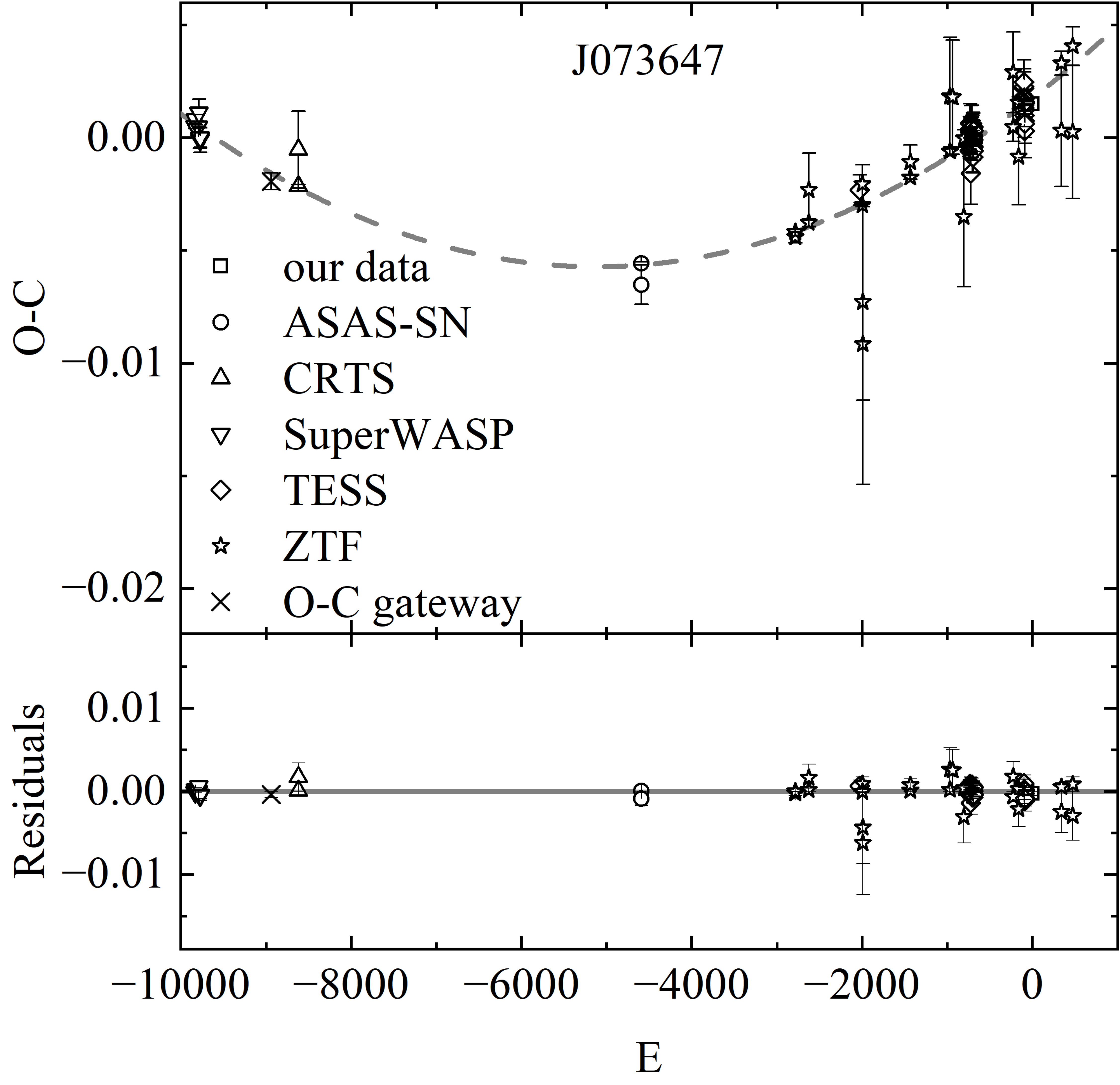}
        \end{minipage}%
        \begin{minipage}[m]{0.3\linewidth}
	    \centering
	    \includegraphics[width=2in]{O-C-J075442}
        \end{minipage}%

	\begin{minipage}[m]{0.3\linewidth}
	    \centering
	      \includegraphics[width=2in]{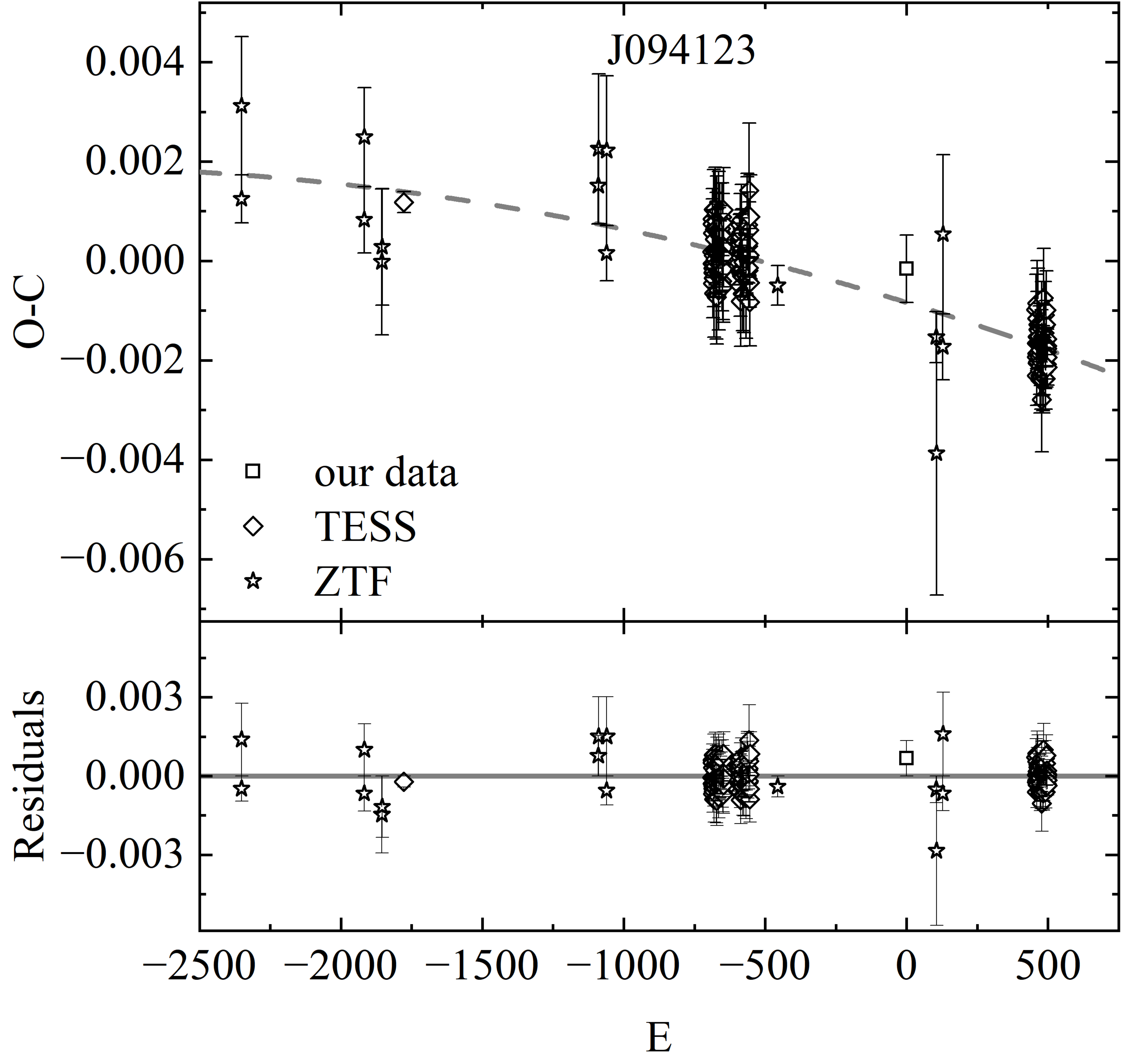}
        \end{minipage}%
        \begin{minipage}[m]{0.3\linewidth}
	    \centering
	    \includegraphics[width=2in]{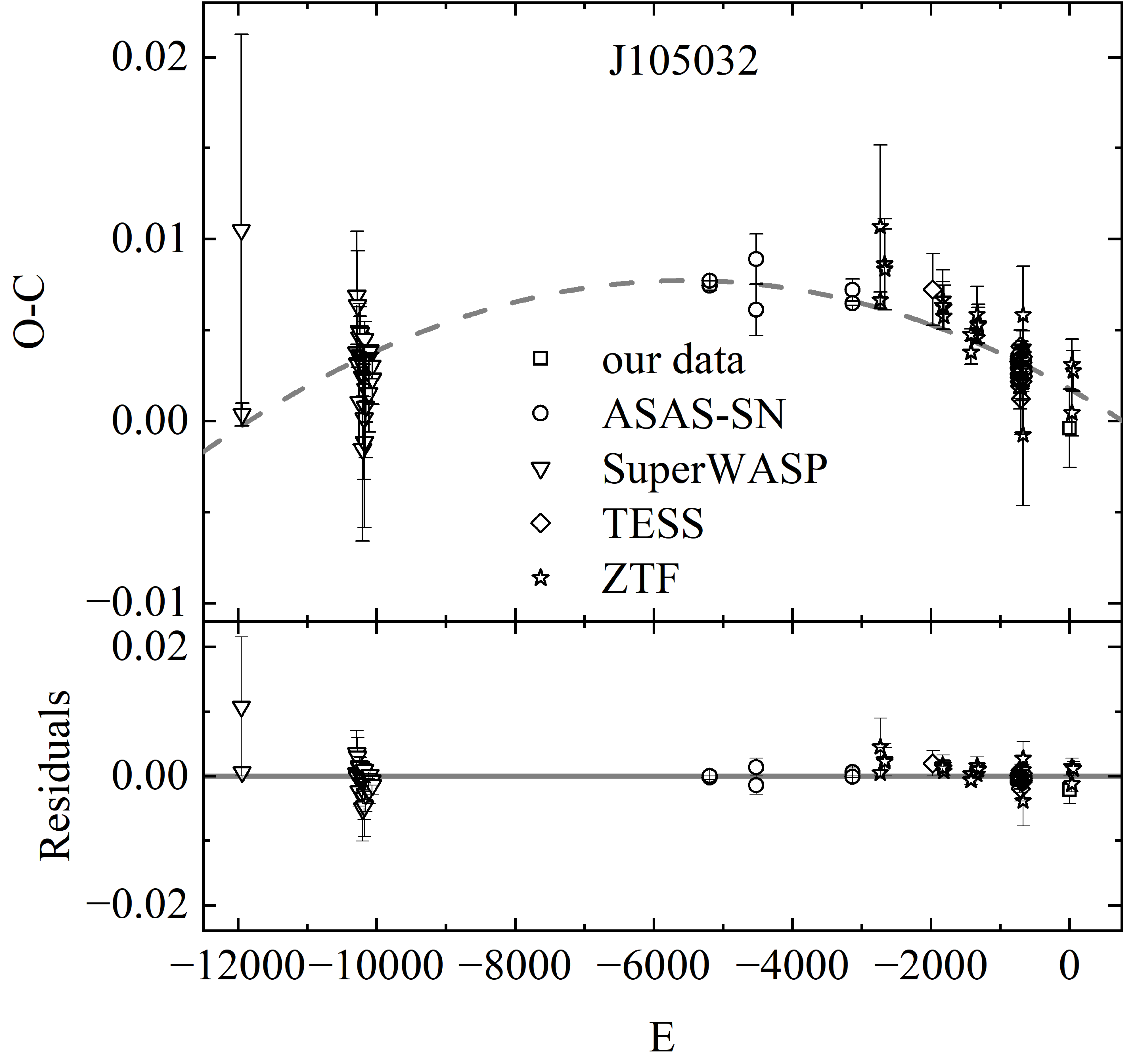}
        \end{minipage}%
        \begin{minipage}[m]{0.3\linewidth}
	    \centering
	    \includegraphics[width=2in]{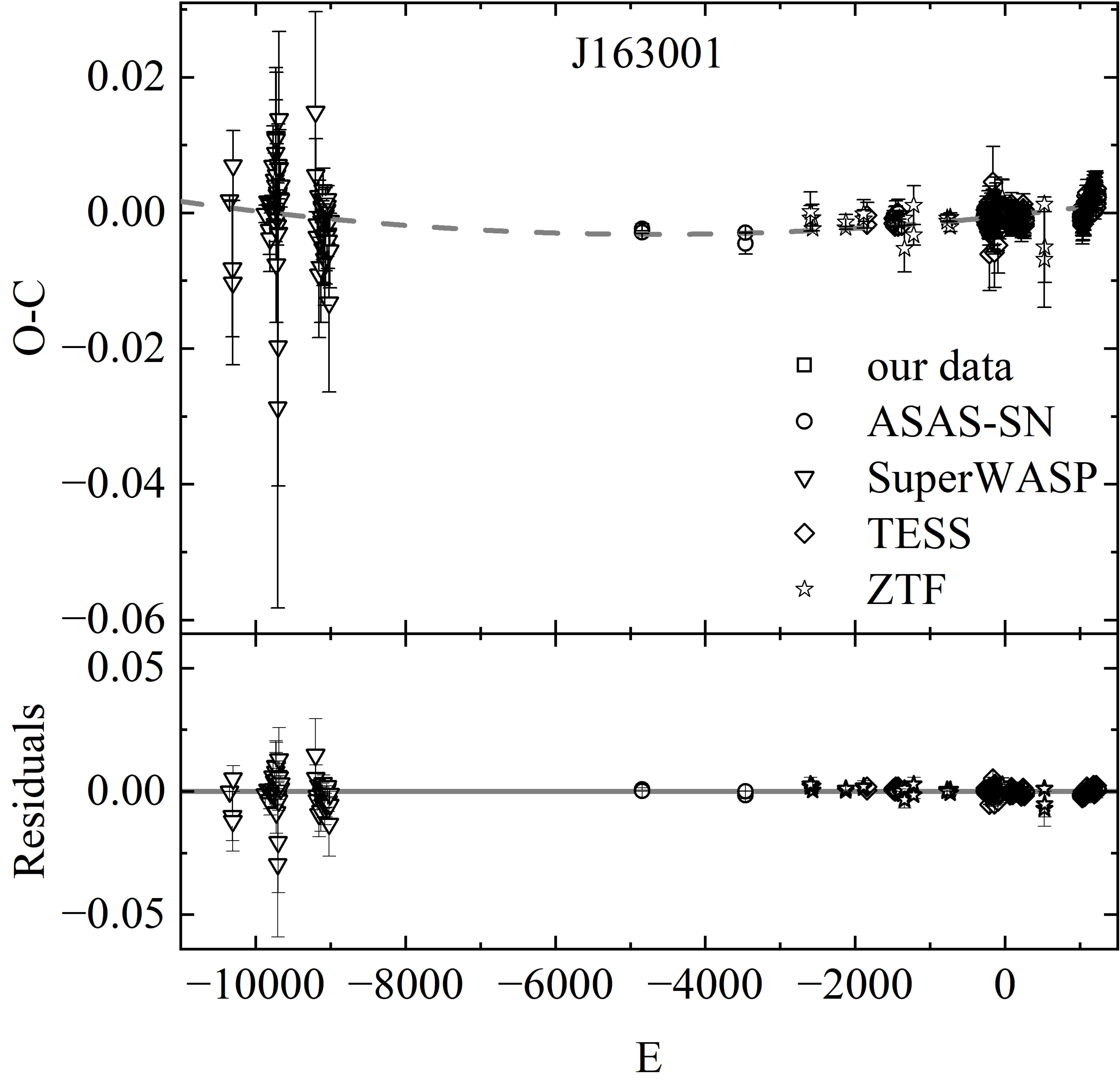}
        \end{minipage}%
    \caption{The $O-C$ diagrams of the six targets. E is the cycle number. The dashed lines in the top panels represents the fitting curves. The bottom panels displays the fitting residuals.}
    \label{fig:Orbital Period Analysis}
\end{figure*}

The secular orbital period variation can indicate a contact binary's dynamical evolution, so we performed the $O-C$ investigation using as many eclipsing times as possible. The eclipsing times were derived directly based on the K-W method in the case of a continuous light curve \citep{1956BAN....12..327K}. On the other hand, we employed the period shift method to calculate the minima for the discrete points of ASAS-SN, CRTS, and TESS with low cadence \citep{2020AJ....159..189L, 2021ApJ...922..122L, 2022AJ....164..202L}. We employed the following equation to calculate $O-C$ values,
\begin{equation}
\label{eq:O-C Calculation}
Min = Min.\uppercase\expandafter{\romannumeral1}_0 + P \times E, 
\end{equation}
where $Min$ is the observed eclipsing time, $Min.\uppercase\expandafter{\romannumeral1}_0$ is the initial epoch tabulated in Table \ref{tab:O-C correction}, P is the orbital period taken from Table \ref{tab:Brief Parametric Introduction}, and E is the cycle number. The $O-C$ curve was fitted with a simple linear equation to correct the initial epoch and orbital period at first. The corrected values, corrected $Min.\uppercase\expandafter{\romannumeral1}_0$ and corrected $P$, are tabulated in Table \ref{tab:O-C correction}. We recalculated the $O-C$ values employing the Equation (\ref{eq:O-C Calculation}) with corrected $Min.\uppercase\expandafter{\romannumeral1}_0$, and corrected $P$. The final eclipsing times and $O-C$ results are listed in Table \ref{tab:Eclipsing Times and O-C values}. Figure \ref{fig:Orbital Period Analysis} displays the six targets' $O-C$ diagrams. Because most targets exhibit parabolic curves, the following equation is applied to fit the $O-C$ results,
\begin{equation}
\label{eq:Quadratic Equation to fit the O-C diagrams}
O-C=\Delta Min.\uppercase\expandafter{\romannumeral1}_0 + \Delta P \times E + \frac{\beta}{2} \times E^2, 
\end{equation}
where $\Delta Min.\uppercase\expandafter{\romannumeral1}_0$ and $\Delta P$ are the correction values for the initial epoch and the orbital period, respectively, and $\beta$ represents the rate of orbital period change. Table \ref{tab:Fitting parameters of O-C diagrams} tabulates the fitting results. Note that the orbital period changing rate of J075442 is of the order $10^{-11} d \; yr^{-1}$, which is so tiny and negligible that we regard its orbital period unchanged. So the coefficients of linear fitting (no $\beta$) are listed as final results in Table \ref{tab:Fitting parameters of O-C diagrams}. The orbital periods of the other five targets change in the long term. Note that the reliability of the $O-C$ analysis should be interpreted with caution, as the time span is not sufficiently long. The long-term variations presented in this paper may represent a part of a periodic variation on a longer timescale. Further observations will be required to confirm these results.

\begin{table*}
\centering
\caption{The values of initial epoch, the corrected initial epoch, and the corrected orbital period for the six targets in $O-C$ analysis.}
\label{tab:O-C correction}
\begin{tabular}{cccc}
\hline
Target & $Min.\uppercase\expandafter{\romannumeral1}_0$& corrected $Min.\uppercase\expandafter{\romannumeral1}_0$ & corrected $P$ \\
       & (BJD, 2400000+)                               &(BJD, 2400000+)                                           &(d)\\
\hline
J063344	&	59947.24234 	&	59947.24213 	&	0.5702047 	\\
J073647	&	59993.37813 	&	59993.37664 	&	0.5584831 	\\
J075442	&	59984.18859 	&	59984.18693 	&	0.5072825 	\\
J094123	&	59977.41425 	&	59977.41441 	&	0.6181130 	\\
J105032	&	60027.25486 	&	60027.25526 	&	0.5769032 	\\
J163001	&	59788.24668 	&	59788.24582 	&	0.5669022 	\\
\hline
\end{tabular}
\end{table*}

\begin{table*}
\begin{threeparttable}
        \centering
	\caption{The Eclipsing times and $O-C$ results of the six targets.}
	\label{tab:Eclipsing Times and O-C values}
	\begin{tabular}{ccccrrc}
\hline
Target	&$Min$	&Error	&E	&\multicolumn{1}{c}{$O-C$}&\multicolumn{1}{c}{Residual}	&Telescope\\
	&(BJD, 2400000+)&(d)	&	&\multicolumn{1}{c}{(d)}	&\multicolumn{1}{c}{(d)}	    & \\
\hline
J063344	&54437.63879 	&0.00182 	&-9662.5	&-0.00043 	&-0.00434 	&SuperWASP\\
        &57618.23080 	&0.00145 	&-4084.5	&-0.01023 	&-0.00533 	&ASAS-SN\\
        &58849.59790 	&0.00048 	&-1925	    &-0.00018 	&0.00353 	&TESS\\
        &59947.24234 	&0.00064	&0	        &0.00021 	&0.00071 	&XL60\\
\hline
\end{tabular}
\begin{tablenotes}[para,flushleft]
\item[1] All calculated eclipsing minima were transformed into Barycentric Julian Date (BJD) on the website \url{https://astroutils.astronomy.osu.edu/time/hjd2bjd.html} \citep{2010PASP..122..935E}.\\
\item[2] This table is available in its entirety in machine-readable form in the online version of this article.\\
\end{tablenotes}
\end{threeparttable}
\end{table*}

\begin{table*}
	\centering
	\caption{Fitting parameters of the $O-C$ diagrams.}
	\label{tab:Fitting parameters of O-C diagrams}
	\begin{tabular}{ccccccccc}
		\hline
Target & $\Delta Min.\uppercase\expandafter{\romannumeral1}_0$ &        error       &     $\Delta P$     &        error       & $\beta$                       & error                         & $dM_1/dt$\\
        &($\times 10^{-4} d$)                                   &($\times 10^{-5} d$)&($\times 10^{-6} d$)&($\times 10^{-8} d$)&($\times 10^{-7} d \; yr^{-1}$)&($\times 10^{-8} d \; yr^{-1}$)&($\times 10^{-8} M_{\sun} \; yr^{-1}$)\\
        \hline 
		  J063344 & -4.98    	&12.7      & 2.20         & 18.4	      & 3.85      &3.08	    &5.22\\
            J073647 & 17.60 	  &15.8      & 2.92   	    & 18.6          & 3.85	    &2.65	  &7.56\\
            J075442 & -16.7   	  &12.5      & 1.91         & 7.25	        & -    	    & -	      &-\\
            J094123 & -8.36  	  &4.58      & -1.76        & 7.96    	    & -3.95	    &9.80	  &-7.62\\
            J105032 & 17.6   	  &31.4      & -2.15     	& 27.9	        & -2.72	    &3.50	  &-5.26\\
            J163001 & -4.54   	  &9.09      & 1.15   	    & 10.7	        & 1.71	    &1.69     &1.94\\

		\hline
	\end{tabular}
\end{table*}

\section{Spectroscopic Analysis}
\label{sec:Spectroscopic Analysis}

\begin{figure*}
        \centering
	\begin{minipage}[m]{0.3\linewidth}
	    \centering
	      \includegraphics[width=2in]{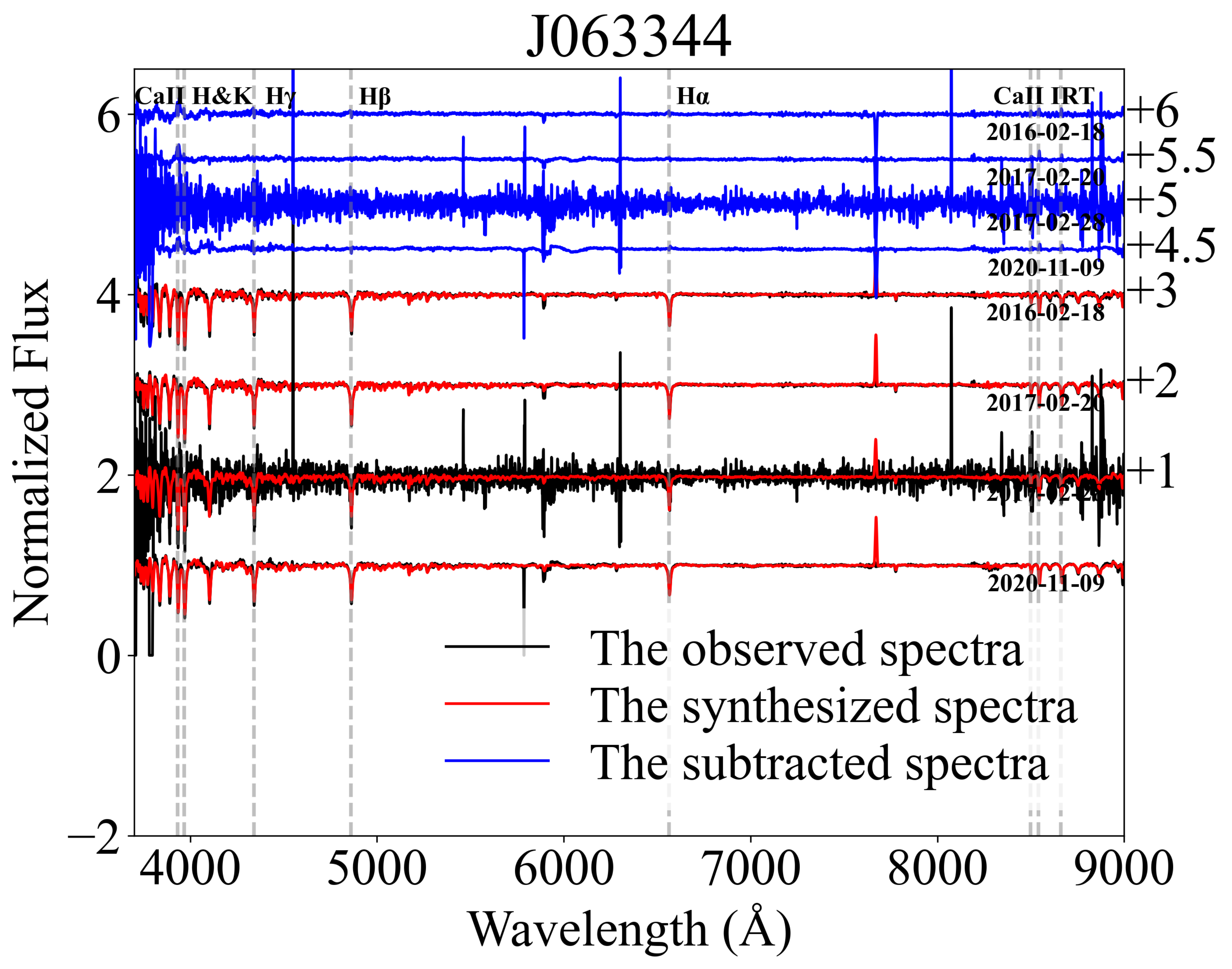}
        \end{minipage}%
        \begin{minipage}[m]{0.3\linewidth}
	    \centering
	    \includegraphics[width=2in]{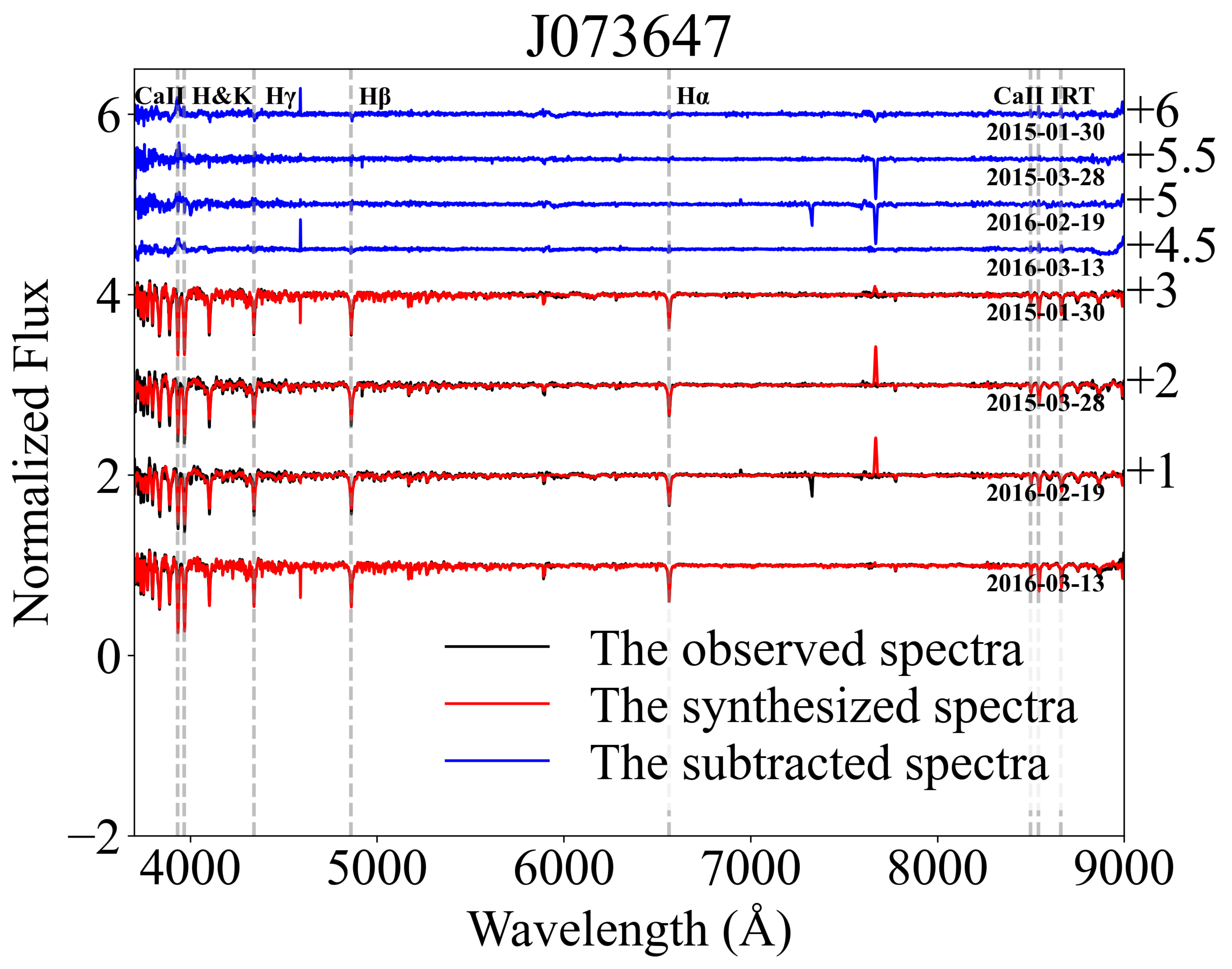}
        \end{minipage}%
        \begin{minipage}[m]{0.3\linewidth}
	    \centering
	    \includegraphics[width=2in]{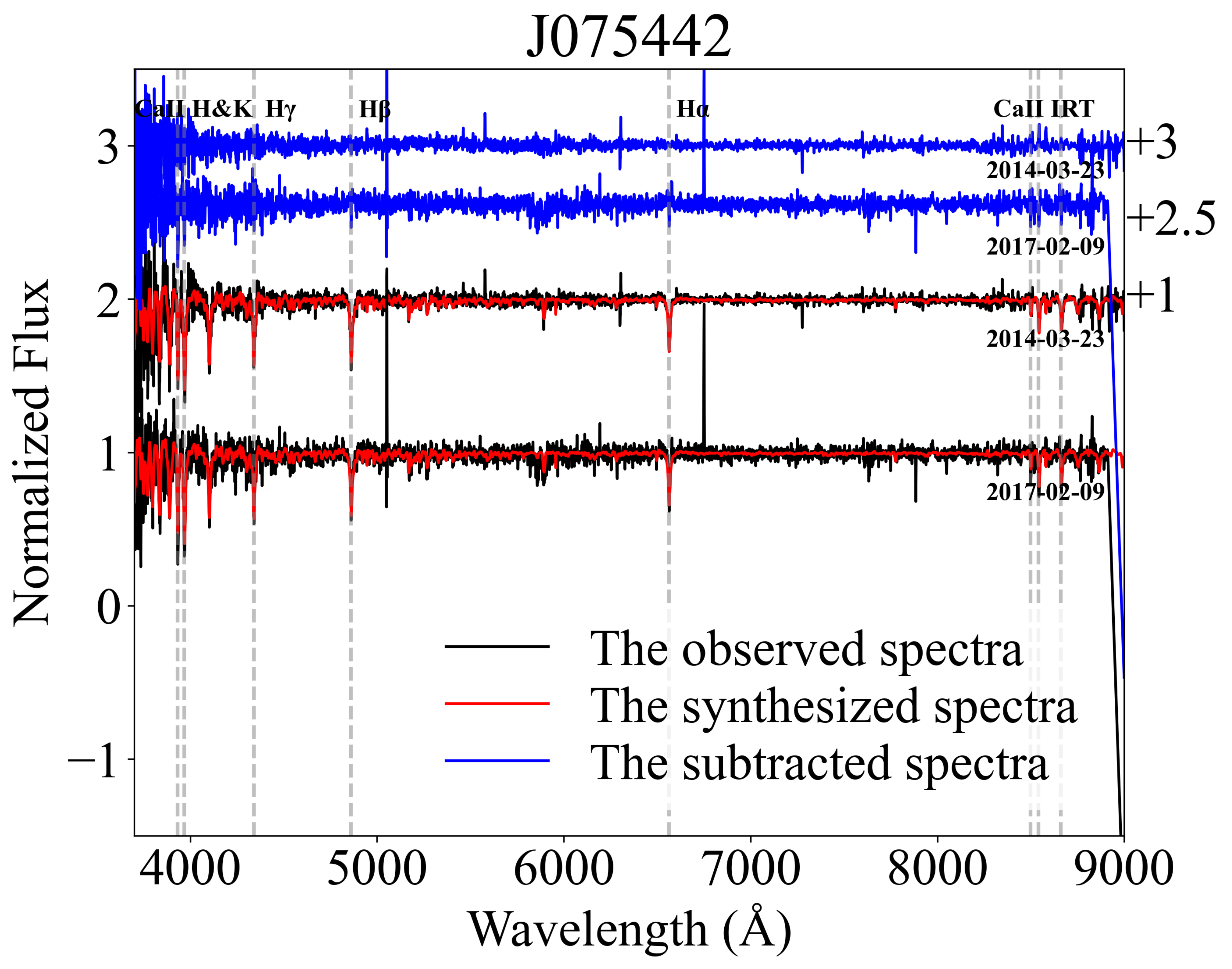}
        \end{minipage}%

	\begin{minipage}[m]{0.3\linewidth}
	    \centering
	      \includegraphics[width=2in]{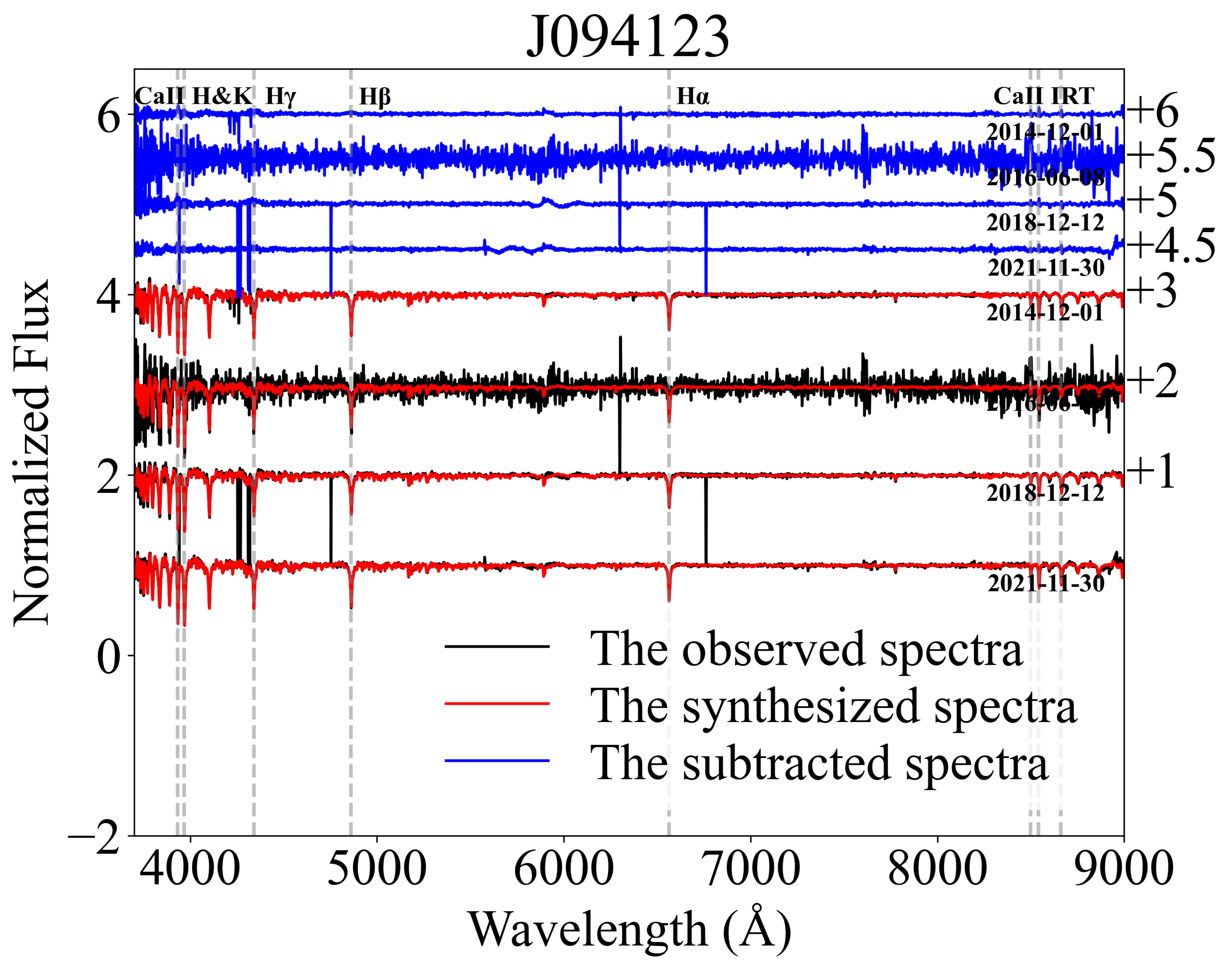}
        \end{minipage}%
        \begin{minipage}[m]{0.3\linewidth}
	    \centering
	    \includegraphics[width=2in]{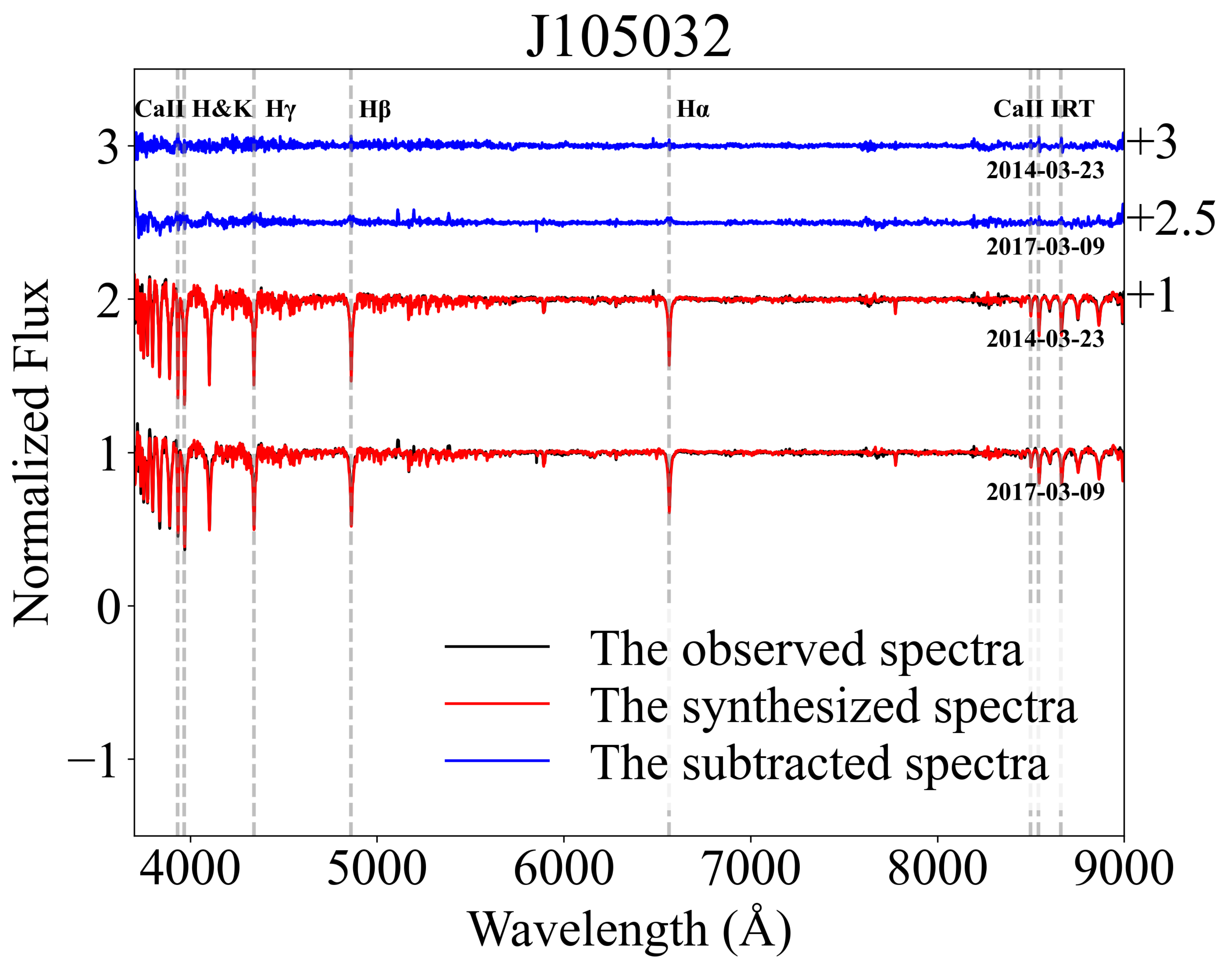}
        \end{minipage}%
        \begin{minipage}[m]{0.3\linewidth}
	    \centering
	    \includegraphics[width=2in]{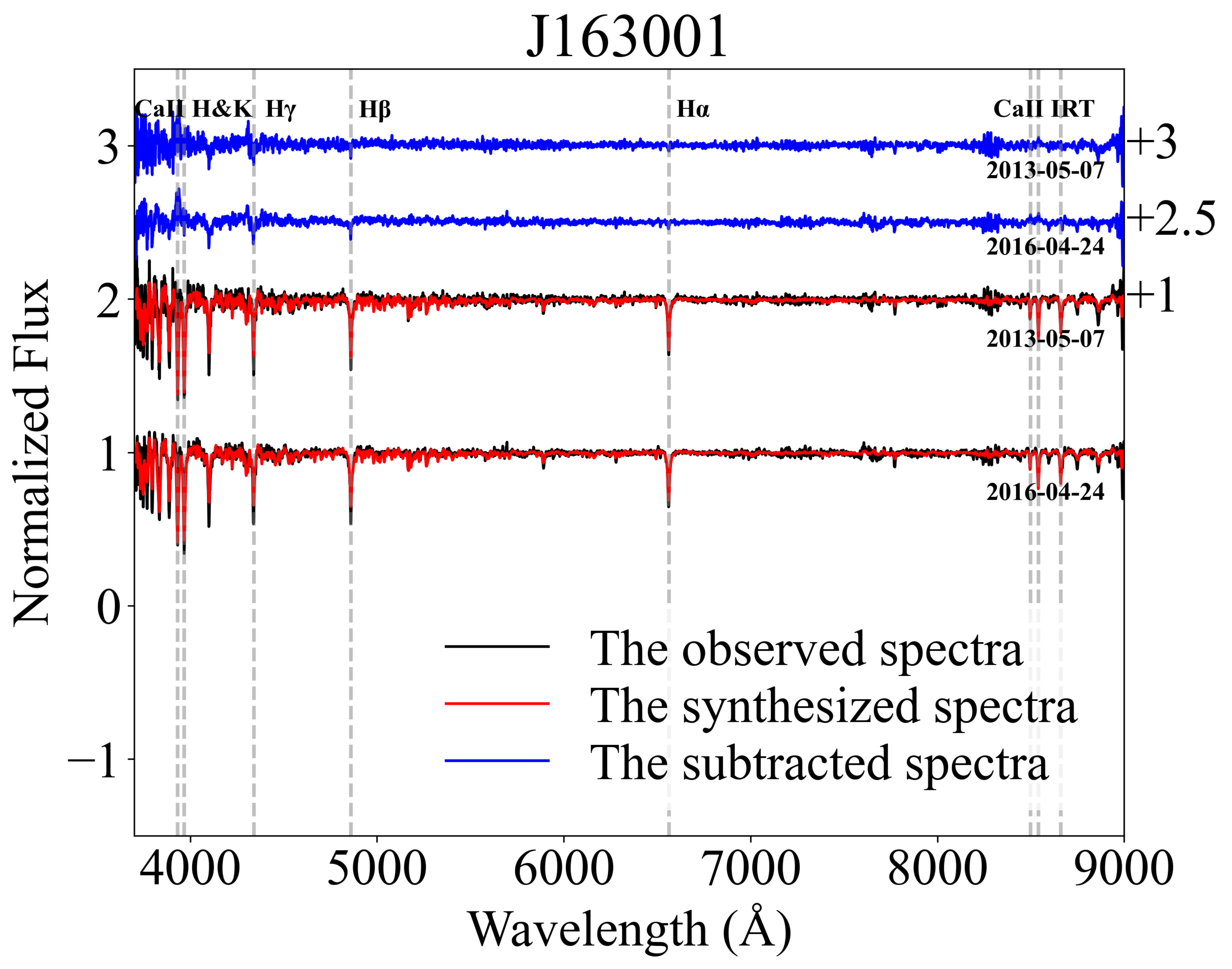}
        \end{minipage}%
    \caption{Spectral analysis of the six targets. The black lines represent the observed spectra of the six targets, the red lines represent the synthesized spectra produced by STARMOD, and the blue lines represent the subtracted spectra. All six targets exhibit no or only very weak chromospheric emission lines.}
    \label{fig:Spectroscopic Investigation}
\end{figure*}

The intensity of the H$\alpha, \beta, \gamma$ and Ca \uppercase\expandafter{\romannumeral2} H\&K and IRT emission lines is supposed to indicate the strength of magnetic activities in the chromosphere. The total flux measured in spectroscopy is a combination of the flux of the photosphere and that of the chromosphere. Therefore, we utilized the spectral subtraction approach to obtain the chromospheric spectra based on the assumption that the flux of the chromosphere does not correlate with that of the photosphere in the local magnetic field regions \citep{1984BAAS...16..893B}. Two spectra of inactive stars are needed for each contact binary and are expected to act as template-worthy spectra. Apart from that, a temperature close to the temperature of each component ($|\Delta T| < 100 K$) and a high signal-to-noise ratio (SNR$_g > 100$) are the two critical criteria when we selected the template spectra from the \citet{2021ApJS..256...14Z}'s catalog. We utilized the STARMOD program to produce synthetic spectra, considering the radial velocities, spin angular velocities, and luminosity ratios of the two inactive components selected \citep{1985ApJ...295..162B}. Then, the chromospheric spectra were obtained by subtracting the synthetic ones from the observed ones, displayed in Figure \ref{fig:Spectroscopic Investigation}. The lack of emission lines at any of the five wavelengths for all six targets suggests no chromospheric activity.

\section{Discussion and Conclusion}

The photometric, spectroscopic, and orbital period analysis of six targets was carried out for the first time. According to the photometric solutions, the mass ratios of the six targets are all smaller than 0.15, which are reliable due to the obvious flat features in the light curves at the vicinity of phase 0.5 \citep{2003CoSka..33...38P, 10.1093/mnras/stw3153, 2021AJ....162...13L}. So we cataloged them as ELMRCBs with $P > 0.5d$. A contact binary is classified as a shallow-contact binary if the contact degree $f < 25\%$, a medium-contact binary if $25\% \leq f < 50\%$, and a deep-contact binary if $f \geq 50\%$. So, J073647, J094123, and J163001 are medium-contact binaries. J063344, J075442, and J105032 are deep-contact binaries. J105032 is a W-subtype contact binary, while the other five targets are A-subtype contact binaries. The analysis of LAMOST spectra shows no or only very weak chromospheric emission lines for all the six targets.

We calculated their absolute physical parameters to discuss the mass transfer between the two components and the evolutionary states of the six contact binaries. \citet{2022AJ....164..202L} revealed the linear relationship between orbital period (P) in the unit of day and semimajor axis (a) in the unit of solar radius, 
\begin{equation}
\label{eq:a - P}
a = 0.501 (\pm 0.063) + 5.621 (\pm 0.138) \times P.
\end{equation}
We derived the semimajor axis $a$ directly according to the orbital period $P$. Then we employed the following equations to calculate their absolute parameters,
\begin{equation}
\label{eq:Absolute parameters}
\begin{split}
&R_i = a \times r_i,\\
&M_{total}=M_1+M_2=\frac{0.0134a^3}{P^2},\\
&q = \frac{M_2}{M_1},\\
&L_i = {T_i}^4 \times {R_i}^2,\\
\end{split}
\end{equation}
where $R_i$ is the radius of one component, $M_i$ is the mass of one component, and $L_i$ is the luminosity of one component. Table \ref{tab:Absolute Parameters of the six targets} lists the absolute physical parameters and errors.

\begin{table*}
	\centering
	\caption{Absolute parameters of the six targets.}
	\label{tab:Absolute Parameters of the six targets}       
	\begin{tabular}{ccccccccc}
		\hline
		Target & a            & R$_1$        & R$_2$        & M$_1$         & M$_2$        & L$_1$            & L$_2$             & $\frac{log(R_2/R_1)}{log(M_2/M_1)}$ \\
                    & (R$_{\sun}$) & (R$_{\sun}$) & (R$_{\sun}$) & (M$_{\sun}$)  & (M$_{\sun}$) & (L$_{\sun}$)     & (L$_{\sun}$)      &                                     \\
		\hline
J063344	&	$	3.71 	\pm	0.10 	$	&	$	2.19 	\pm	0.06 	$	&	$	0.85 	\pm	0.04 	$	&	$	1.90 	\pm	0.15 	$	&	$	0.20 	\pm	0.02 	$	&	$	10.44 	\pm	0.60 	$	&	$	1.47 	\pm	0.28 	$	&	$	0.43 	\pm	0.01 	$	\\
J073647	&	$	3.64 	\pm	0.10 	$	&	$	2.06 	\pm	0.06 	$	&	$	0.92 	\pm	0.04 	$	&	$	1.80 	\pm	0.15 	$	&	$	0.27 	\pm	0.02 	$	&	$	8.14 	\pm	0.47 	$	&	$	1.52 	\pm	0.13 	$	&	$	0.43 	\pm	0.01 	$	\\
J075442	&	$	3.35 	\pm	0.09 	$	&	$	2.05 	\pm	0.06 	$	&	$	0.72 	\pm	0.03 	$	&	$	1.81 	\pm	0.15 	$	&	$	0.15 	\pm	0.01 	$	&	$	8.66 	\pm	0.52 	$	&	$	0.96 	\pm	0.94 	$	&	$	0.42 	\pm	0.01 	$	\\
J094123	&	$	3.98 	\pm	0.11 	$	&	$	2.26 	\pm	0.06 	$	&	$	0.94 	\pm	0.04 	$	&	$	1.94 	\pm	0.16 	$	&	$	0.26 	\pm	0.02 	$	&	$	10.15 	\pm	0.58 	$	&	$	1.61 	\pm	0.22 	$	&	$	0.44 	\pm	0.01 	$	\\
J105032	&	$	3.74 	\pm	0.10 	$	&	$	2.19 	\pm	0.06 	$	&	$	0.89 	\pm	0.03 	$	&	$	1.89 	\pm	0.15 	$	&	$	0.22 	\pm	0.02 	$	&	$	11.35 	\pm	0.65 	$	&	$	2.41 	\pm	0.20 	$	&	$	0.42 	\pm	0.01 	$	\\
J163001	&	$	3.69 	\pm	0.10 	$	&	$	2.20 	\pm	0.06 	$	&	$	0.80 	\pm	0.03 	$	&	$	1.91 	\pm	0.16 	$	&	$	0.18 	\pm	0.02 	$	&	$	9.46 	\pm	0.55 	$	&	$	1.00 	\pm	0.12 	$	&	$	0.43 	\pm	0.01 	$	\\
		\hline
	\end{tabular}
        \begin{tablenotes}
         \item Errors were estimated regarding error propagation formulas. 
        \end{tablenotes}
\end{table*}

By analyzing the orbital period changes, we derived that the orbital periods of J063344, J073647, and J163001 are increasing secularly, while J094123 and J105032 are experiencing long-term decreases. The increase is responsible for the mass transfer from the less massive component to the heavier component. The mass transfer rate is estimated using the following equation,
\begin{equation}
\label{eq:the mass transfer rate}
\frac{dM_1}{dt}=\frac{M_1M_2}{3P(M_1-M_2)} \times \frac{dP}{dt},
\end{equation}
where $\frac{dP}{dt}$ corresponds to the value of $\beta$ in Table \ref{tab:Fitting parameters of O-C diagrams}. The final values are displayed in Table \ref{tab:Fitting parameters of O-C diagrams}. The secular decrease of the orbital period can be attributed typically to angular momentum loss and mass transfer from the more massive component to the less massive component. Considering the angular momentum loss, the most credible explanation is magnetic braking due to stellar wind \citep{1976ApJ...209..829W}. We calculated the orbital period decrease rate according to the standard magnetic braking model \citep{1988ASIC..241..345G},
\begin{equation}
\dot{P}_{\text{AML}} \approx -1.1 \times 10^{-8} \cdot \frac{(1+q)^2}{q} \cdot \frac{(k_1^2 M_1 R_1^4 + k_2^2 M_2 R_2^4)}{(M_1 + M_2)^{5/3} P^{7/3}},
\label{eq:the orbital period decrease rate dP/dt}
\end{equation}
where ${k_1}^2={k_2}^2=0.06$ denote the dimensionless gyration radii of the two components \citep{2006MNRAS.369.2001L}. So the results are $\dot{P}_{\text{AML}} = -2.72 \times 10^{-7} d \; yr^{-1}$ for J094123 and $\dot{P}_{\text{AML}} = -2.29 \times 10^{-7} d \; yr^{-1}$ for J105032. Compared with the observed period decrease rate ($\dot{P}_{obs} = \beta$) in Table \ref{tab:Fitting parameters of O-C diagrams}, the $\dot{P}_{\text{AML}}$ constitutes a significant portion of $\dot{P}_{obs}$ both for J094123 and J105032. Indeed, the orbital period decrease of J094123 and J105032 may result from angular momentum loss and mass transfer on both sides. However, angular momentum loss is supposed to be the dominant factor in yielding the secular orbital period decrease of J094123 and J105032. 

\begin{figure*}
  \centering
	\begin{minipage}[m]{0.48\linewidth}
	    \centering
	      \includegraphics[width=3.4in]{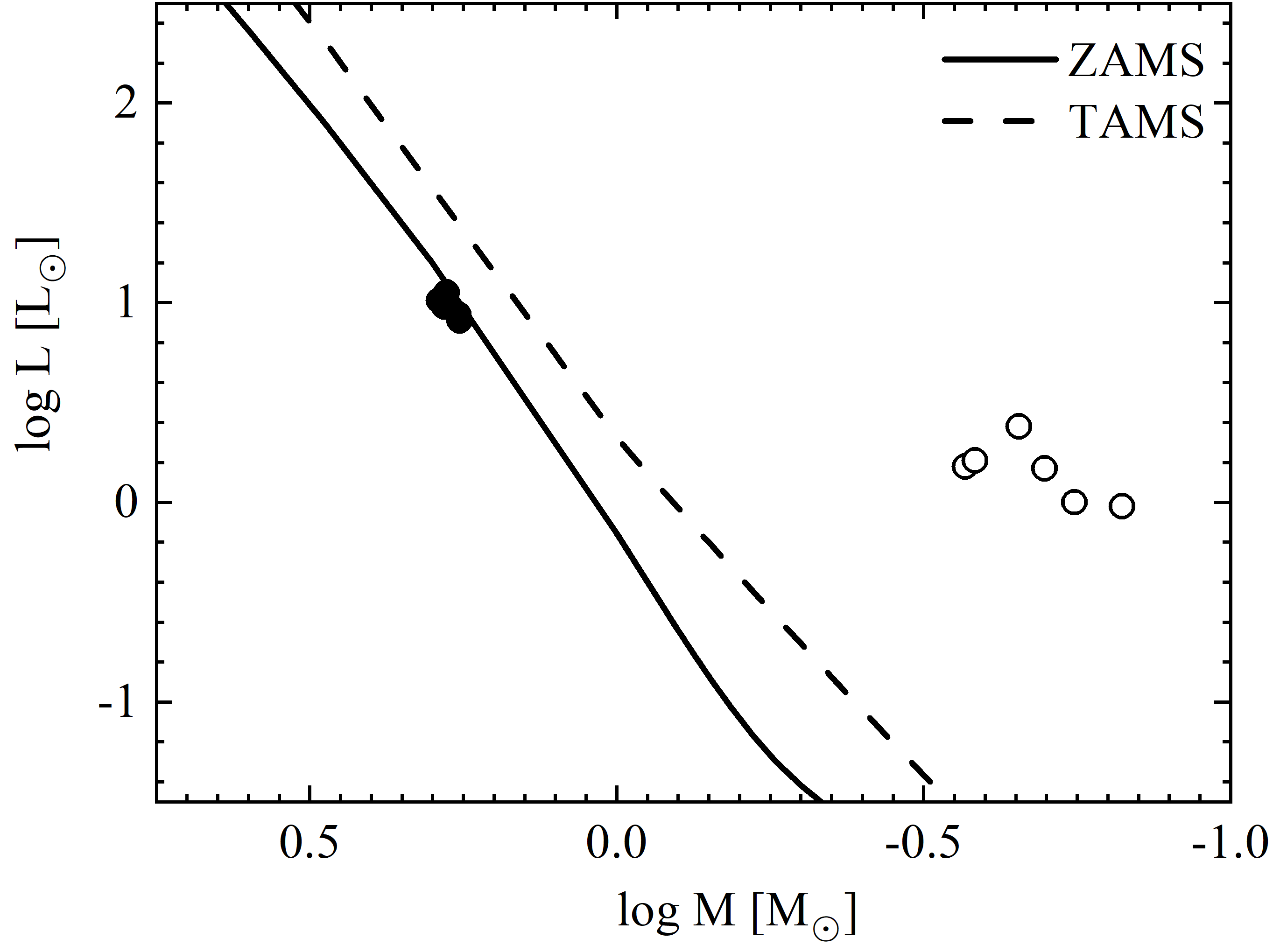}
        \end{minipage}%
        \begin{minipage}[m]{0.48\linewidth}
	    \centering
	    \includegraphics[width=3.4in]{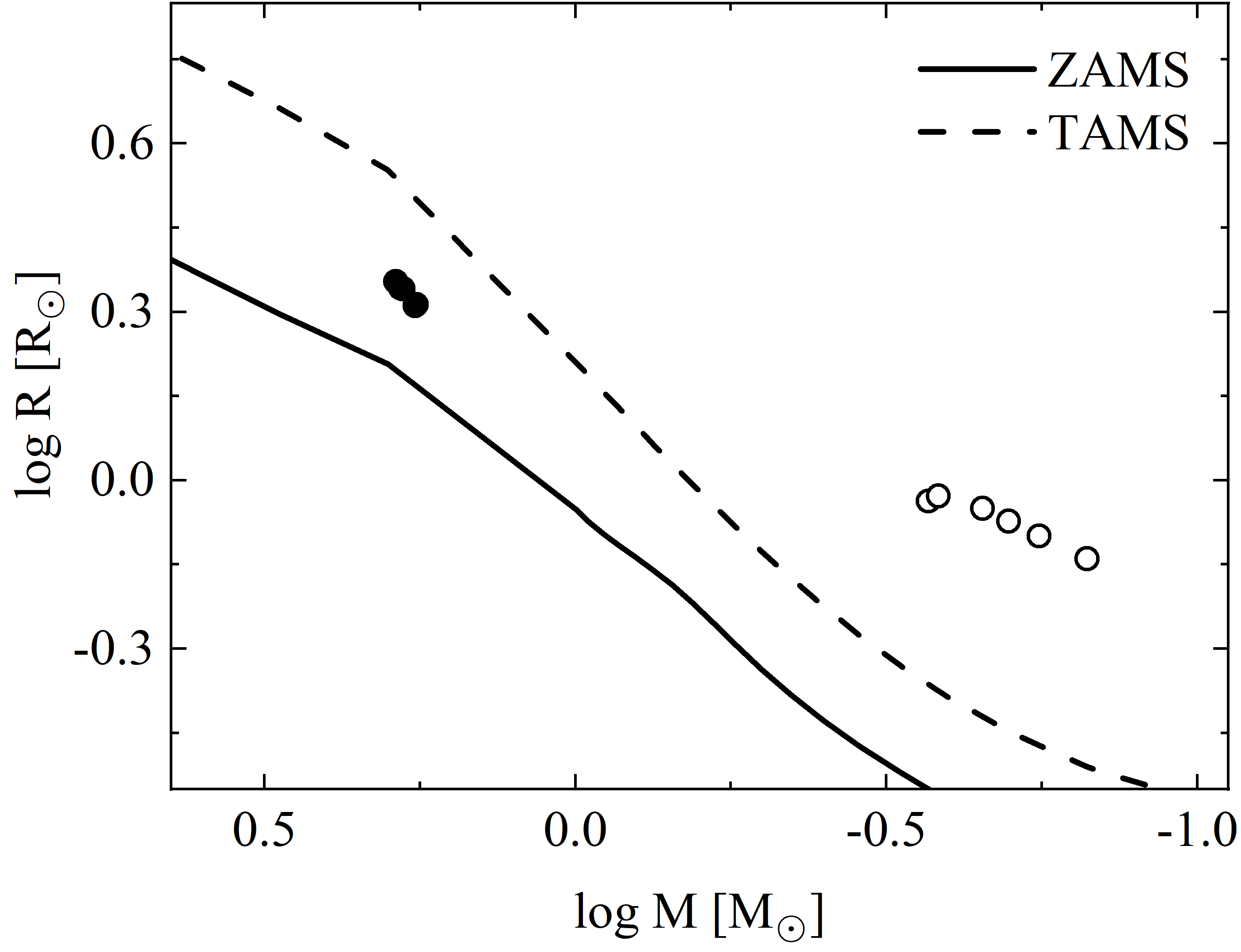}
        \end{minipage}%
\caption{Diagrams to demonstrate the evolutionary state. In the two diagrams, the solid dots denote the primary components, while the hollow circles denote the secondary components. Colored dotted lines denote the evolutionary tracks for solar chemical compositions \citep{2000A&AS..141..371G}. The black solid and dashed lines denote ZAMS and TAMS respectively \citep{2000A&AS..141..371G}.}
  \label{fig:M-L & M-R}
\end{figure*}

The masses and radii of both components follow the relation when they are zero-age main-sequence (ZAMS) stars \citep{1973asqu.book.....A},
\begin{equation}
\label{eq:log(R2/R1)/ log(M2/M1) = 0.64}
\frac{log(R_2/R_1)}{log(M_2/M_1)}=0.64.
\end{equation}
So, we calculated the value of $log(R_2/R_1)/log(M_2/M_1)$ of the six targets to discuss the evolutionary stage of the six targets. As is shown in Table \ref{tab:Absolute Parameters of the six targets}, they are all less than 0.64, indicating that the secondary component is more evolved than the primary component for each target. Apart from that, the mass-luminosity (M-L) and mass-radius (M-R) diagrams are displayed in Figure \ref{fig:M-L & M-R}. The main sequence is between the zero-age main sequence (ZAMS) and the terminal-age main sequence (TAMS). All the primary components around the ZAMS indicate that they are non-evolved or little-evolved main-sequence stars. The less massive secondary components are above the TAMS, suggesting that they have evolved from the main sequence. Therefore, the results obtained from these two methods are consistent.

The initial mass ($M_{1i, 2i}$) and the initial mass ratios ($q_{ini}$) were calculated by the following equations \citep{10.1093/mnras/stt028},
\begin{equation}
\label{eq:initial mass and mass ratio}
\begin{split}
&\gamma = 0.664 \frac{M_{lost}}{\delta M},\\
&M_{1i} = M_1-(\delta M-M_{lost}) = M_1-\delta M(1-\gamma),\\
&M_L = (L_2/1.49)^{\frac{1}{4.216}},\\
&M_{2i} = M_2+\delta M = M_2+2.50(M_L-M_2-0.07)^{0.64},\\
&q_{ini} = \frac{M_{2i}}{M_{1i}},\\
\end{split} 
\end{equation}
where $M_{lost}$ denotes the mass loss of a contact binary. $M_1$ and $M_2$ are the current mass of the primary and secondary components, respectively. $M_L$ denotes the mass of a single star with the same luminosity as the secondary component of a contact binary. $\delta M$ is the mass loss of the secondary component. As is listed in Table \ref{tab:Relevant Physical Parameters of the six targets}, the results correspond to the reference ranges proposed by \citet{10.1093/mnras/stt028} regarding the deviations of estimation. $M_{2i} > M_{1i}$ for all the six targets, opposite to the current mass situation $M_1 > M_2$. The initial secondary component, whose mass was smaller, expanded and filled its Roche Lobe by acquiring mass from the initial primary component. The mass transfer between the two components makes the initial primary component evolve into the current secondary component. In contrast, the initial secondary component is the current primary component. And the common envelope of a contact binary system takes shape with the long-term expansion \citep{10.1093/mnras/stt028}. The ages of the six targets, $t$, were calculated based on the model of \citet{2014MNRAS.437..185Y} in which the following equations were proposed,
\begin{equation}
\begin{aligned}
\label{eq:the age of each component}
&\overline{M_2}=\frac{M_{2i}+M_L}{2},\\
&t_{MS}={\frac{10}{({\frac{M}{M_{\sun}}})^{4.05}}} \times {(5.60 \times {10^{-3}({\frac{M}{M_{\sun}}}+3.993)^{3.16}+0.042)}},\\   
&t \approx t_{MS}(M_{2i})+t_{MS}(\overline{M_2}),\\
\end{aligned}
\end{equation}
where $t_{MS}$ is a star's main sequence lifetime. As is tabulated in Table \ref{tab:Relevant Physical Parameters of the six targets}, the ages of the six targets are about 2 or 3 Gyr. The significant decrease in the mass ratios of the six targets indicates the vigorous mass transfer between the two components in the process of evolution.

\begin{figure*}
  \centering
        \begin{minipage}[m]{0.48\linewidth}
	    \centering
	    \includegraphics[width=3.3in]{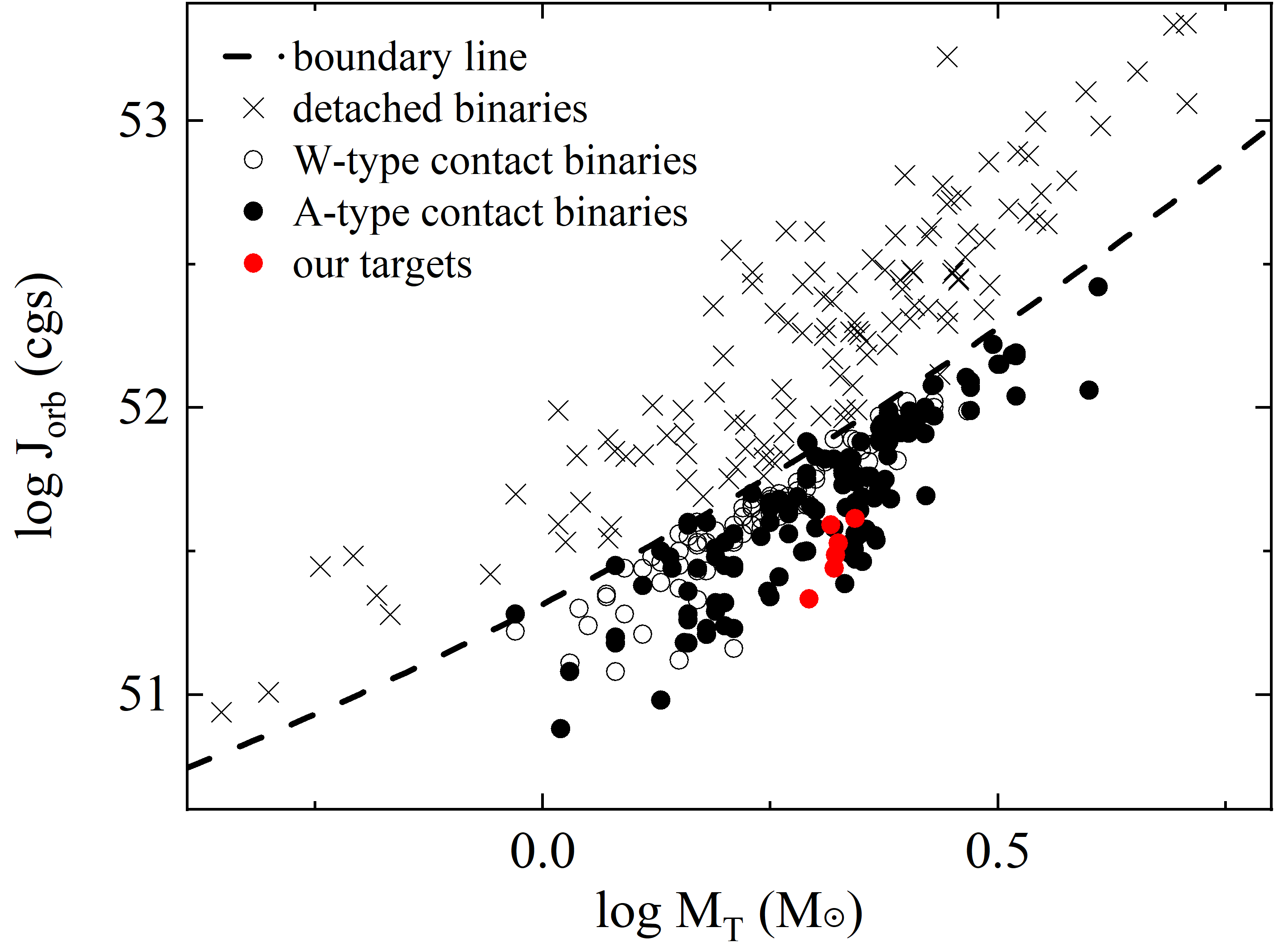}
        \end{minipage}%
        \begin{minipage}[m]{0.48\linewidth}
	    \centering
	    \includegraphics[width=3.3in]{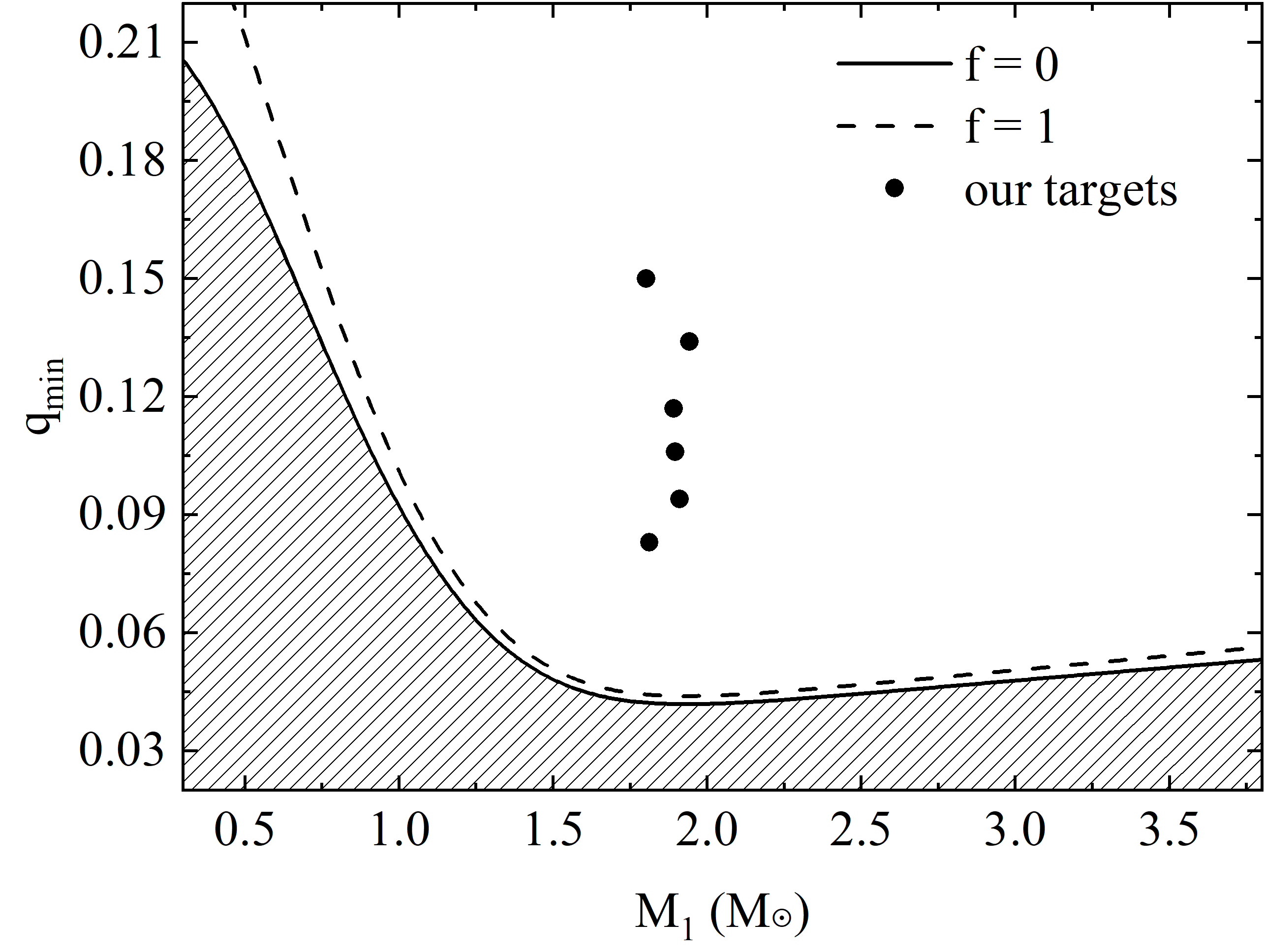}
        \end{minipage}%
\caption{Diagrams to demonstrate the formation and stability of the six targets. In the first diagram, the boundary line ($J_{lim}$) and the detached binaries represented by black cross were taken from \citet{10.1111/j.1365-2966.2006.11073.x}. The solid and hollow black circles represent the contact binaries collected by \citet{2021AJ....162...13L} and \citet{2022PASP..134k4202G}. The solid red circles are our six targets. In the second diagram, the solid and dashed lines correspond to the cutoff mass ratio ($q_{min}$) in the two cases of the contact degree, $f = 0$ and $f = 1$, respectively \citep{2024SerAJ.208....1A}.}
  \label{fig:J & qmin}
\end{figure*}

The orbital angular momentum ($J_{orb}$) was calculated to study the formation of contact binary by the following equation \citep{Christopoulou_2013},
\begin{equation}
\label{eq:J_o}
J_{orb} = 1.24 \times 10^{52} \times M_{total}^{\frac{5}{3}} \times P^{\frac{1}{3}} \times {\frac{q}{(1+q)^2}}.
\end{equation}
The results are listed in Table \ref{tab:Relevant Physical Parameters of the six targets}. Then we obtained the diagram of $J_{orb}$ and total mass ($M_{total}=M_1+M_2$) in Figure \ref{fig:J & qmin} and figured out that $J_{orb}$ is smaller than $J_{lim}$ with the same total mass. The six targets may evolve from short-period detached binaries through angular momentum loss, which results from stellar winds \citep{2007AJ....134.1769Q, 2015ApJ...798L..42Q}. In addition, the orbital angular momenta of the six targets are smaller than those of other contact binaries with the same total mass. So, the six targets are in late evolutionary states.

In terms of the stability and subsequent evolution of the contact binary, tidal gravity makes the two components unable to rotate simultaneously. A contact binary is getting to merge when the ratio of spin angular momentum to orbital angular momentum is greater than 1/3 ($J_{spin}/J_{orb} > 1/3$) \citep{1980A&A....92..167H}. So we calculated the $J_{spin}/J_{orb}$ employing the following equation \citep{Yang_2015},
\begin{equation}
\label{eq:Js/Jo}
\frac{J_{spin}}{J_{orb}}={\frac{1+q}{q}}[(k_1r_1)^2+(k_2r_2)^2q].
\end{equation}
The results are displayed in Table \ref{tab:Relevant Physical Parameters of the six targets}, and they are all smaller than 1/3, which means that all the six targets are currently dynamically stable. The instability mass ratio of a contact binary is related to the mass of the primary component \citep{2021MNRAS.501..229W, 2024MNRAS.535.2494W, 2024SerAJ.208....1A}. We calculated the instability mass ratios ($q_{min}$) of our six targets based on the Equation (10) of \citet{2024SerAJ.208....1A}. During the calculations, the gyration radius of the primary component ($k_1$) was determined by the Equation (15) of \citet{2024SerAJ.208....1A}. $k_2^2 = 0.205$ was used because of the less massive secondary component and its fully convective attribute \citep{2007MNRAS.377.1635A}. The results are $q_{min} \sim 0.042-0.044$ for all the six targets. The diagram of the instability mass ratio versus primary mass is shown in Figure \ref{fig:J & qmin}. The current mass ratio is greater than the instability mass ratio for each target by comparison, indicating that all six targets are stable at present. Therefore, the results obtained from the two methods are consistent.

\begin{table*}
\centering
\caption{Relevant physical parameters of the six targets.}
\label{tab:Relevant Physical Parameters of the six targets}       
\begin{tabular}{cccccccc}
\hline
Target  & $M_{1i}$	 & $M_{2i}$	  &$q_{ini}$ & t	    &$J_{orb}$              & $\frac{J_{spin}}{J_{orb}}$& $\beta_{etp}$ \\
        &(M$_{\sun}$)&(M$_{\sun}$)&          & (Gyr)    & ($\times 10^{51} cgs$)&                           &          \\
\hline
J063344	&	1.63 	&	2.24 	&	1.37 	&	2.64 	&	3.064 	&	0.223 	&	0.893 	\\
J073647	&	1.56 	&	2.19 	&	1.41 	&	2.74 	&	3.902 	&	0.151 	&	0.860 	\\
J075442	&	1.56 	&	2.11 	&	1.35 	&	3.28 	&	2.152 	&	0.295 	&	0.917 	\\
J094123	&	1.69 	&	2.23 	&	1.32 	&	2.62 	&	4.107 	&	0.168 	&	0.874 	\\
J105032	&	1.59 	&	2.44 	&	1.53 	&	1.98 	&	3.367 	&	0.201 	&	0.849 	\\
J163001	&	1.67 	&	2.10 	&	1.26 	&	3.28 	&	2.755 	&	0.251 	&	0.916 	\\
\hline
\end{tabular}
\end{table*}

Statistical analysis is performed to obtain the distribution characteristics of physical parameters of contact binaries in different categories. We collected 218 contact binaries with $P < 1d$, which were analyzed with radial velocity curves to guarantee the reliability of mass ratios. They are categorized into four major groups: contact binaries with $P < 0.5d$ and $q < 0.15$, contact binaries with $P > 0.5d$ and $q < 0.15$, contact binaries with $P < 0.5d$ and $q > 0.15$, and contact binaries with $P > 0.5d$ and $q > 0.15$. Their absolute parameters are tabulated in Table 11.

\begin{table*}
\centering
\label{tab:Staristics} 
\rotatebox{90}{
\begin{minipage}{\textheight}
\flushleft
\caption{The 218 contact binaries with radial velocity analysis.}
\begin{tabular}{lcccccccccccccccc}
\hline 
Name	&	Type	&	P	&	f	&	q	&	$T_1$	&	$T_2$	&	a	&	$M_1$	&	$M_2$	&	$R_1$	&	$R_2$	&	$L_1$	&	$L_2$	&	$(L_2/L_1)_{bol}$	&	$\beta$	&	Reference	\\
	&		&	(day)	&	(\%)	&		&	(K)	&	(K)	&	($R_{\sun}$)	&	($M_{\sun}$)	&	($M_{\sun}$)	&	($R_{\sun}$)	&	($R_{\sun}$)	&	($L_{\sun}$)	&	($L_{\sun}$)	&		&		&		\\
\hline
1SWASP J034501.24+493659.9	&	W	&	0.37653 	&	25 	&	0.421 	&	6494	&	6514	&	2.14 	&	0.65 	&	0.28 	&	1.01 	&	0.69 	&	1.63 	&	0.78 	&	0.47 	&	0.704 	&	(1)	\\
1SWASP J093010.78+533859.5	&	A	&	0.22771 	&	17 	&	0.397 	&	4700	&	4700	&	1.67 	&	0.86 	&	0.34 	&	0.79 	&	0.52 	&	0.27 	&	0.12 	&	0.44 	&	0.715 	&	(2)	\\
1SWASP J150822.80-054236.9	&	A	&	0.26006 	&	16 	&	0.510 	&	4500	&	4500	&	2.01 	&	1.07 	&	0.55 	&	0.90 	&	0.68 	&	0.30 	&	0.17 	&	0.57 	&	0.688 	&	(3)	\\
1SWASP J160156.04+202821.6	&	A	&	0.22653 	&	13 	&	0.670 	&	4500	&	4500	&	1.76 	&	0.86 	&	0.57 	&	0.75 	&	0.63 	&	0.21 	&	0.15 	&	0.71 	&	0.700 	&	(3)	\\
ASASSN-V J225203.51+325424.4	&	W	&	0.31827 	&	16 	&	0.952 	&	5754	&	6009	&	2.45 	&	0.95 	&	1.00 	&	0.95 	&	1.02 	&	1.06 	&	1.02 	&	0.94 	&	0.849 	&	(4)	\\
ASASSN-V J225325.14+354532.5	&	W	&	0.35425 	&	10 	&	0.283 	&	5304	&	5602	&	2.19 	&	0.25 	&	0.88 	&	0.61 	&	1.09 	&	0.33 	&	0.84 	&	2.39 	&	0.723 	&	(4)	\\
\hline
\end{tabular}
\vspace{6pt} 
\fontsize{8}{9}\selectfont
\begin{tablenotes}[para]
\item Reference. (1) \citet{2019AJ....157...73K}, (2) \citet{2015A&A...578A.103L}, (3) \citet{2014A&A...563A..34L}, (4) \citet{2024arXiv241200377W}, (5) \citet{2008AJ....136..586R}, (6) \citet{2011A&A...525A..66D}, (7) \citet{2007AJ....133..255L}, (8) \citet{2010NewA...15..155Y}, (9) \citet{2017ApJ...840....1M}, (10) \citet{2016PASA...33...43S}, (11) \citet{2018AN....339..472K}, (12) \citet{2011NewA...16..242G}, (13) \citet{2010A&A...514A..36C}, (14) \citet{2024A&A...692L...4L}, (15) \citet{2020MNRAS.493.1565D}, (16) \citet{2014NewA...30...64L}, (17) \citet{2010AcA....60..305S}, (18) \citet{2019RAA....19...10Y}, (19) \citet{2004AcA....54..195B}, (20) \citet{2015IBVS.6134....1N}, (21) \citet{2011NewA...16...12C}, (22) \citet{2015NewA...36..100G}, (23) \citet{2024NewA..11302291L}, (24) \citet{2016RAA....16....2L}, (25) \citet{2011MNRAS.412.1787D}, (26) \citet{2001AJ....122.1974R}, (27) \citet{2013NewA...21...46L}, (28) \citet{2012NewA...17..143O}, (29) \citet{2007A&A...465..943S}, (30) \citet{2018AcA....68..159A}, (31) \citet{2015NewA...39....9G}, (32) \citet{1991AJ....102..262L}, (33) \citet{1993A&AS..101..253L}, (34) \citet{2003AJ....125..322K}, (35) \citet{2005AcA....55..123G}, (36) \citet{2004AcA....54..299Z}, (37) \citet{2017AJ....154...99L}, (38) \citet{1989A&A...211...81H}, (39) \citet{2010IBVS.5951....1N}, (40) \citet{2008NewA...13..468S}, (41) \citet{2013AJ....146..157C}, (42) \citet{2006AcA....56..127G}, (43) \citet{2015NewA...41...26G}, (44) \citet{1991ApJ...370..677M}, (45) \citet{2012AJ....144..149C}, (46) \citet{2011AJ....142...99C}, (47) \citet{2011AJ....141..147L}, (48) \citet{1983MNRAS.203....1M}, (49) \citet{2005AJ....130.2252Y}, (50) \citet{2002IBVS.5258....1P}, (51) \citet{2019PASP..131h4202L}, (52) \citet{2003AJ....125.3258R}, (53) \citet{2004AJ....128.2997S}, (54) \citet{2004MNRAS.347..307Z}, (55) \citet{2016ApJ...817..133Z}, (56) \citet{2005PASJ...57..983Y}, (57) \citet{1976PASP...88..936L}, (58) \citet{1993ApJ...407..237B}, (59) \citet{1984A&AS...58..405M}, (60) \citet{2004AJ....127.1712P}, (61) \citet{2010JASS...27...69O}, (62) \citet{2018A&A...612A..91M}, (63) \citet{2016NewA...46...31G}, (64) \citet{2011AN....332..626K}, (65) \citet{2019IBVS.6266....1N}, (66) \citet{2018MNRAS.479.3197A}, (67) \citet{2014NewA...29...57N}, (68) \citet{2012NewA...17..673Z}, (69) \citet{2015AJ....149..194L}, (70) \citet{2018RAA....18...20T}, (71) \citet{2014NewA...31...14O}, (72) \citet{2016NewA...47...57G}, (73) \citet{2003A&A...412..465K}, (74) \citet{2000A&AS..147..243W}, (75) \citet{2002AJ....124.1738R}, (76) \citet{2019PASJ...71...34S}, (77) \citet{1986AJ.....92..666K}, (78) \citet{2013PASJ...65....1P}, (79) \citet{2012RAA....12..419Y}, (80) \citet{2012PASJ...64...48L}, (81) \citet{2014AJ....148..126C}, (82) \citet{1996A&AS..118..453L}, (83) \citet{1988AJ.....95..894Y}, (84) \citet{1999A&AS..136..139Y}, (85) \citet{2007AJ....133..169D}, (86) \citet{2013SerAJ.186...47C}, (87) \citet{2007AJ....133.1977P}, (88) \citet{2009AJ....138..540Y}, (89) \citet{2015Ap&SS.357...59P}, (90) \citet{2016NewA...46...73E}, (91) \citet{2006NewA...12..192E}, (92) \citet{2009NewA...14..321E}, (93) \citet{2002A&A...387..240O}, (94) \citet{2016AdAst2016E...7Z}, (95) \citet{2018Ap&SS.363...34S}, (96) \citet{2004PASP..116..826Y}, (97) \citet{2010MNRAS.408..464Z}, (98) \citet{2009NewA...14..461O}, (99) \citet{2016AstBu..71...64G}, (100) \citet{2018PASP..130c4201L}, (101) \citet{1994A&AS..103...39N}, (102) \citet{1999AJ....118.2451R}, (103) \citet{2015AJ....150...70L}, (104) \citet{2014PASP..126..121C}, (105) \citet{2012NewA...17...46U}, (106) \citet{2013AJ....145...80D}, (107) \citet{2005AcA....55..389Z}, (108) \citet{2012PASJ...64...85H}, (109) \citet{2015PASJ...67...98Z}, (110) \citet{2005AJ....130..224Q}, (111) \citet{1998A&A...336..920R}, (112) \citet{2004A&A...417..725Y}, (113) \citet{2016PASJ...68..102X}, (114) \citet{1990A&A...231..365L}, (115) \citet{2015AJ....149...62X}, (116) \citet{1994A&A...289..827G}, (117) \citet{2003AJ....126.1555K}, (118) \citet{2009AJ....137.3655P}, (119) \citet{1999IBVS.4702....1G}, (120) \citet{2005ARBl...20..193G}, (121) \citet{2015PASP..127..742A}, (122) \citet{2020NewA...7801354K}, (123) \citet{2015NewA...41...17L}, (124) \citet{2001A&A...367..840D}, (125) \citet{2005NewA...10..163A}, (126) \citet{2012NewA...17..603E}, (127) \citet{2011AN....332..690G}, (128) \citet{2014AJ....147...91L}, (129) \citet{2017AJ....154..260P}, (130) \citet{1978Ap&SS..58..301N}, (131) \citet{2004MNRAS.348.1321B}, (132) \citet{2009AJ....138.1465H}, (133) \citet{2013AJ....146...35Y}, (134) \citet{1995ApJ...455..300H}, (135) \citet{2019RAA....19...97Y}, (136) \citet{2014NewA...31...56U}, (137) \citet{2018AJ....156..199S}, (138) \citet{2018IBVS.6256....1A}, (139) \citet{2015NewA...34..262H}, (140) \citet{2003AJ....126.1960Y}, (141) \citet{2016Ap&SS.361...63L}, (142) \citet{1986MNRAS.223..581H}, (143) \citet{2001CoSka..31....5C}, (144) \citet{2011PASP..123...34L}, (145) \citet{2016MNRAS.457..836E}, (146) \citet{2011RAA....11.1158Y}, (147) \citet{2015NewA...34..271Y}, (148) \citet{2007ASPC..362...82O}, (149) \citet{2000A&A...356..603C}, (150) \citet{2005Ap&SS.296..305S}, (151) \citet{2010ASPC..435..111K}, (152) \citet{2016AJ....151...67Z}, (153) \citet{2015AJ....149..120L}, (154) \citet{1981MNRAS.196..305H}, (155) \citet{2011AN....332..607P}, (156) \citet{1993AJ....106..361L}, (157) \citet{2001AJ....122..402L}, (158) \citet{2013AJ....145...39Z}.\\
(This table is available in its entirety in machine-readable form.)
\end{tablenotes}
\end{minipage}
}
\end{table*}

(a) Mass transfer between the two components of a contact binary is accompanied by energy transfer which is a function of mass and bolometric luminosity ratio \citep{2004A&A...426.1001C}. The energy transfer parameter ($\beta$) can be calculated with the following equation, 
\begin{equation}
\label{eq:beta}
\beta = \frac{L_{1,obs}}{L_{1,ZAMS}} = \frac{1 + q^{4.6}}{1 + q^{0.92} (\frac{T_2}{T_1})^4},
\end{equation}
where $L_{1,obs}$ is the observed luminosity of the primary component and $L_{1,ZAMS}$ is the primary luminosity corresponding to ZAMS. The calculated $\beta$ of our six targets are listed in Table \ref{tab:Relevant Physical Parameters of the six targets}, where it's marked with $\beta_{etp}$. Additionally, the minimum value of energy transfer parameter ($\beta_{min}$) was estimated using the following equations \citep{2004A&A...426.1001C},
\begin{equation}
\begin{aligned}
\label{eq:beta_min}
&\left(\frac{L_2}{L_1}\right)_{bol} = \frac{L_2}{L_1} \times 10^{0.4 \times (BC_1 - BC_2)},\\
&\beta_{min} = \frac{1}{1 + \left(\frac{L_2}{L_1}\right)_{bol}},
\end{aligned}
\end{equation}
where $(L_2/L_1)_{bol}$ is the bolometric luminosity ratio and $BC_{1,2}$ are the bolometric corrections of primary and secondary components obtained from \citet{Pecaut_2013}'s table\footnote{\url{http://www.pas.rochester.edu/~emamajek/EEM_dwarf_UBVIJHK_colors_Teff.txt}}, respectively. For the 218 collected contact binaries, the results of $\beta$ and $(L_2/L_1)_{bol}$ are listed in Table 11. We draw the distribution of $(L_2/L_1)_{bol}$, $\beta$, $\beta_{min}$, and mass ratio $q$ shown in Figure \ref{fig:f, beta, q, L} (a).

\citet{2004A&A...426.1001C} derived $\beta - \beta_{min} = 0.52q^{4.1}$ indicating that the difference between $\beta$ and $\beta_{min}$ decreases when the mass ratio gets smaller. Therefore, the diagram confirms the conclusion. Besides, we can figure out that the energy transfer parameter for ELMRCBs is generally higher than that of contact binaries with $q > 0.15$. We focus more on the ELMRCBs, especially the ELMRCBs with $P > 0.5d$, to explore their features in the pre-merger stage. For targets with extremely low mass ratios, the energy transfer parameter for systems with $P > 0.5d$ is generally higher than that for systems with $P < 0.5d$. Since the $\beta$ is closer to $\beta_{min}$ for ELMRCBs with $P < 0.5d$ than ELMRCBs with $P > 0.5d$, the energy transfer between the two components of ELMRCBs with $P > 0.5d$ is more intense than ELMRCBs with $P < 0.5d$.

\begin{figure*}
        \centering
	\begin{minipage}[m]{0.32\linewidth}
	    \centering
	      \includegraphics[width=2.3in]{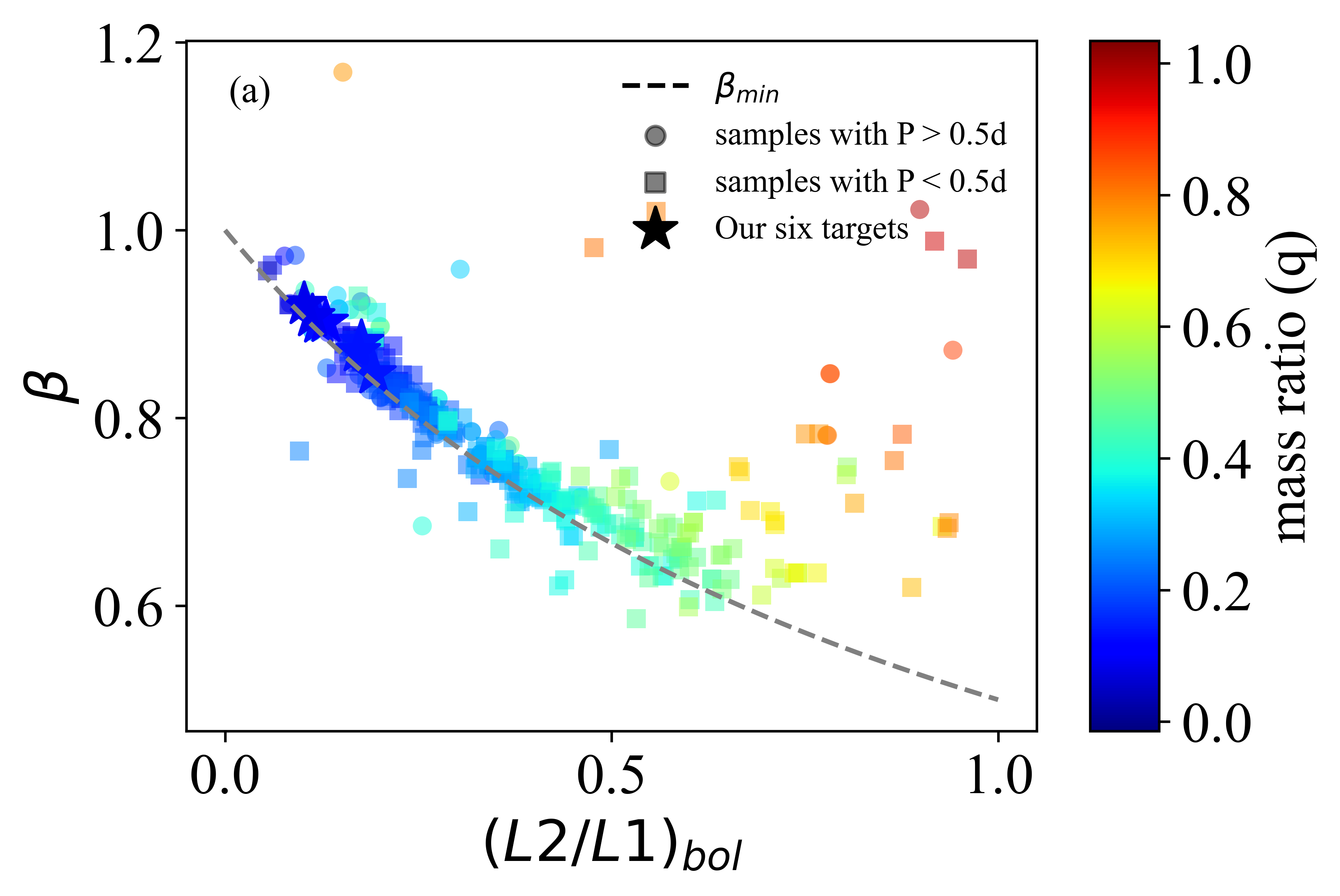}
        \end{minipage}
        \begin{minipage}[m]{0.32\linewidth}
	    \centering
	    \includegraphics[width=2.3in]{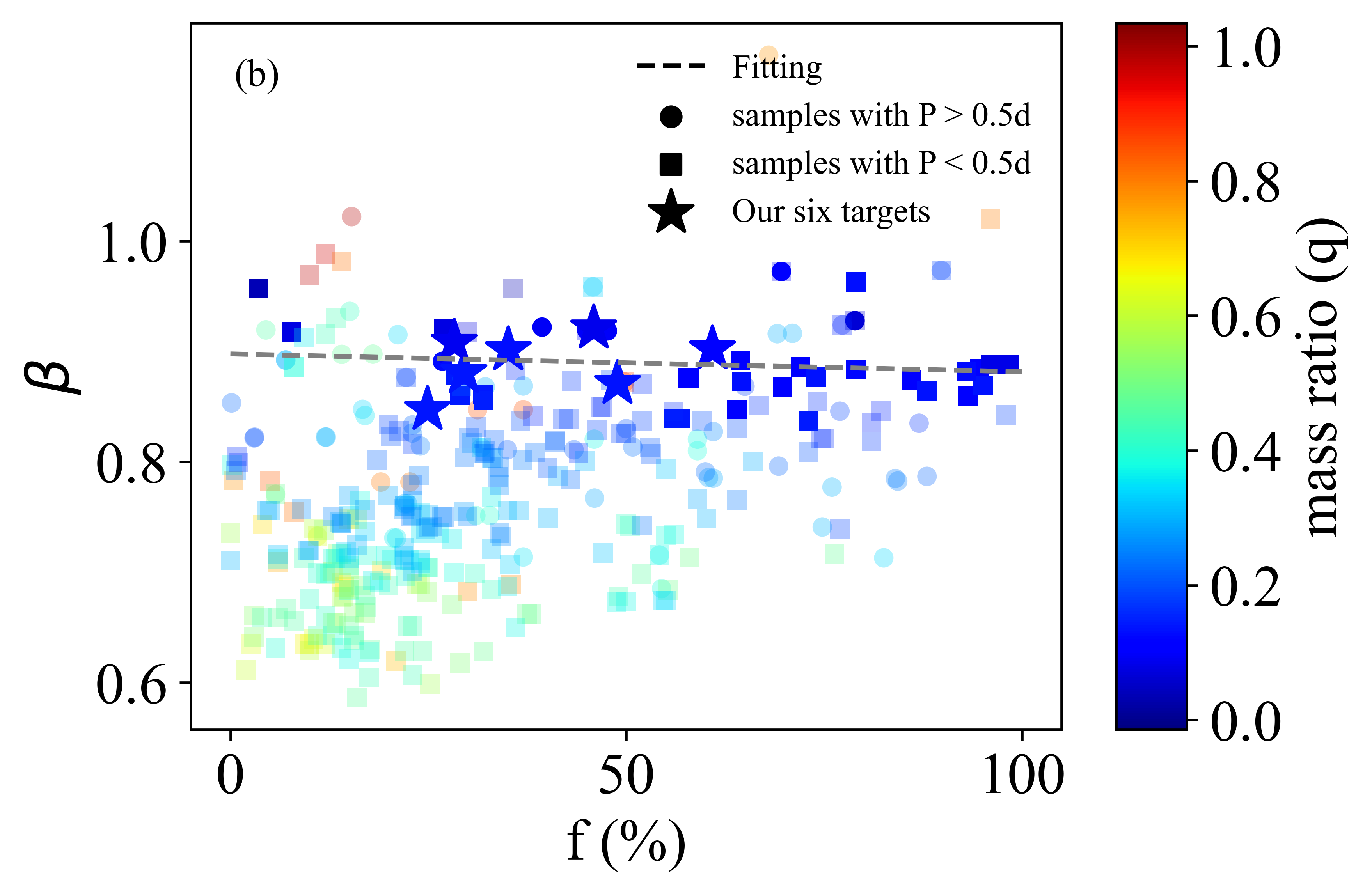}
        \end{minipage}
        \begin{minipage}[m]{0.32\linewidth}
	    \centering
	    \includegraphics[width=2.3in]{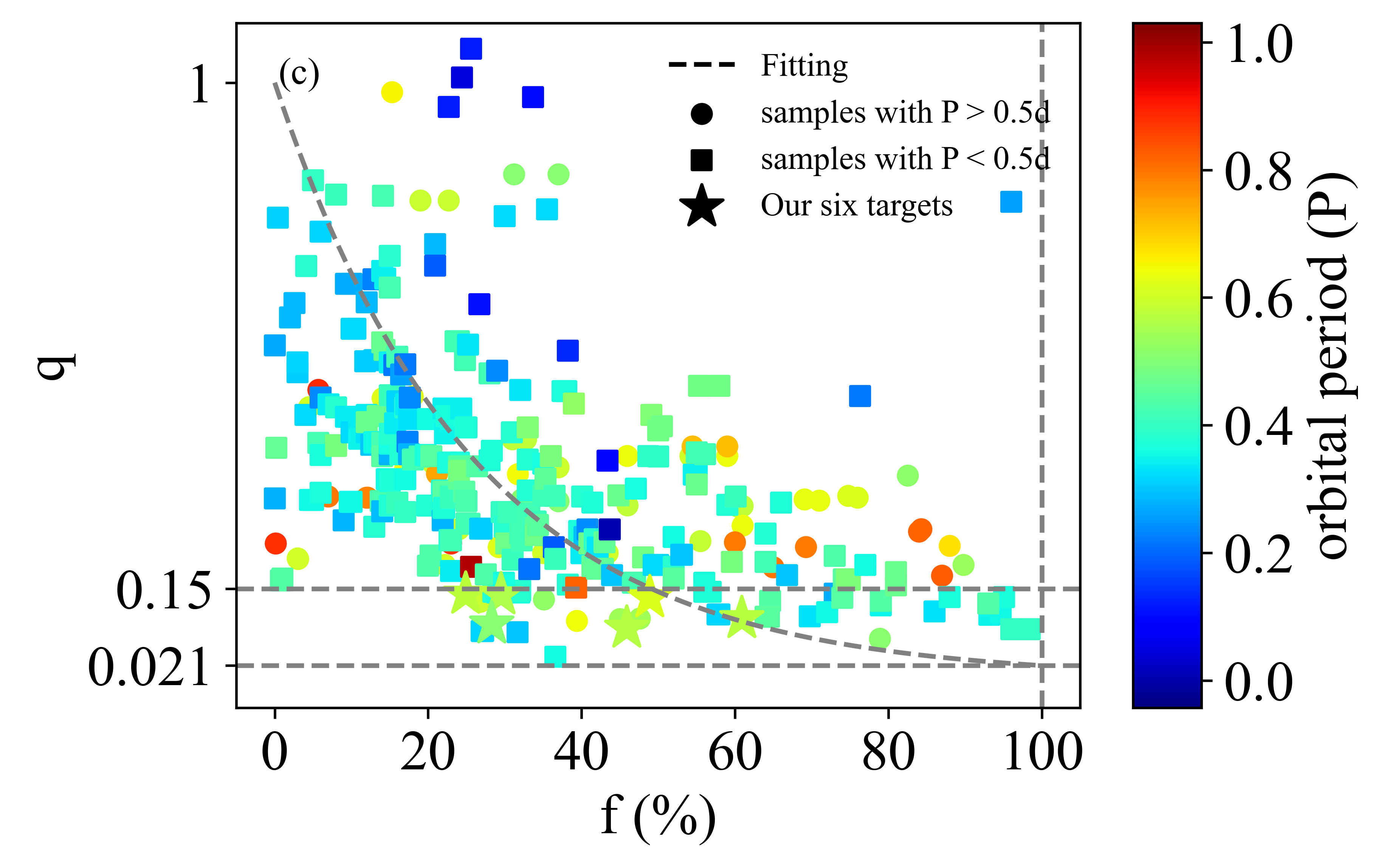}
        \end{minipage}
    \caption{Distributions of bolometric luminosity ratio $(L_2/L_1)_{bol}$, energy transfer parameter $\beta$, contact degree $f$, mass ratio $q$, and orbital period $P$ for the collected targets.}
    \label{fig:f, beta, q, L}
\end{figure*}

(b) The diagram of $f$, $\beta$, and $q$ is shown in Figure \ref{fig:f, beta, q, L} (b). It is found that the energy transfer parameter for ELMRCBs is, on the whole, approximately equivalent compared to contact binaries with $q > 0.15$. Thus, we performed a linear fit of $f$ versus $\beta$ for ELMRCBs, obtaining a slope of $-1.6 \times 10^{-4} \sim 0$. Consequently, we estimate that the energy transfer parameter between the two components of ELMRCBs is independent of the contact degree, hovering around a value of approximately 0.89. However, the sample size of ELMRCBs is limited, and further observations are required to validate this hypothesis.

(c) The extremely high contact degree and the extremely low mass ratio are criteria for merging conditions. The diagram of $f$, $q$, and $P$ is shown in Figure \ref{fig:f, beta, q, L} (c). The contact degrees of the large mass-ratio contact binaries do not exceed $40\%$, and the contact degrees of ELMRCBs exceed $20\%$ generally. $q$ and $f$ exhibit a roughly inverse correlation, corresponding to the result of \citet{2022MNRAS.512.1244C}. We fit it using a non-linear equation,
\begin{equation}
\label{eq:q-f}
q = e^{-3.86137 (\pm 0.1762) f}.
\end{equation}
Then, we obtained a possible pre-merge mass ratio $q_{min} = 0.021 \pm 0.004$ with substituting $f = 100\%$. The correlation between $q$ and $f$ is weak regarding the orbital period. More observations and studies of ELMRCBs with $P > 0.5d$ are needed to discuss the effect of contact degree on searching for merging candidates.

In conclusion, we studied six ELMRCBs with $P > 0.5d$ for the first time, including photometric, spectroscopic, and orbital period analysis. J073647, J094123, and J163001 are all A-subtype and medium-contact binaries. J063344 and J075442 are A-subtype and deep-contact binaries. J105032 is a W-subtype and deep-contact binary. We determined their absolute parameters with the photometric solutions and the linear relationship between the orbital period and the semimajor axis. For the six targets, the primary components are all little-evolved stars, while all the secondary components evolve away from the main sequence. The initial mass, age, energy transfer parameter, instability parameters, and evolutionary state were estimated. More observations in the future, especially radial velocity observations for ELMRCBs, are needed to obtain more precise absolute parameters, which are critical for investigating the ultimate fate of contact binaries and searching for contact binary merger candidates.

\section*{Acknowledgements}
Great thanks to the referee for the very constructive comments and suggestions that greatly improved our manuscript.

This work is supported by National Natural Science Foundation of China (NSFC) (No. 12273018), and the Joint Research Fund in Astronomy (No. U1931103) under cooperative agreement between NSFC and Chinese Academy of Sciences(CAS), and by the Young Data Scientist Project of the National Astronomical Data Center, and by the Qilu Young Researcher Project of Shandong University, and by Young Scholars Program of Shandong University, Weihai (Nos. 20820171006) and by the Cultivation Project for LAMOST Scientific Payoff and Research Achievement of CAMS-CAS, and by the Chinese Academy of Science Interdisciplinary Innovation Team. The calculations in this work were carried out at Supercomputing Center of Shandong University, Weihai.

We acknowledge the support of the staff of the Xinglong 60cm telescope. This work was partially supported by National Astronomical Observatories, Chinese Academy of Sciences.

The spectral data were provided by Guoshoujing Telescope (the Large Sky Area Multi-Object Fiber Spectroscopic Telescope LAMOST) is a National Major Scientific Project built by the Chinese Academy of Sciences. Funding for the project has been provided by the National Development and Reform Commission. LAMOST is operated and managed by the National Astronomical Observatories, Chinese Academy of Sciences.

This work makes use of data provided by Transiting Exoplanet Survey Satellite mission (TESS), whose funding is supported by NASA Science Mission Directorate. We acknowledge the TESS team for its data source in this work.

We thank LasCumbres Observatory and its staff for their continued support of ASAS-SN. ASAS-SN is funded in part by the Gordon and Betty Moore Foundation through grant nos. GBMF5490 and GBMF10501 to the Ohio State University, and also funded in part by the Alfred P. Sloan Foundation grant no. G-2021-14192.

This paper makes use of observation from the Two Micron All Sky Survey(MASS), a joint project of the University of Massachusetts and the Infrared Processing and Analysis Center/California Institute of Technology. Funding of MASS is provided by the National Aeronautics and Space Administration and the National Science Foundation.

This paper makes use of data from the DR1 of the WASP data \citep{2010A&A...520L..10B} as provided by the WASP consortium, and computational resources supplied by the project "e-Infrastruktura CZ" (e-INFRA CZ LM2018140) supported by the Ministry of Education, Youth and Sports of the Czech Republic.

\section*{Data Availability}

The ASAS-SN data are publicly available at \href{https://asas-sn.osu.edu/variables/lookup}{https://asas-sn.osu.edu/variables/lookup}, and the CRTS data are publicly available at\href{http://crts.caltech.edu}{http://crts.caltech.edu}, and the SWASP data are publicly available at \href{https://wasp.cerit-sc.cz/search}{https://wasp.cerit-sc.cz/search}, and the TESS data are publicly available at \href{http://archive.stsci.edu/tess/bulk downloads.html}{http://archive.stsci.edu/tess/bulk downloads.html}, and the ZTF data are publicly available at\href{https://irsa.ipac.caltech.edu/cgi-bin/ZTF}{https://irsa.ipac.caltech.edu/cgi-bin/ZTF}.


\bibliographystyle{mnras}
\bibliography{reference.bib} 


\appendix

\setcounter{table}{0}
\renewcommand{\thetable}{A\arabic{table}}

\begin{table*}
\begin{threeparttable}
\centering
\caption{Our photometric data of XL60 and NEXT.}
\label{tab:our photometric data of XL60 and NEXT}
\begin{tabular}{cccccccc}
\hline
Target  &Telescope& HJD-V & $\Delta m$-V & HJD-R & $\Delta m$-R & HJD-I & $\Delta m$-I\\
        &         & (2400000+)  &  (mag) & (2400000+)  &  (mag) & (2400000+) &    (mag)    \\
\hline
J063344 &XL60     &59946.92843&	1.0195	   &59946.92931	&1.0432	       &59946.93002	&0.4555\\
        &         &59946.93092&	1.0218	   &59946.93181	&1.0463	       &59946.93252	&0.4777\\
        &         &59946.93342&	1.0283	   &59946.93430	&1.0571	       &59946.93501	&1.1053\\
        &         &59946.93591&	1.0285	   &59946.93680	&1.0496	       &59946.93751	&1.1034\\
        &         &59946.93841&	1.0450	   &59946.93929	&1.0555	       &59946.94001	&1.0928\\
\hline
\end{tabular}
\begin{tablenotes}[para,flushleft]
\item[1] This table is available in its entirety in machine-readable form in the online version of this article.\\
\end{tablenotes}
\end{threeparttable}
\end{table*}

\begin{table*}
\begin{threeparttable}
        \caption{Photometric solutions of J063344.}
	\centering
	\label{tab:Photometric solutions of J063344}  
	\begin{tabular}{lccc} 
\hline
J063344  	        &ASAS-SN	     &XL60            &ZTF \\
\hline
$T_1$(K)	&	$	7019	\pm	80	$	&	$	7025	\pm	84	$	&	$	7019	\pm	82	$	\\
$T_2$(K)	&	$	6839	\pm	102	$	&	$	6847	\pm	141	$	&	$	6884	\pm	125	$	\\
$q$($M_2$/$M_1$)	&	$	0.075 	\pm	0.002 	$	&	$	0.100 	\pm	0.001 	$	&	$	0.100 	\pm	0.001 	$	\\
i($^{\circ}$)	&	$	74.5 	\pm	1.0 	$	&	$	75.9 	\pm	0.3 	$	&	$	77.6 	\pm	0.5 	$	\\
$\Omega$	&	$	1.873 	\pm	0.010 	$	&	$	1.920 	\pm	0.004 	$	&	$	1.933 	\pm	0.007 	$	\\
$r_1$	&	$	0.605 	\pm	0.004 	$	&	$	0.598 	\pm	0.001 	$	&	$	0.592 	\pm	0.002 	$	\\
$r_2$	&	$	0.192 	\pm	0.022 	$	&	$	0.227 	\pm	0.008 	$	&	$	0.217 	\pm	0.012 	$	\\
$(L_2/L_1)_{r^{\prime}}$	&			-			&			-			&	$	0.125 	\pm	0.002 	$	\\
$(L_2/L_1)_V$	&	$	0.070 	\pm	0.004 	$	&	$	0.126 	\pm	0.001 	$	&			-			\\
$(L_2/L_1)_R$	&			-			&	$	0.129 	\pm	0.001 	$	&			-			\\
$(L_2/L_1)_I$	&			-			&	$	0.131 	\pm	0.001 	$	&			-			\\
$f$($\%$)	&	$	11.3 	\pm	20.6 	$	&	$	60.9 	\pm	5.6 	$	&	$	40.0 	\pm	10.7 	$	\\	
\hline
\end{tabular}
\begin{tablenotes}[para]
\item[1] The difference in contact degree $f$ between different telescopes may result from the different shapes of light curves.\\
\end{tablenotes}
\end{threeparttable}
\end{table*}

\begin{table*}
\begin{threeparttable}
    \centering
    \caption{Photometric solutions of J073647.}
    \label{tab:Photometric solutions of J073647}
\begin{tabular}{lccccc} 
        \hline
        J073647  	&ASAS-SN	     &CRTS             &NEXT &SuperWASP        &ZTF \\
        \hline
$T_1$(K)            &$6811\pm 22$    &$6818\pm 32$     &$6834 \pm 21$&$6808\pm 25$    &$6811\pm 21$\\
$T_2$(K)            &$6633\pm 33$    &$6573\pm 95$     &$6518 \pm 27$&$6581\pm 51$    &$6592\pm 25$\\
$q$($M_2$/$M_1$)    &$0.112\pm 0.002$&$0.112\pm 0.004$ &$0.140 \pm 0.001$&$0.118\pm 0.003$&$0.130\pm 0.001$\\
i($^{\circ}$)       &$77.1\pm 0.6$   &$76.8\pm 1.2$    &$82.0 \pm 0.2$&$74.7\pm 1.2$   &$79.1\pm 0.4$\\ 
$\Omega$            &$1.975\pm 0.008$&$1.984\pm 0.017$ &$2.050 \pm 0.003$&$1.990\pm 0.010$&$2.015\pm 0.004$\\
$r_1$               &$0.580\pm 0.003$&$0.575\pm 0.005$ &$0.563 \pm 0.001$&$0.577\pm 0.003$&$0.571\pm 0.001$\\
$r_2$               &$0.220\pm 0.012$&$0.219\pm 0.027$ &$0.238 \pm 0.004$&$0.225\pm 0.017$&$0.236\pm 0.008$\\
$(L_2/L_1)_{g^{\prime}}$ &-               &-                &$0.140 \pm 0.001$&-               &$0.143\pm 0.002$\\
$(L_2/L_1)_{r^{\prime}}$ &-               &-                &$0.149 \pm 0.001$&-               &$0.149\pm 0.001$\\
$(L_2/L_1)_{i^{\prime}}$ &-               &-                &$0.154 \pm 0.001$&-               &$0.152\pm 0.002$\\
$(L_2/L_1)_V$       &$0.130\pm 0.003$&$0.122\pm 0.008$ &-&-               &-\\
$(L_2/L_1)$ for SuperWASP&-               &-                &-&$0.132\pm 0.005$&-\\
$f$($\%$)           &$28.1 \pm11.0$&$14.9 \pm23.9$     &$29.5 \pm3.6$&$30.5 \pm12.6$  &$38.6 \pm4.8$\\
Cool Spot             &-&-&-&Star 2   &-\\
$\theta$($^{\circ}$)  &-&-&-&$91\pm{1}$&-\\
$\lambda$($^{\circ}$) &-&-&-&$32\pm{3}$&-\\
$r_{spot}$($^{\circ}$)&-&-&-&$45\pm{3}$&-\\
$T_{spot}$            &-&-&-&$0.802\pm{0.005}$&-\\
\hline
\end{tabular}
\end{threeparttable}
\end{table*}

\begin{table*}
\begin{threeparttable}
    \centering
    \caption{Photometric solutions of J075442.}
    \label{tab:Photometric solutions of J075442}
\begin{tabular}{lccc} 
        \hline
        J075442  &ASAS-SN	        &NEXT            &ZTF \\
        \hline
$T_1$(K)         &$6920\pm 252$    &$6921 \pm 251$    &$6920\pm 251$ \\   
$T_2$(K)         &$6777\pm 410$    &$6778 \pm 442$    &$6790\pm 450$  \\ 
$q$($M_2$/$M_1$) &$0.075\pm 0.004$ &$0.091 \pm 0.001$ &$0.079\pm 0.001$\\
i($^{\circ}$)    &$77.1\pm 2.0$    &$85.2 \pm 0.4$    &$77.3\pm 0.5$   \\
$\Omega$         &$1.850\pm 0.021$ &$1.914 \pm 0.001$ &$1.856\pm 0.001$\\
$r_1$            &$0.618\pm 0.008$ &$0.595 \pm 0.002$ &$0.616\pm 0.001$\\
$r_2$            &$0.205\pm 0.058$ &$0.208 \pm 0.012$ &$0.216\pm 0.003$\\
$(L_2/L_1)_{g^{\prime}}$ &-        &$0.110 \pm 0.001$ &$0.107\pm 0.002$\\       
$(L_2/L_1)_{r^{\prime}}$ &-        &$0.113 \pm 0.001$ &$0.109\pm 0.002$\\      
$(L_2/L_1)_{i^{\prime}}$ &-        &$0.115 \pm 0.001$ &$0.111\pm 0.002$\\       
$(L_2/L_1)_V$       &$0.100\pm 0.005$ &-              &-        \\       
$f$($\%$)           &$59.5 \pm40.4$&$28.3 \pm11.1$    &$72.2 \pm1.8$\\
\hline
\end{tabular}
\end{threeparttable}
\end{table*}

\begin{table*}
\begin{threeparttable}
    \centering
    \caption{Photometric solutions of J094123.}
    \label{tab:Photometric solutions of J094123}
\begin{tabular}{lccc} 
        \hline
        J094123  &ASAS-SN	        &NEXT            &ZTF \\
        \hline
$T_1$(K)         &$6870\pm 74$     &$6861 \pm 78$     &$6869\pm 78$ \\   
$T_2$(K)         &$6684\pm 99$     &$6757 \pm 136$    &$6703\pm 124$  \\ 
$q$($M_2$/$M_1$) &$0.129\pm 0.001$ &$0.135 \pm 0.001$ &$0.137\pm 0.002$\\
i($^{\circ}$)    &$80.6\pm 0.6$    &$80.5 \pm 0.2$    &$78.3\pm 0.4$   \\
$\Omega$         &$2.029\pm 0.006$ &$2.018 \pm 0.003$ &$2.039\pm 0.007$\\
$r_1$            &$0.565\pm 0.002$ &$0.572 \pm 0.001$ &$0.565\pm 0.002$\\
$r_2$            &$0.227\pm 0.006$ &$0.244 \pm 0.005$ &$0.238\pm 0.009$\\
$(L_2/L_1)_{g^{\prime}}$ &-        &$0.164 \pm 0.001$ &$0.154\pm 0.003$\\     
$(L_2/L_1)_{r^{\prime}}$ &-        &$0.167 \pm 0.001$ &$0.159\pm 0.002$\\    
$(L_2/L_1)_{i^{\prime}}$ &-        &$0.169 \pm 0.001$ &$0.161\pm 0.003$\\       
$(L_2/L_1)_V$       &$0.144\pm 0.003$ &-              &   -        \\       
$f$($\%$)           &$18.6 \pm7.2$ &$48.9 \pm3.7$     &$31.3 \pm7.8$\\
\hline
\end{tabular}
\end{threeparttable}
\end{table*}

\begin{table*}
\begin{threeparttable}
    \centering
    \caption{Photometric solutions of J105032.}
    \label{tab:Photometric solutions of J105032}
\begin{tabular}{lcccc} 
        \hline
        J105032  &ASAS-SN	        &NEXT&SuperWASP                     &ZTF \\
        \hline
$T_1$(K)         &$7145\pm 69$    &$7118 \pm 30$    & $7150\pm 55$     &$7152\pm 33$\\
$T_2$(K)         &$7138\pm 84$    &$7284 \pm 48$    & $7103\pm 100$    &$7096\pm 35$\\
$q$($M_2$/$M_1$) &$0.117\pm 0.003$&$0.141 \pm 0.001$& $0.120\pm 0.003$ &$0.132\pm 0.001$\\
i($^{\circ}$)    &$71.2\pm 0.9$   &$75.3 \pm 0.3$   & $75.2\pm 1.0$    &$73.2\pm 0.3$\\ 
$\Omega$         &$1.962\pm 0.012$&$2.055 \pm 0.004$& $1.994\pm 0.010$ &$2.008\pm 0.005$\\
$r_1$            &$0.588\pm 0.004$&$0.561 \pm 0.001$& $0.577\pm 0.003$ &$0.575\pm 0.001$\\
$r_2$            &$0.238\pm 0.026$&$0.236 \pm 0.006$& $0.230\pm 0.017$ &$0.245\pm 0.008$\\
$(L_2/L_1)_{g^{\prime}}$ &-&$0.198 \pm 0.002$&-&$0.169\pm 0.002$\\
$(L_2/L_1)_{r^{\prime}}$ &-&$0.190 \pm 0.002$&-&$0.170\pm 0.002$\\
$(L_2/L_1)_{i^{\prime}}$ &-&$0.186 \pm 0.002$&-&$0.171\pm 0.002$\\
$(L_2/L_1)_V$       &$0.160\pm 0.006$&-&-&-\\
$(L_2/L_1)$ for SuperWASP&-&-&$0.152\pm 0.006$&-\\
$f$($\%$)           &$63.6 \pm16.4$&$24.9 \pm4.4$&$32.5 \pm12.2$&$52.9 \pm5.3$\\
\hline
\end{tabular}
\end{threeparttable}
\end{table*}

\begin{table*}
\begin{threeparttable}
        \caption{Photometric solutions of J163001.}
	\centering
	\label{tab:Photometric solutions of J163001}  
	\begin{tabular}{lcccc} 
        \hline
        J163001  &ASAS-SN	        &NEXT           &SuperWASP          &ZTF \\
        \hline
$T_1$(K)         &$6834\pm 55$    &$6831 \pm 43$    & $6831\pm 45$      &$6826\pm 50$\\
$T_2$(K)         &$6482\pm 65$    &$6530 \pm 62$    & $6534\pm 60$      &$6571\pm 56$\\
$q$($M_2$/$M_1$) &$0.080\pm 0.002$&$0.084 \pm 0.001$& $0.091\pm 0.005$  &$0.086\pm 0.001$\\
i($^{\circ}$)    &$79.3\pm 0.8$   &$83.8 \pm 0.3$   & $81.4\pm 0.4$     &$83.0\pm 0.5$\\ 
$\Omega$         &$1.871\pm 0.005$&$1.884 \pm 0.003$& $1.904\pm 0.005$  &$1.891\pm 0.004$\\
$r_1$            &$0.609\pm 0.002$&$0.605 \pm 0.001$& $0.599\pm 0.002$  &$0.604\pm 0.001$\\
$r_2$            &$0.207\pm 0.017$&$0.210 \pm 0.007$& $0.214\pm 0.009$  &$0.209\pm 0.012$\\
$(L_2/L_1)_{g^{\prime}}$ &-&$0.094 \pm 0.001$&-&$0.098\pm 0.003$\\
$(L_2/L_1)_{r^{\prime}}$ &-&$0.100 \pm 0.001$&-&$0.103\pm 0.002$\\
$(L_2/L_1)_{i^{\prime}}$ &-&$0.103 \pm 0.001$&-&$0.106\pm 0.002$\\
$(L_2/L_1)_V$       &$0.089\pm 0.003$&-&-&-\\
$(L_2/L_1)$ for SuperWASP&-&-&$0.103\pm 0.001$&-\\
$f$($\%$)           &$44.5 \pm9.8$&$45.9 \pm5.3$&$43.8 \pm8.8$&$41.8 \pm6.2$\\
        \hline
	\end{tabular}
\end{threeparttable}
\end{table*}

\bsp	
\label{lastpage}
\end{CJK}
\end{document}